# Stability and flow fields' structure for interfacial dynamics with interfacial mass flux

Daniil V. Ilyin (1), Yasuhide Fukumoto (2), Sergei I. Anisimov (3), William A. Goddard III (1),
Snezhana I. Abarzhi (4)*

California Institute of Technology, United States of Americana (1); Kyushu University, Japan (2);
Landau Institute for Theoretical Physics of the Academy of Sciences, Russia (2);
The University of Western Australia, Australia (4)

* corresponding author, snezhana.abarzhi@gmail.com

We analyze from a far field the evolution of an interface that separates ideal incompressible fluids of different densities and has an inertial mass flux. We develop and apply the general matrix method to rigorously solve the boundary value problem involving the governing equations in the fluid bulk and the boundary conditions at the interface and at the outside boundaries of the domain. We find the fundamental solutions for the linearized system of equations, and analyze the interplay of interface stability with flow fields' structure, by directly linking rigorous mathematical attributes to physical observables. New mechanisms are identified of the interface stabilization and destabilization. We find that interfacial dynamics is stable when it conserves the fluxes of mass, momentum and energy; the stabilization is due to inertial effects causing small oscillations of the interface velocity. In the classic Landaus' dynamics, the postulate of perfect constancy of the interface velocity leads to the development of the Landau-Darrieus instability. This destabilization is also associated with the imbalance of the perturbed energy at the interface, in full consistency with the classic results. We identify extreme sensitivity of the interface dynamics to the interfacial boundary conditions, including formal properties of fundamental solutions and qualitative and quantitative properties of the flow fields. This provides new opportunities for studies, diagnostics, and control of multiphase flows in a broad range of processes in nature and technology.

1.   **Introduction**

Interfacial mixing and transport control a broad range of natural and technological processes in fluids, plasmas, and materials, from celestial to atomistic scales [Abarzhi et al. 2018, Abarzhi et al. 2013, Zeldovich & Raiser 2002, Landau & Lifshitz 1987]. Examples include supernovae and fusion, planetary convection and reactive flows, formation of phases in super-critical fluids and material transformations under impact [Abarzhi 2010, Azechi et al. 2007, Bell et al. 2004, Remington et al. 2000, Stein & Norlund 2000, Bodner et al. 1998, Rast 1998, Arnett 1996, Prosperetti & Plesset 1984]. In realistic circumstances material transport is often characterized by sharply and rapidly changing flow fields and by relatively small effects of dissipation and diffusion. This may lead to formation of discontinuities (interfaces) separating the flow non-uniformities (phases) at continuous (macroscopic) scales [Abarzhi et al. 2015, Anisimov et al. 2013, Abarzhi 2010, Kadau et al. 2010, Meshkov 2006, Zeldovich & Raizer 2002]. Here we consider from a far field the evolution of a hydrodynamic discontinuity that separates ideal incompressible fluids of different densities and has the interfacial mass flow [Landau & Lifshitz 1987]. We develop the general matrix method to rigorously solve the linearized boundary value problem and directly link the interface stability to the flow fields' structure. We find extreme sensitivity of the flow dynamics to the interfacial boundary conditions, and identify new mechanisms of the interface stabilization and destabilization, thus providing new opportunities for studies of interfacial dynamics in a broad range of processes in nature and technology [Abarzhi et al. 2018].

In hydrodynamics, the evolution of an interface with mass flow across it is usually studied in premixed combustion and in fluid vaporization [Ilyin et al. 2018, Prosperretti 2017, Peters 2000, Mayo et al 1990, Sivashinsky 1983, Williams 1965, Zeldovich 1944]. Similar problems are considered in plasmas (stability of ablation fronts in inertial confinement fusion), astrophysics (thermonuclear flashes on surface of compact stars), material science (melting and evaporation of material under high pressure and high strain rate), industrial applications (atomization of liquid jets in scramjets) [Hurricane et al. 2014, Anisimov et al. 2013, Abarzhi 2010, Kadau et al 2010, Arnett et al. 2009, Azechi et al. 2007, Marmottant & Villermaux 2004, Bell et al. 2004, Piriz & Portogues 2003, Cherkov & Yakhot 1998, Wu et al. 1997, Remington et al. 1995, Sanz 1994]. The classic theoretical framework for the problem has been developed by Landau 1944. It analyzes the dynamics of a discontinuous interface separating ideal incompressible fluids of different densities. By balancing at the interface the fluxes of mass and momentum and by implementing a special condition for the perturbed mass flux, Landau 1944 finds that the interface is unconditionally unstable: small perturbations of the interface grow exponentially with time leading to the development of the Landau-Darrieus instability (LDI).

To connect the classical framework [Landau 1944] to realistic environments, several theoretical approaches have been developed. In high energy density plasmas, significant departures have been detected between the growth-rates of ablative Rayleigh-Taylor (RT) and the Landau-Darrieus (LD) instabilities in gravity field [Hurricane et al. 2014, Sanz 1944, Kull & Anisimov 1985]. In reactive and super-critical fluids, stabilizing influences have been found of dissipation, diffusion, and finite interface thickness on the interfacial dynamics at small scales [Peters 2010, Peters 2000, Clavin 1985, Prosperetti & Plesset 1984, Williams 1971, Markstein 1951, Zeldovich 1944]. Recently, it has been found that the interface stability is sensitive to the flux of energy fluctuations produced by the perturbed interface [Abarzhi et al. 2015]. While these theories and models successfully expanded the classical framework [Landau 1944, Zeldovich 1944] to explain the observations [Hurricane et al. 2014, Aglitsky et al. 2010. Arnett et l. 2009, Azechi et al. 2007, Bell et al. 2004, Remington et al. 1995, Peters 2000, Williams 1971, Buckmaster 1979, Markstein 1951, Zeldovich 1944], some fundamental challenges remain.

First, the general theoretical framework [Landau 1944] describes the evolution of a phase boundary and is relevant to phenomena far beyond the processes with gradually changing flow fields [Hurricane et al. 2014, Anisimov et al. 2013, Abarzhi 2010, Arnett et al. 2009, Kadau et al. 2010, Meshkov 2006]. We still have to understand whether the interface is stable when the flow quantities experience sharp changes and the effects of dissipation and diffusion are negligible.

Second, direct diagnostics of various physical effects on dynamics of a multi-phase flow requires detailed information of the interface structure. Such information is often a challenge to directly obtain in experiments and simulations [Hurricane et al. 2014, Azechi et al. 2007, Bell et al. 2004, Peters 2000, Rast 1998, Remington et al. 1995, Wu et al. 1997, Clavin 1985, Williams 1971]. Besides, one usually observes the flow evolution from a far field and at time scales and length scales that are substantially greater than characteristic scales induced by, e.g., diffusion, dissipation, and finite thickness [Abarzhi et al. 2015, Kadau et al. 2010, Remington et al. 2000, Landau & Lifshitz 1987, Sivashinsky 1983]. Thus, there is a need to directly connect the interface stability to the flow fields' structure, and to associate the mechanism of (de)stabilization of the interfacial dynamics with qualitative and quantitative properties of the vector and scalar fields of the flow [Abarzhi et al. 2018].

Third, there are still a few formal questions to clarify in the classic theory of Landau 1944 for ideal incompressible fluids. For instance, the Landau's framework involves consideration of four independent variables: velocity potential of the heavy fluid, velocity potential of the light fluid, vortical potential of the light fluid, and interface perturbation [Landau & Lifshitz 1987, Landau 1944]. These variables obey the principles of conservation of mass and normal and tangential components of momentum and a special condition for the perturbed mass flow at the interface. However, only two eigenvalues of the linearized dynamics are usually derived [Peters 2010, Landau & Lifshitz 1987,

Zeldovich 1994, Clavin 1985, Williams 1971, Landau 1944]. These eigenvalues are the unstable positive real eigenvalue (that leads to theLDI) and its stable negative counterpart. It is unclear whether the other eigenvalues exist, and, for each of the fundamental solutions, what the typical flow fields are, what the link of these fields is to the interface stability, and what the effect of fluctuations is on the flow stability and fields. These questions should be answered to quantify the flow sensitivity to boundary conditions at the interface.

In this work, we address these challenges by systematically studying in a far field approximation the evolution of the discontinuous interface separating ideal incompressible fluids of different densities and having the inertial mass flux. By expanding the classic framework [Landau & Lifshitz 1987], we develop and apply the general matrix method to solve the boundary value problem involving the governing equations in the fluid bulk and the boundary conditions at the interface and at the outside boundaries of the domain. We find formal fundamental solutions of the dynamics, analyze the interplay of interface stability with flow fields' structure, and directly link rigorous mathematical attributes to physical observables [Abarzhi et al. 2015]. New mechanisms are identified of the interface stabilization and destabilization. We find that the interfacial dynamics is stable when it conserves the fluxes of mass, momentum and energy; the stabilization is due to inertial effects, leading to small oscillations of the interface velocity. This mechanism is absent in the classic Landau's dynamics due to the postulated constancy of the interface velocity, causing the LDI to develop. The interface destabilization is also associated with the imbalance of the perturbed energy at the interface, which is fully consistent with the classic results [Landau 1944]. We identify the extreme sensitivity of the interface dynamics to the interfacial boundary conditions, including formal properties of fundamental solutions and qualitative and quantitative properties of the flow fields, thus providing new opportunities for diagnostics and control of multiphase flows in a broad range of applications [Abarzhi et al. 2018].

## 2. Method

In this section we outline the equations governing the interface dynamics, formulate the boundary value problem, and develop the general matrix method for the problem solution.

### 2.1. Governing equations

The flow dynamics is governed by the conservation of mass, momentum, and energy. In an inertial frame of reference

$$\frac{\partial \rho}{\partial t}+\frac{\partial}{\partial x_i}\rho v_i = 0, \quad \frac{\partial \rho v_i}{\partial t}+\sum_{j=1}^{3}\frac{\partial}{\partial x_i}\rho v_i v_j + \frac{\partial P}{\partial x_i} = 0, \quad \frac{\partial E}{\partial t}+\frac{\partial (E+P)v_i}{\partial x_i}=0 \qquad (1)$$

where $x_i$ are the spatial coordinates with $(x_1, x_2, x_3) = (x, y, z)$, $t$ is time, $(\rho, \mathbf{v}, P, E)$ are the fields of density $\rho$, velocity $\mathbf{v}$, pressure $P$ and energy $E = \rho(e + \mathbf{v}^2/2)$, and $e$ is the specific internal energy in its physics definition [Landau & Lifshitz 1987]. The specific internal energy refers to the energy per unit mass that is contained in the system, while excluding the kinetic energy of motion of the system as a whole and the potential energy of the system as a whole due to external force fields [Landau & Lifshitz 1987]. The inertial frame of reference refers to a frame of reference moving with a constant velocity $\mathbf{V}_0 = (0, 0, V_0)$ [Landau & Lifshitz 1987].

We introduce a continuous local scalar function $\theta(x, y, z, t)$, whose derivatives $\dot{\theta}$ and $\nabla\theta$ exist (with dot marking a partial time-derivative), such that $\theta = 0$ at the interface, and the heavy (light) fluid is located in the region $\theta > 0$ ($\theta < 0$) [Abarzhi et al. 2015]. In the bulk of the heavy (light) fluid the flow fields are $(\rho, \mathbf{v}, P, E) \to (\rho, \mathbf{v}, P, E)_{h(l)}$, and they are marked with subscript $h(l)$. We represent the flow fields in the entire domain as $(\rho, \mathbf{v}, P, E) = (\rho, \mathbf{v}, P, E)_h H(\theta) + (\rho, \mathbf{v}, P, E)_l H(-\theta)$ by using the Heaviside step-function $H(\theta)$ [Abarzhi et al. 2018]. The balances of fluxes of mass and normal and tangential components of momentum and energy across the interface obey the boundary conditions [Landau & Lifshitz 1987]:

$$[\tilde{\mathbf{j}} \cdot \mathbf{n}] = 0, \quad \left[\left(P + (\tilde{\mathbf{j}} \cdot \mathbf{n})^2/\rho\right)\mathbf{n}\right] = 0, \quad \left[(\tilde{\mathbf{j}} \cdot \mathbf{n})(\tilde{\mathbf{j}} \cdot \boldsymbol{\tau})/\rho)\boldsymbol{\tau}\right] = 0, \quad \left[(\tilde{\mathbf{j}} \cdot \mathbf{n})(W + \tilde{\mathbf{j}}^2/2\rho^2)\right] = 0 \qquad (2)$$

Here $[...] = 0$ denotes the jump of functions across the interface; $\mathbf{n}$ and $\boldsymbol{\tau}$ are the unit normal and tangential vectors at the interface with $\mathbf{n} = \nabla\theta/|\nabla\theta|$ and $(\mathbf{n} \cdot \boldsymbol{\tau}) = 0$; $\tilde{\mathbf{j}} = \rho(\mathbf{n}\dot{\theta}/|\nabla\theta| + \mathbf{v})$ is the mass flux across the moving interface; and $W = e + P/\rho$ is the specific enthalpy in its physics definition [Landau & Lifshitz 1987]. Particularly, the value of specific enthalpy $W$ includes the enthalpy of formation [Landau & Lifshitz 1987].

For readers' convenience, we provide some details on how the interfacial boundary conditions are derived from the governing equations. As a sample case, we consider first of these conditions – the condition for mass conservation; other conditions are derived likewise [Abarzhi et al. 2018].

In the entire domain the density of the fluid is represented as $\rho = \rho_h H(\theta) + \rho_l H(-\theta)$ leading to

$$\frac{\partial \rho}{\partial t} = \left(\frac{\partial \rho_h}{\partial t}\right) H(\theta) + \left(\frac{\partial \rho_l}{\partial t}\right) H(-\theta) + \delta(\theta)(\rho_h - \rho_l)\left(\frac{\partial \theta}{\partial t}\right)$$

Here we account for that $\partial H(\theta)/\partial \theta = \delta(\theta)$ and $\delta(\theta) = \delta(-\theta)$, with $\delta(\theta)$ being the Dirac delta-function. The other term is calculated as

$$\frac{\partial \rho v_i}{\partial x_i} = \left(\frac{\partial (\rho v_i)_h}{\partial x_i}\right) H(\theta) + \left(\frac{\partial (\rho v_i)_l}{\partial x_i}\right) H(-\theta) + \delta(\theta)\left((\rho v_i)_h - (\rho v_i)_l\right)\left(\frac{\partial \theta}{\partial x_i}\right)$$

We substitute these expressions in the equation for mass conservation, introduce the unit normal and tangential vectors at the interface $\mathbf{n}$ and $\mathbf{\tau}$ as

$$\frac{\partial \theta}{\partial x_i} = \nabla \theta = \mathbf{n}|\nabla \theta|, \quad \mathbf{n} = \frac{\nabla \theta}{|\nabla \theta|}, \quad (\mathbf{n} \cdot \mathbf{\tau}) = 0$$

and find

$$\frac{\partial \rho}{\partial t} + \frac{\partial \rho v_i}{\partial x_i} = \left(\frac{\partial \rho}{\partial t} + \frac{\partial \rho v_i}{\partial x_i}\right)_h H(\theta) + \left(\frac{\partial \rho}{\partial t} + \frac{\partial \rho v_i}{\partial x_i}\right)_l H(-\theta) + \delta(\theta)|\nabla \theta|\left[\tilde{\mathbf{j}} \cdot \mathbf{n}\right] = 0$$

The expression is further reduced to the conditions in the fluids' bulk and at the interface

$$\left(\frac{\partial \rho}{\partial t} + \frac{\partial \rho v_i}{\partial x_i}\right)_h = 0, \quad \left(\frac{\partial \rho}{\partial t} + \frac{\partial \rho v_i}{\partial x_i}\right)_l = 0, \quad \left[\tilde{\mathbf{j}} \cdot \mathbf{n}\right] = 0$$

We consider a spatially extended two-dimensional flow propagating in the $z$ direction, being periodic in the $x$ direction, $x + \lambda \to x$, and having no motion in the $y$ direction, Figure 1. In the inertial frame of reference, the boundary conditions at the outside boundaries of the domain are

$$\mathbf{v}_h\big|_{z \to -\infty} = \mathbf{V}_h = (0, 0, V_h), \qquad \mathbf{v}_l\big|_{z \to +\infty} = \mathbf{V}_l = (0, 0, V_l), \tag{3}$$

with constant $V_{h(l)}$, Figure 1. Initial conditions are the initial flow fields at the interface. They define the characteristic length-scale and time-scale of the dynamics [Zeldovich 1944].

Note that the governing equations and the boundary conditions Eqs.(1-3) are derived in an inertial frame of reference that moves with a constant velocity $\mathbf{V}_0$, Figure 1. Some discussion is required to better understand the relation between the velocity of the inertial frame of reference $\mathbf{V}_0$ and the velocity $\tilde{\mathbf{V}}$ of the interface as a whole. For a planar and steady interface the interface velocity $\tilde{\mathbf{V}}$ is constant, $\tilde{\mathbf{V}} = \tilde{\mathbf{V}}_0$, and it may be equal to the velocity $\mathbf{V}_0$ of the inertial frame of reference, $\tilde{\mathbf{V}}_0 = \mathbf{V}_0$. In general case of a non-steady non-planar interface $\tilde{\mathbf{V}} \neq \tilde{\mathbf{V}}_0$, the interface velocity $\tilde{\mathbf{V}}$ and the velocity the inertial frame of reference are distinct values, $\tilde{\mathbf{V}} \neq \tilde{\mathbf{V}}_0$, and the interface velocity $\tilde{\mathbf{V}}$ obeys the condition $\tilde{\mathbf{V}}\mathbf{n} = -\mathbf{v}\mathbf{n}\big|_{\theta \to 0^+} = -(\tilde{\mathbf{j}}/\rho)\mathbf{n}\big|_{\theta \to 0^+}$ [Peters 2011, Landau & Lifshitz 1987]. There is also an important particular case, in which the velocity of the inertial frame of references $\mathbf{V}_0$ is the same as the velocity of the steady planar interface, $\mathbf{V}_0 = \tilde{\mathbf{V}}_0$, and in which the heavy fluid is at rest in the laboratory frame of reference. This leads to $\widetilde{\mathbf{V}}_0 = -\mathbf{V}_h$ with $\widetilde{\mathbf{V}}_0 = \mathbf{V}_0 = (0,0,-V_h)$ [Peters 2011, Landau & Lifshitz 1987, Clavin 1985, Williams 1971].

In our general theoretical framework, we consider the velocity $\mathbf{V}_0$ as the constant velocity of the inertial frame of reference. To study the effect of the interfacial boundary conditions on the interface velocity, we set the velocity of the steady planar interface equal to the velocity of the inertial frame of reference, as $\widetilde{\mathbf{V}}_0 = \mathbf{V}_0$.

### 2.2. Linearized dynamics

We define the local scalar function $\theta$ as:

$$\theta = -z + z^*(x,t),\ \dot{\theta} = \partial z^*/\partial t,\ \nabla\theta = (\partial z^*/\partial x, 0, -1),\ |\nabla\theta| = \sqrt{1 + (\partial z^*/\partial x)^2} \qquad (4).$$

We slightly perturb the flow fields as $\widetilde{\mathbf{j}} = \mathbf{J} + \mathbf{j}$, $P = P_0 + p$, $W = W_0 + w$, $|\mathbf{j}| << |\mathbf{J}|$, $|p| << |P_0|$, $|w| << |W_0|$, and $\mathbf{v} = \mathbf{V} + \mathbf{u}$ with $|\mathbf{u}| << |\mathbf{V}|$. We slightly perturb the interface, $|\dot{\theta}/|\nabla\theta|| << |\mathbf{V}|$ and $|\partial z^*/\partial x| << 1$. leading to $\mathbf{n} = \mathbf{n}_0 + \mathbf{n}_1$ and $\boldsymbol{\tau} = \boldsymbol{\tau}_0 + \boldsymbol{\tau}_1$, where $|\mathbf{n}_1| << |\mathbf{n}_0|$ and $|\boldsymbol{\tau}_1| << |\boldsymbol{\tau}_0|$. In the laboratory frame of reference, the perturbed velocity of the interface is $\widetilde{\mathbf{V}} = \widetilde{\mathbf{V}}_0 + \widetilde{\mathbf{v}}$, with $|\widetilde{\mathbf{v}}| << |\widetilde{\mathbf{V}}_0|$.

To the leading order, the mass flux is $\mathbf{J} = \rho\mathbf{V}$. The flow fields are uniform in the bulk, $(\rho,\mathbf{v},P,W)_{h(l)} = (\rho,\mathbf{V},P_0,W_0)_{h(l)}$ and obey conditions at the outside boundaries of the domain. The interface is planar, $\theta = -z$, $\mathbf{n} = \mathbf{n}_0$ and $\boldsymbol{\tau} = \boldsymbol{\tau}_0$, with $\mathbf{n}_0 = (0,0,-1)$ and $\boldsymbol{\tau}_0 = (1,0,0)$. The normal component of mass flux is $J_n = \mathbf{J}\cdot\mathbf{n}_0$. The boundary conditions at the interface are:

$$[J_n] = 0,\ [(P_0 + J_n^2/\rho)\mathbf{n}_0] = 0,\ [J_n((\mathbf{J}\cdot\boldsymbol{\tau}_0)/\rho)\boldsymbol{\tau}_0] = 0,\ [J_n(W_0 + \mathbf{J}^2/2\rho^2)] = 0. \qquad (5)$$

The boundary conditions Eqs.(2,3,5) are general and are applicable for either compressible or incompressible dynamics [Landau & Lifshitz 1987].

Here we consider the limiting case of incompressible dynamics, in which the speed(s) of sound of the fluid(s) is substantially greater than other velocity scales, and the effect of density variations is negligible. For ideal gases, the speed of sound is $c = \sqrt{\gamma P/\rho}$, where $\gamma$ is the fluid's adiabatic index. For $(c/V)_{h(l)} \to \infty$, the values approach $(P_0 + J_n^2/\rho)_{h(l)} \to (P_0)_{h(l)}$ and $(W_0 + \mathbf{J}^2/2\rho^2)_{h(l)} \to (W_0)_{h(l)}$. Eqs.(5) is transformed to

$$[J_n] = 0,\ [P_0\mathbf{n}_0] = 0,\ [J_n((\mathbf{J}\cdot\boldsymbol{\tau}_0)/\rho)\boldsymbol{\tau}_0] = 0,\ [J_nW_0] = 0 \qquad (6)$$

By introducing temperature $(T)_{h(l)}$ and the specific heat at constant pressure $(c_P)_{h(l)}$, we present the physics enthalpy as $(W_0)_{h(l)} = (\overline{W}_0 + c_P T)_{h(l)}$. The enthalpy $\overline{W}_0$ is often used in eingneering applications, and it has a jump at the interface, with $[\overline{W}_0] = \overline{Q}$, where $\overline{Q}$ is the heat release per unit mass. In reactive fluids, the value $\overline{Q}$ is considered as the specific heat release of a chemical reaction at absolute zero temperature [Landau & Lifshitz 1987].

Note the difference between the definitions of enthalpy $W_0$ in physics sense accounting for the enthalpy of formation, and the enthalpy $\overline{W}_0$ commonly applied in reactive fluids and engineering. Their difference is discussed in details in [Landau & Lifshitz 1987]. In full consistency with this discussion, in the incompressible limit the physics enthalpy $W_0$ is continuous at the interface, $[W_0] = 0$, whereas the enthalpy $\overline{W}_0$ is discontinuous at the interface and its jump at the interface is $[\overline{W}_0] = \overline{Q}$.

To the first order, the boundary conditions at the interface take the form

$$[j_n] = 0, \; [(p + 2J_n j_n/\rho)\mathbf{n}_0] = 0, \; [J_n(\mathbf{J}\cdot\boldsymbol{\tau}_1 + \mathbf{j}\cdot\boldsymbol{\tau}_0)/\rho] = 0, \; [J_n(w + (\mathbf{J}\cdot\mathbf{j})/\rho^2)] = 0 \quad (7)$$

where the first order perturbation of the mass flux is $\mathbf{j} = \rho(\mathbf{u} + \mathbf{n}_0 \dot{\theta})$; its normal component is $j_n = \mathbf{j}\cdot\mathbf{n}_0$. The first order terms for the normal and tangential vectors of the interface are $\mathbf{n}_1 = (\partial z^*/\partial x, 0, 0)$ and $\boldsymbol{\tau}_1 = (0, 0, \partial z^*/\partial x)$.

In addition to the boundary conditions at the interface, the governing equations in the bulk and at the outside boundaries of the domain to the first order are

$$\nabla\cdot\mathbf{u}_{h(l)} = 0, \; \dot{\mathbf{u}}_{h(l)} + (\mathbf{V}_{h(l)}\cdot\nabla)\mathbf{u}_{h(l)} + \nabla p_{h(l)}/\rho_{h(l)} = 0, \; \mathbf{u}_h|_{z\to-\infty} = 0, \; \mathbf{u}_l|_{z\to+\infty} = 0 \quad (8)$$

In the laboratory frame of reference, the interface velocity is $\widetilde{\mathbf{V}} = \widetilde{\mathbf{V}}_0 + \widetilde{\mathbf{v}}$, such that the velocity perturbation is $\widetilde{\mathbf{v}}$ and

$$\widetilde{\mathbf{V}} = \widetilde{\mathbf{V}}_0 + \widetilde{\mathbf{v}}, \; |\widetilde{\mathbf{v}}| \ll |\widetilde{\mathbf{V}}_0|, \; \widetilde{\mathbf{v}}\mathbf{n}_0 = -(\mathbf{u}_h \mathbf{n}_0 + \dot{\theta})|_{\theta=0^+} \quad (9)$$

### 2.3 Fundamental solutions
#### 2.3.1 Solution structure

We seek solutions for the boundary value problem Eqs.(7,8) with the structure, in which the velocity field of the heavy fluid is potential, and the velocity field of the light fluid has both potential and vortical components [Landau & Lifshitz 1987]:

$$\mathbf{u}_h = \nabla \Phi_h, \quad \mathbf{u}_l = \nabla \Phi_l + \nabla \times \mathbf{\Psi}_l, \tag{10.1}$$

The rationale for this structure is the following: In the sub-domain of the heavy fluid, the flow is potential in accordance with the Kelvin theorem. In the sub-domain of the light fluid, the flow can be a superposition of the potential and vortical components. This structure is applied in theoretical framework [Landau 1944], and agrees with observations [Peters 2000, Zeldovich 1944]. It is expected that the solution structure is established for any initial conditions [Landau 1944, Zeldovich 1944].

The scalar and vector potentials of the fluid and the interface perturbation are:

$$\Phi_h = \Phi \, exp(ikx + kz + \Omega t), \; \Phi_l = \widetilde{\Phi} \, exp(ikx - kz + \Omega t),$$

$$\mathbf{\Psi}_l = (0, \Psi_l, 0), \; \Psi_l = \Psi \, exp(ikx - \widetilde{k}z + \Omega t), \; z^* = Z exp(ikx + \Omega t) \tag{10.2}$$

Here $\Omega$ is the growth-rate, i.e., the characteristic frequency or the eigenvalue of the system of linear differential equations, $k = 2\pi/\lambda$ is the wavevector, and $\lambda$ is the wavelength (spatial period). The pressure perturbations $p_{h(l)}$ and the vortex length-scale $\widetilde{\lambda} = 2\pi/\widetilde{k}$ are identified from the equations Eqs.(9) as follows:

$$\nabla(\dot{\Phi}_{h(l)} + V_{h(l)}(\partial \Phi_{h(l)}/\partial z) + p_{h(l)}/\rho_{h(l)}) = 0, \quad (\partial/\partial t + V_l(\partial/\partial z))(\nabla \times \mathbf{\Psi}_l) = 0, \tag{10.3}$$

leading to

$$p_{h(l)} = -\rho_{h(l)}(\dot{\Phi}_{h(l)} + V_{h(l)} \partial \Phi_{h(l)}/\partial z), \quad \widetilde{k} = \Omega/V_l \tag{10.4}$$

Note that the vortical field does not contribute to pressure [Landau & Lifshitz 1987].

In the system of equations Eqs.(7,8), the initial conditions set the characteristic length-scale $1/k$ and time-scale $1/kV_h$. We use dimensionless values for the growth-rate $\omega = \Omega/kV_h$ and for the density ratio $R = \rho_h/\rho_l$ with $R \geq 1$. This leads to $V_l/V_h = R$ and $\widetilde{k}/k = \omega/R$.

### 2.3.2 Eigenvalues, eigenvectors and fundamental solutions

The system of differential equations governing the interface dynamics Eqs.(7,8) is reduced to the linear system $\mathbf{Mr} = 0$, where vector $\mathbf{r}$ is $\mathbf{r} = (\Phi_h, \Phi_l, V_h z^*, \Psi_l)^T$ in dimensional units, and the matrix $\mathbf{M}$ is defined by the boundary conditions at the interface. In the dimensionless form, its elements are some functions of the growth-rate (eigenvalue) $\omega$ and the density ratio $R$, $\mathbf{M} = \mathbf{M}(\omega, R)$. Traditionally, the linear system of differential equations Eqs.(7,8) is solved by applying the compatibility condition: Some eigenvalues $\omega$ are derived, and the (in)stability of the associated solutions is identified by using the criterion $Re[\omega(R)] < 0 \, (>0)$ [Peters 2010, Peters 2000, Landau & Lifshitz 1987, Clavin 1985, Williams 1971, Landau 1944, Zeldovich 1944]. In this work, we proceed in a more formal way. Particularly, we develop and apply the general matrix method to analyze the interfacial dynamics, Eqs.(1-10). We find the fundamental solutions for the system of linear ordinary differential equations, including their eigenvalues and eigenvectors, and analyze for each of these solutions the stability and the structure of the flow fields. This formal analysis is required to better understand the interfacial dynamics. To our knowledge, it has not been done before.

We derive the eigenvalues $\omega_i$ by using the condition $det\,M(\omega_i, R) = 0$ and identify the associated eigenvectors $\mathbf{e}_i$ by reducing the matrix $M(\omega_i, R)$ to row-echelon form. Matrix M is 4×4. For a non-degenerate 4×4 matrix, the expected rank is 4, and the number of eigenvalues $\omega_i$ and associated eigenvectors $\mathbf{e}_i$ is 4, with $i = 1...4$, corresponding to 4 degrees of freedom and 4 independent variables obeying 4 equations, Eqs.(6,10).

Solution $\mathbf{r}$ for system given by Eqs.(6,10) is a linear combination of fundamental solutions $\mathbf{r}_i$

$$\mathbf{r} = \sum_{i=1}^{4} C_i \mathbf{r}_i \tag{11}$$

with $\mathbf{r}_i = \left( \Phi_i e^{ikx+kz+\omega_i kV_h t}, \widetilde{\Phi}_i e^{ikx-kz+\omega_i kV_h t}, V_h Z_i e^{ikx+\omega_i kV_h t}, \Psi_i e^{ikx-\tilde{k}z+\omega_i kV_h t} \right)^T$ in the dimensional units. In the dimensionless units, by introducing $\varphi_i = \Phi_i/(V_h/k)$, $\widetilde{\varphi}_i = \widetilde{\Phi}_i/(V_h/k)$, $\bar{z} = kZ$, $\psi_i = \Psi_i/(V_h/k)$ and with $kx \to x, kz \to z, kV_h t \to t$, the solution is $\mathbf{r}_i = \left( \varphi_i e^{ix+z+\omega_i t}, \widetilde{\varphi}_i e^{ix-z+\omega_i t}, \bar{z}_i e^{ix+\omega_i t}, \psi_i e^{ix-(\tilde{k}/k)z+\omega_i t} \right)^T$ and $\mathbf{r}_i = \mathbf{r}_i(\omega_i, \mathbf{e}_i)$, with $\mathbf{e}_i = (\varphi_i, \widetilde{\varphi}_i, \bar{z}_i, \psi_i)^T$ and with $C_i$ being integration constants.

The linear system $M\mathbf{r} = 0$ with $M = M(\omega, R)$ results from a linear system $P\dot{\mathbf{r}} = S\mathbf{r}$, where P, S are 4×4 matrixes, $P = P(R), S = S(R)$, under assumption that vector $\mathbf{r}$ varies in time as $\mathbf{r} \sim e^{\omega t}$. This leads to $M = (S - \omega P)$. In a non-degenerate case, the inverse $P^{-1}$ exists, and the system $P\dot{\mathbf{r}} = S\mathbf{r}$ can be reduced to a standard form $\dot{\mathbf{r}} = P^{-1}S\mathbf{r}$. The eigenvalues $\omega_i$ of the dynamics can then be found from condition $det(P^{-1}S - \omega I) = 0$, $\omega = \{\omega_i\}$, where the unit matrix is I, and index $i$ marks the independent degrees of freedom. Equations $det(P^{-1}S - \omega I) = 0$ and $det\,M(\omega, R) = 0$ have the same eigenvalues $\omega_i$ because a linear combination of differential equations preserves the dynamic properties [Beklemishev 1971].

## 3. Results

In this section we find the fundamental solutions for the boundary value problems in the conservative system, and the classic and dynamic Landau's systems. We identify the stability, the flow fields' structure, the interface velocity, and the formal properties of these fundamental solutions.

### 3.1. Conservative system

#### 3.1.1 Fundamental solutions

We consider properties of the dynamics balancing the fluxes of mass, normal and tangential components of momentum and energy across the interface, Eqs.(6). We call this dynamics the conservative dynamics. For the conservative dynamics, the matrix $M$ has the form $M = M$:

$$M = \begin{pmatrix} -R & -1 & -\omega + R\omega & i \\ 1 & -1 & 1 - R & i\omega/R \\ R - R\omega & R + \omega & 0 & -2iR \\ \omega & -\omega & \omega - R\omega & iR \end{pmatrix} \quad (12.1)$$

Its rank is 4. Its determinant is $det M = i((R-1)^2/R)(\omega - R)(\omega + R)(\omega^2 + R)$, and the eigenvalues $\omega_i$ and eigenvectors $\mathbf{e}_i$ (derived as described in the foregoing) are:

$$\omega_1 = i\sqrt{R}, \mathbf{e}_1 = \left(i\frac{-1+R}{i+\sqrt{R}}, -\frac{-1+R}{i+\sqrt{R}}\sqrt{R}, 1, 0\right)^T; \quad \omega_2 = \omega_1^*, \mathbf{e}_2 = \mathbf{e}_1^*; \quad (12.2)$$

$$\omega_3 = R, \mathbf{e}_3 = (0, i, 0, 1)^T; \quad \omega_4 = -R, \mathbf{e}_4 = \left(\frac{2i}{1+R}, -\frac{i(-1+R)}{1+R}, 0, 1\right)^T.$$

#### 3.1.2 Stability of fundamental solutions

For the conservative system Eqs.(12), there are four fundamental solutions. Solutions $\mathbf{r}_{1(2)}(\omega_{1(2)}, \mathbf{e}_{1(2)})$ are stable, with $\omega_1 = i\sqrt{R}$, $\omega_2 = \omega_1^*$ and $\text{Re}[\omega_{1(2)}] = 0$. Solution $\mathbf{r}_3(\omega_3, \mathbf{e}_3)$ is unstable, $\omega_3 = R$ and $\text{Re}[\omega_3] > 0$. Solution $\mathbf{r}_4(\omega_4, \mathbf{e}_4)$ is stable, $\omega_4 = -R$ and $\text{Re}[\omega_4] < 0$. Figure 2 shows the dependence of real and imaginary parts of the eigenvalues on the density ratio.

#### 3.1.3 Flow fields of fundamental solutions

By substituting fundamental solutions $\mathbf{r}_i = \mathbf{r}_i(\omega_i, \mathbf{e}_i)$ from Eqs.(12) to expressions Eqs.(10), we find the flow fields for the conservative dynamics. For each of the fundamental solutions, Figures 3 – 6 illustrate the interface perturbation $z^*$, the perturbed velocity fields $\mathbf{u}_{h(l)}$, the perturbed velocity streamlines $\mathbf{s}_{h(l)}$ defined as $(d\mathbf{s}_{h(l)}/dt) \times \mathbf{u}_{h(l)} = 0$, the contour plots of the perturbed pressure $p_{h(l)}$, and (when applicable) the vortical field of the light fluid velocity, including the velocity vortical component $\nabla \times \mathbf{\Psi}_l$ and the contour plot of vorticity $\nabla \times \mathbf{u}_l$ with $\nabla \times \mathbf{u}_l = \left(0, \left(k^2 - \tilde{k}^2\right)\Psi, 0\right)$. Real parts of functions are shown at some density ratio and some instance of time in the motion plane. In contour plots, red (blue) color marks positive (negative) values. Each plot has its own range of values.

Fundamental solution $\mathbf{r}_1(\omega_1, \mathbf{e}_1)$ in Eqs.(12) describes stable oscillations of the flow fields in a vicinity of the interface which decay away from the interface. For this solution, the velocity fields are potential, and all boundary conditions are satisfied, including balances of mass, momentum and energy at the interface, and the conditions at the outside boundaries of the domain. Note that for solution $\mathbf{r}_1(\omega_1, \mathbf{e}_1)$ the velocity field of the light fluid $\mathbf{u}_l$ has a phase shift $\pi/2$ relative to that of the heavy fluid $\mathbf{u}_h$. That is, the field $\mathbf{u}_l$ is shifted relative to $\mathbf{u}_h$ by $\lambda/4$ along the $x$ direction of translation, Eqs.(12). The velocity fields are also shifted relative the interface perturbation. These shifts are needed to balance the perturbed mass flux across the perturbed interface in the incompressible approximation, Figure 3.

For solution $\mathbf{r}_1$, the perturbations of pressure $p_{h(l)}$ oscillate in the vicinity of the interface and decay away from the interface. Pressure fields $p_{h(l)}$ are shifted relative to the interface perturbation. The fields of pressure $p_h$ and $p_l$ are in anti-phase with one another thus causing stable oscillations in the $z$ direction. These fields span the same range of values with $|p_l|_{max(min)} = |p_h|_{max(min)}$. For solution $\mathbf{r}_1(\omega_1, \mathbf{e}_1)$ the vortical field is zero, $\Psi_l = 0, \nabla \times \mathbf{\Psi}_l = 0, \nabla \times \mathbf{u}_l = 0$, Figure 3.

Properties of the fundamental solution $\mathbf{r}_2(\omega_2, \mathbf{e}_2)$ in Eqs.(12) are similar to those of $\mathbf{r}_1(\omega_1, \mathbf{e}_1)$. This is because these solutions are complex conjugates, with $\omega_2 = \omega_1^*, \mathbf{e}_2 = \mathbf{e}_1^*, C_2 = C_1^*$, Figure 3.

Note that by their very meaning the solutions $\mathbf{r}_{1(2)}$ are traveling waves. The interference of these traveling waves results in the appearance of standing waves, whose flow fields experience stable oscillations, particularly, solutions $\mathbf{r}_{CD} = (\mathbf{r}_1 + \mathbf{r}_2)/2$ and $\tilde{\mathbf{r}}_{CD} = (\mathbf{r}_1 - \mathbf{r}_2)/2$. Figure 4 illustrates the flow fields for solution $\mathbf{r}_{CD}$, including the perturbed velocity, the perturbed velocity streamlines, the pressure perturbations and the interface perturbation at some density ratio and some instance of time in the motion plane.

Fundamental solution $\mathbf{r}_3(\omega_3, \mathbf{e}_3)$ obeys all the boundary conditions and is thus a physical solution. Because $\text{Re}[\omega_3(R)] > 0$ for any density ratio $R \geq 1$, it appears at first glance to describe unstable dynamics. However, a more detailed consideration finds that for this solution, with $\omega_3 = R$ and $\mathbf{e}_3 = (0, i, 0, 1)$, the velocity fields of the light and heavy fluids are identically zero, with $\mathbf{u}_l = 0$ and $\mathbf{u}_h = 0$. Furthermore, the interface perturbation and the pressure fields are also zero, $z^* = 0$ and $p_{h(l)} = 0$, Figure 5. Note that while for this solution the vortical velocity component is $\nabla \times \mathbf{\Psi}_l \neq 0$, its vorticity is zero, $\nabla \times \mathbf{u}_l = 0$, because for the vorticity field with $\nabla \times \mathbf{u}_l = \left(0, \left(k^2 - \tilde{k}^2\right)\Psi, 0\right)$ the values are $\left(\tilde{k}/k\right)^2 = \left(\omega_3/R\right)^2 = 1$, Figure 5. These properties hold true in the entire domain, at any time, and for any integration constant $C_3$. Therefore, solution $\mathbf{r}_3$ is unstable, because its eigenvalue is $\omega_3 > 0$, and it corresponds to unperturbed fields of the velocity, pressure, and the interface, because its eigenvector is $\mathbf{e}_3 = (0, i, 0, 1)$, Figure 5.

Fundamental solution $\mathbf{r}_4(\omega_4, \mathbf{e}_4)$ appears at first glance to be stable and physical, Figure 6. A more detailed consideration finds that the required choice of $C_4$ is $C_4 = 0$ for this solution. Indeed, according to expression for $\omega_4, \mathbf{e}_4$, at the outside boundaries of the domain, the velocity field of the heavy fluid decays $\mathbf{u}_h|_{z \to -\infty} = 0$ at any time. For the velocity field of the light fluid $\mathbf{u}_l$ the potential component is vanishing away from the interfaces, $\nabla \Phi_l|_{z \to +\infty} = 0$, whereas its vortical component $\nabla \times \mathbf{\Psi}_l$, while decaying in time, as $\sim e^{-Rt}$, increases away from the interface, as $\sim e^{kz}$. Emphasize that while for this solution the vortical component of velocity is $\nabla \times \mathbf{\Psi}_l \neq 0$, its vorticity is zero, $\nabla \times \mathbf{u}_l = 0$, because for the vorticity field with $\nabla \times \mathbf{u}_l = \left(0, \left(k^2 - \tilde{k}^2\right)\Psi, 0\right)$ the value is $\left(\tilde{k}/k\right)^2 = 1$. Finally note that for this solution the pressure fields $p_{h(l)}$ are anti-phase with one another and decay away from the interface. Thus, in order for solution $\mathbf{r}_4$ to satisfy boundary condition $\mathbf{u}_l|_{z \to +\infty} = 0$, we must have $C_4 = 0$, Figure 6.

### 3.1.4 Interface velocity for fundamental solutions

For fundamental solutions $\mathbf{r}_{1(2)}$, which describe stable traveling waves, as well as for solutions $\mathbf{r}_{CD}, \tilde{\mathbf{r}}_{CD}$, which describe stable standing waves resulted from interference of the traveling waves $\mathbf{r}_{1(2)}$, the interface velocity is $\tilde{\mathbf{V}} = \tilde{\mathbf{V}}_0 + \tilde{\mathbf{v}}$ in the laboratory frame of reference with $\tilde{\mathbf{v}}\mathbf{n}_0 = -\left(\mathbf{u}_h \mathbf{n}_0 + \dot{\theta}\right)\big|_{\theta = 0^+}$. Because of oscillatory terms $\left(\mathbf{u}_h \mathbf{n}_0 + \dot{\theta}\right) \sim e^{\pm i \sqrt{R} t}$, the interface velocity experiences stable oscillations near the steady value $\tilde{\mathbf{V}}_0$. The oscillations' amplitude is small, $|\tilde{\mathbf{v}}| \ll |\tilde{\mathbf{V}}_0|$ and is set by the initial conditions and the integration constant(s) $C_{1(2)}$, whereas the oscillations period $1/\sqrt{R}$ is substantially smaller than the characteristic time-scale of the dynamics.

For fundamental solutions $\mathbf{r}_3$ in the laboratory frame of reference the interface velocity is constant, $\tilde{\mathbf{V}} = \tilde{\mathbf{V}}_0$, because its perturbed velocity fields and perturbed interface are zero for any integration constant $C_3$. For solution $\mathbf{r}_4$, in the laboratory frame of reference the interface velocity is constant $\tilde{\mathbf{V}} = \tilde{\mathbf{V}}_0$ because its integration constant is zero, $C_4 = 0$.

### 3.1.5 Formal properties of the conservative system

To study formal properties of the conservative system, we find for matrix $\mathbf{M} = M$ the associated matrices $\mathbf{S} = S_M$ and $\mathbf{P} = P_M$:

$$S_M = \begin{pmatrix} -R & -1 & 0 & i \\ 1 & -1 & 1-R & 0 \\ R & R & 0 & -2iR \\ 0 & 0 & 0 & iR \end{pmatrix}, \quad P_M = \begin{pmatrix} 0 & 0 & 1-R & 0 \\ 0 & 0 & 0 & -i/R \\ R & -1 & 0 & 0 \\ -1 & 1 & -1+R & 0 \end{pmatrix} \quad (12.3)$$

Solutions of equations $det(P_M^{-1} S_M - \omega I) = 0$ and $det M = 0$ yield the same eigenvalues: $\omega_{1(2)} = \pm i\sqrt{R}$ and $\omega_3 = R$, $\omega_4 = -R$. The conservative dynamics has 4 eigenvalues, 4 fundamental solutions, and 4 independent degrees of freedom. It is non-degenerate.

### 3.1.6 Summary of properties of solutions for the conservative system

For the conservative system, there are four fundamental solutions $\mathbf{r}_i(\omega_i, \mathbf{e}_i)$, $i = 1,...,4$, Eqs.(12), Figure 2-6, Table 1. Solutions $\mathbf{r}_1$ and $\mathbf{r}_2$ with some integration constant $C_1, C_2 = C_1^*$ describe stable oscillations of the flow fields. These solutions are stable traveling waves; their interference results in stably oscillating standing waves $\mathbf{r}_{CD}$ and $\tilde{\mathbf{r}}_{CD}$. For these solutions, the interface velocity $\tilde{\mathbf{V}}$ experiences small stable oscillations near the steady value $\tilde{\mathbf{V}}_0$.

Solution $\mathbf{r}_3$, corresponds to unperturbed fields of the velocity, pressure and the interface at any constant $C_3$. For solution $\mathbf{r}_4$ the value $C_4$ must be zero to obey all the boundary conditions, as discussed in the foregoing, $C_4 = 0$. For solutions $\mathbf{r}_{3(4)}$, the interface velocity is constant, $\tilde{\mathbf{V}} = \tilde{\mathbf{V}}_0$.

### 3.2. Classic Landau's system
#### 3.2.1. Fundamental solutions

Landau 1944 analyzed stability of the interface by considering the boundary conditions balancing the transports of mass and normal and tangential components of momentum at the interface, and by omitting the equation for the energy balance at the interface [Peters 2010, Landau 1944, Zeldovich 1944, Mallard & Le Chatelier 1883]. Since four degrees of freedom and four independent variables require four conditions, Landau employed additional condition for the continuity of normal component of the perturbed velocity at the interface $[\mathbf{u} \cdot \mathbf{n}_0] = 0$, or $j_n = 0$ at the interface [Landau 1944].

With this condition, the boundary conditions have the form
$$[j_n] = 0, \; [(p + 2J_n j_n/\rho)\mathbf{n}_0] = [p\mathbf{n}_0] = 0, \; [J_n(\mathbf{J} \cdot \boldsymbol{\tau}_1 + \mathbf{j} \cdot \boldsymbol{\tau}_0)/\rho] = 0, \; [\mathbf{u} \cdot \mathbf{n}_0] = 0 \qquad (13)$$

Recall that the perturbed mass flux is $\mathbf{j} = \rho(\mathbf{u} + \mathbf{n}_0 \dot{\theta})$ with $j_n = \mathbf{j} \cdot \mathbf{n}_0 = \rho((\mathbf{u} \cdot \mathbf{n}_0) + \dot{\theta})$. The condition $j_n = 0$ leads to $(\mathbf{u} \cdot \mathbf{n}_0) = -\dot{\theta}$ for any $\rho$, and thus to $[\mathbf{u} \cdot \mathbf{n}_0] = 0$ in Eqs.(13). Due to the condition $j_n = 0$, the condition for the normal component of momentum is transformed to $[(p + 2J_n j_n/\rho)\mathbf{n}_0] = [p\mathbf{n}_0] = 0$, and the field of pressure perturbation is continuous at the interface, $[p\mathbf{n}_0] = 0$.

For the classic Landau's system, matrix $\mathbf{M}$ is $\mathbf{M} = \mathbf{L}$:

$$L = \begin{pmatrix} -R & -1 & -\omega + R\omega & i \\ 1 & -1 & 1-R & i\omega/R \\ R - R\omega & R + \omega & 0 & -2iR \\ -1 & -1 & 0 & i \end{pmatrix} \qquad (14.1)$$

Its rank is 4. Its determinant is $det L = i((R-1)/R)(\omega - R)((R+1)\omega^2 + 2R\omega - R(R-1))$. Its eigenvalues $\omega_i$ and eigenvectors $\mathbf{e}_i$ are:

$$\omega_1 = \frac{-R + \sqrt{-R + R^2 + R^3}}{1 + R},$$

$$\mathbf{e}_1 = \left( -i\frac{(1 + \sqrt{R(-1 + R + R^2)})}{(-1 + R)(1 + R)}, i\frac{(R^2 + \sqrt{R(-1 + R + R^2)})}{(-1 + R)(1 + R)}, -i\frac{(R^2 + \sqrt{R(-1 + R + R^2)})}{(-1 + R)^2 R}, 1 \right)^T;$$

$$\omega_2 = \frac{-R - \sqrt{-R + R^2 + R^3}}{1 + R}$$

$$\mathbf{e}_2 = \left( i\frac{(-1 + \sqrt{R(-1 + R + R^2)})}{(-1 + R)(1 + R)}, i\frac{(R^2 - \sqrt{R(-1 + R + R^2)})}{(-1 + R)(1 + R)}, -i\frac{(R^2 - \sqrt{R(-1 + R + R^2)})}{(-1 + R)^2 R}, 1 \right)^T;$$

$$\omega_3 = R$$
$$\mathbf{e}_3 = (0, i, 0, 1)^T \qquad (14.2)$$

### 3.2.2. Stability of fundamental solutions

For the classic Landau dynamics, there are three fundamental solutions. Solution $\mathbf{r}_1(\omega_1, \mathbf{e}_1)$ is unstable, $\text{Re}[\omega_1] > 0$. Solution $\mathbf{r}_2(\omega_2, \mathbf{e}_2)$ is stable, $\text{Re}[\omega_2] < 0$. Solution $\mathbf{r}_3(\omega_3, \mathbf{e}_3)$ is unstable, $\text{Re}[\omega_3] > 0$, and is identical to that for the conservative system. Figure 7 illustrates the dependence of eigenvalues on the density ratio.

### 3.2.3. Flow fields of fundamental solutions

By substituting solutions $\mathbf{r}_i = \mathbf{r}_i(\omega_i, \mathbf{e}_i)$ in Eqs.(14) to expressions Eqs.(10) for each $i$, we find the flow fields for fundamental solutions describing the classic Landau's dynamics.

For each of fundamental solutions $\mathbf{r}_i(\omega_i, \mathbf{e}_i)$ for the classic Landau's dynamics, Figures 8–9 illustrate the interface perturbation $z^*$, the perturbed velocity fields $\mathbf{u}_{h(l)}$, the perturbed velocity streamlines $\mathbf{s}_{h(l)}$ defined as $(d\mathbf{s}_{h(l)}/dt) \times \mathbf{u}_{h(l)} = 0$, the contour plots of the perturbed pressure fields $p_{h(l)}$, and the vortical field of the light fluid velocity, including the vortical velocity component $\nabla \times \mathbf{\Psi}_l$ and the contour plot of vorticity $\nabla \times \mathbf{u}_l$ with $\nabla \times \mathbf{u}_l = (0, (k^2 - \widetilde{k}^2)\Psi, 0)$. Real parts of functions are shown at some density ratio and some instance of time in the motion plane. In contour plots, red (blue) color marks positive (negative) values. Each plot has its own range of values

Fundamental solution $\mathbf{r}_1(\omega_1, \mathbf{e}_1)$ in Eqs.(14) corresponds to the Landau-Darrieus instability [Landau 1944], Figure 8. This solution, although unstable, is a physical solution. It satisfies all boundary conditions, including the assigned boundary conditions at the interface and the conditions at the outside boundaries of the domain. This solution describes dynamics, in which the interface perturbation and the vortical and potential components are strongly coupled. The perturbed velocity fields $\mathbf{u}_{h(l)}$ and the perturbed velocity streamlines $\mathbf{s}_{h(l)}$ illustrate the formation of vortical structures near the interface and in the bulk of the light fluid. For solution $\mathbf{r}_1(\omega_1, \mathbf{e}_1)$ the vortical component, $\nabla \times \mathbf{\Psi}_l$, and the vorticity $\nabla \times \mathbf{u}_l$, while increasing in time, decay away from the interface. The fields of pressure perturbations $p_h$ and $p_l$ are equal one another at the interface and decay away from the interface. They are in phase with one another and in anti-phase with the interface perturbation. Fields of pressure $p_{h(l)}$ are symmetric, and span the same range of values with $|p_l|_{max(min)} = |p_h|_{max(min)}$, Figure 8.

The vortical field for solution $\mathbf{r}_1(\omega_1, \mathbf{e}_1)$ in Eqs.(14) has the following properties, Figure 8. (1) Consistent with expressions in Eqs.(14), the potential components of $\mathbf{u}_h$ and $\mathbf{u}_l$ are in anti-phase with one another, whereas vortical component of velocity $\mathbf{u}_l$ is shifted relative to potential components of velocities $\mathbf{u}_h$ and $\mathbf{u}_l$ and the interface perturbation $z^*$. (2) For the length-scale of the vortical field, we have $\tilde{k}/k = \omega_1/R$, $\tilde{k}/k = \left(-R + \sqrt{-R + R^2 + R^3}\right)/R(1+R)$. The length-scale of the vortical field is large, with $\tilde{k}/k \sim (R-1)/2 \to 0$ for $R \to 1^+$ and with $\tilde{k}/k \sim R^{-1/2} \to 0$ for $R \to \infty$. The maximum value of $\tilde{k}/k$ (minimum value of $\tilde{\lambda}/\lambda$) is achieved at $R = 2 + \sqrt{5} \approx 4.24$. (3) For $R \to 1^+$ the values are $|\Psi/\Phi| \sim (R-1)$ and $|\Psi/\tilde{\Phi}| \sim (R-1)$, whereas for $R \to \infty$ the values are $|\Psi/\Phi| \sim R^{3/2}$ and $|\Psi/\tilde{\Phi}| \sim 1$. Therefore, the contribution of the vortical field is quantitatively large for fluids with very different densities, and is quantitatively small for fluids with similar densities. (4) To quantify the strength of the vorticity field for given $\Psi$, we notice that $|\nabla \times \mathbf{u}_l|/|\Psi|k^2 = \left(1 - (\tilde{k}/k)^2\right)$. For solution $\mathbf{r}_1(\omega_1, \mathbf{e}_1)$, this value is $\sim 1$ for all $R \geq 1$.

The other fundamental solution in Eqs.(14) is solution $\mathbf{r}_2(\omega_2, \mathbf{e}_2)$, Figure 9. Solution $\mathbf{r}_2$ describes the dynamics for which the interface perturbation and the vortical and potential components are strongly coupled. It appears stable and physical at a first glance. Its fields of pressure $p_{h(l)}$ are equal one another and are in-phase at the interface, and decay away from the interface. For this solution, the field of the perturbed velocity of the heavy fluid vanishes at the outside boundaries of the domain $\mathbf{u}_h|_{z \to -\infty} = 0$ at all times. The potential component of the perturbed velocity of the light fluid $\mathbf{u}_l$ also vanishes away from the interface, $\nabla \Phi_l|_{z \to +\infty} = 0$. Its vortical component, $\nabla \times \Psi_l$, decays exponentially in time but increases far from the interface. The strength of the vorticity field $\nabla \times \mathbf{u}_l$ for this solution also increases far from the interface. In order for fundamental solution $\mathbf{r}_2(\omega_2, \mathbf{e}_2)$ in Eqs.(14) to satisfy the boundary condition $\mathbf{u}_l|_{z \to +\infty} = 0$ at any time, the value of integration constant $C_2$ should be zero, $C_2 = 0$.

Fundamental solution $\mathbf{r}_3(\omega_3, \mathbf{e}_3)$ in Eqs.(14) is identical to that in Eqs.(12), Figure 5. Note that solution $\mathbf{r}_3$ exists in the classic framework [Landau & Lifshitz 1987, Landau 1944] and can be found similarly to Eqs.(13,14).

### 3.2.4  Interface velocity for fundamental solutions

For fundamental solution $\mathbf{r}_1$, in the laboratory frame of reference the interface velocity is $\tilde{\mathbf{V}} = \tilde{\mathbf{V}}_0 + \tilde{\mathbf{v}}$ with $\tilde{\mathbf{v}}\mathbf{n}_0 = -\left(\mathbf{u}_h \mathbf{n}_0 + \dot{\theta}\right)_{\theta=0^+}$. While in this expression each of the terms $\mathbf{u}_h$ and $\dot{\theta}$ grows exponentially in time, $\mathbf{u}_h, \dot{\theta} \sim e^{\omega_1 t}$, the exact balance in the boundary conditions $[\mathbf{u} \cdot \mathbf{n}_0] = 0$ in leads to $\left(\mathbf{u}_h \mathbf{n}_0 + \dot{\theta}\right)\Big|_{\theta=0^+} = 0$ and $\tilde{\mathbf{v}} = 0$, and thus to the constancy of the interface velocity $\tilde{\mathbf{V}} = \tilde{\mathbf{V}}_0$. The classic Landau's dynamics is the perfect match!

For fundamental solutions $\mathbf{r}_{2(3)}$, in the laboratory frame of reference the interface velocity is constant $\tilde{\mathbf{V}} = \tilde{\mathbf{V}}_0$, similarly to the foregoing.

### 3.2.5 Formal properties of the classic Landau's system

To study formal properties of the classic Landau's system, we find for matrix $M = L$ the associated matrices $S = S_L$ and $P = P_L$:

$$S_L = \begin{pmatrix} -R & -1 & 0 & i \\ 1 & -1 & 1-R & 0 \\ R & R & 0 & -2iR \\ -1 & -1 & 0 & i \end{pmatrix}, \quad P_{L_G} = \begin{pmatrix} 0 & 0 & 1-R & 0 \\ 0 & 0 & 0 & -i/R \\ R & -1 & 0 & 0 \\ 0 & 0 & 0 & 0 \end{pmatrix} \quad (14.3)$$

In matrix $P_L$, the fourth row only has null elements and $det\, P_L = 0$. Thus the inverse matrix $P_L^{-1}$ does not exist. This suggests the degeneracy of the classical Landau's system, and a singular character of the Landau-Darrieus instability.

### 3.2.6 Summary of properties of solutions for the classic Landau's system

For the classic Landau's system, there are three fundamental solutions $\mathbf{r}_i(\omega_i, \mathbf{e}_i)$, $i = 1,2,3$. Eqs.(13,14), Figure 5,7-9, Table 2. Solution $\mathbf{r}_1(\omega_1, \mathbf{e}_1)$ with $C_1 \neq 0$ describes the Landau-Darrieus instability. For solution $\mathbf{r}_2(\omega_2, \mathbf{e}_2)$, the integration constant must be zero, $C_2 = 0$. Solution $\mathbf{r}_3(\omega_3, \mathbf{e}_3)$ corresponds to unperturbed fields of velocity and pressure at any $C_3$. For each of these solutions, in the laboratory frame of reference the interface velocity is constant $\tilde{\mathbf{V}} = \tilde{\mathbf{V}}_0$.

### 3.3. Dynamic Landau's system

#### 3.3.1. Fundamental solutions

The paradox of the classic Landau's system (to our knowledge it has not been discussed in details before) is that it has a smaller than expected number of fundamental solutions. On the side of physics, this paradox can be understood from the following considerations. The boundary conditions in the classic Landau's system Eqs.(13) imply dynamic conditions for the conservation of mass and normal and tangential components of momentum at the interface, because they contain [atrial time-derivatives of components of vector $\mathbf{r}$. Landau's special condition in Eqs.(13) is a 'static' condition of continuity of normal component of perturbed velocity at the interface $[\mathbf{u} \cdot \mathbf{n}_0] = 0$. This condition implies that the perturbed mass flux is a constant value at the interface, $j_n = const$ and, furthermore, this value is zero, $j_n = 0$. In order to resolve the paradox, we modify the special Landau's condition $j_n = 0$ to a more general form $j_n = const$ and implement it in a dynamic condition, $\partial j_n / \partial t = 0$ or $[\partial(\mathbf{u} \cdot \mathbf{n}_0)/\partial t] = 0$ at the interface. While other implementations are possible, we note that the condition of constant mass flux across the interface is met in realistic environments [Azechi et al. 2007, Peters 2000].

For this dynamic implementation of Landau's system, the modified boundary conditions are

$$[j_n] = 0, \ [(p + 2J_n j_n / \rho)\mathbf{n}_0] = 0, \ [J_n(\mathbf{J} \cdot \boldsymbol{\tau}_1 + \mathbf{j} \cdot \boldsymbol{\tau}_0)/\rho] = 0, \ [\partial(\mathbf{u} \cdot \mathbf{n}_0)/\partial t] = 0 \qquad (15)$$

For the dynamic Landau's system, matrix $M$ has the form $M = \widetilde{L}$ :

$$\widetilde{L} = \begin{pmatrix} -R & -1 & -\omega + R\omega & i \\ 1 & -1 & 1 - R & i\omega/R \\ R - R\omega & R + \omega & 0 & -2iR \\ -\omega & -\omega & 0 & i\omega \end{pmatrix} \qquad (16.1)$$

Its rank is 4. Its determinant is $det\widetilde{L} = i((R-1)/R)\omega(\omega - R)((R+1)\omega^2 + 2R\omega - R(R-1))$. It has 4 eigenvalues $\omega_i$ and 4 eigenvectors $\mathbf{e}_i$. For $i = 1,2,3$ the eigenvalues $\omega_i$ and eigenvectors $\mathbf{e}_i$ are the same as in Eqs.(14). For $i = 4$ the fundamental solution is $\mathbf{r}_4^o(\omega_4^o, \mathbf{e}_4^o)$ with

$$\omega_4^o = 0, \ \mathbf{e}_4^o = \left(-i\frac{1}{-1+R}, i\left(\frac{-1+2R}{-1+R}\right), -i\frac{2R}{(-1+R)^2}, 1\right)^T \qquad (16.2)$$

Symbol 'o' in $\mathbf{r}_4^o(\omega_4^o, \mathbf{e}_4^o)$ emphasizes the presence of the null eigenvalue.

### 3.3.2. Stability of fundamental solution

Fundamental solutions $\mathbf{r}_i(\omega_i, \mathbf{e}_i)$ with $i = 1,2,3$ of Eqs.(16) are the same as solutions for system Eqs.(14) and have the same stability properties. Fundamental solution $\mathbf{r}_4^o(\omega_4^o, \mathbf{e}_4^o)$ is neutrally stable because of the null eigenvalue $\omega_4^o = 0$, Figure 7.

### 3.3.3. Flow fields of fundamental solutions

Fundamental solutions $\mathbf{r}_i(\omega_i, \mathbf{e}_i)$ with $i = 1,2,3$ in Eqs.(16) are the same as solutions for system Eqs.(14) and have the same corresponding fields of the velocity, the pressure and the interface perturbation, Figure 5,8,9. For fundamental solution $\mathbf{r}_4^o(\omega_4^o, \mathbf{e}_4^o)$, Figure 10 represent the fields of the perturbed velocity $\mathbf{u}_{h(l)}$, the perturbed streamlines $\mathbf{s}_{h(l)}$, the perturbed pressure $p_{h(l)}$, the perturbed vortical field $\nabla \times \boldsymbol{\Psi}_l$, the vorticity $\nabla \times \mathbf{u}_l = \left(0, (k^2 - \widetilde{k}^2)\Psi, 0\right)$, and the interface perturbation $z^*$ at some instance of time for some density ratio in the motion plane. Each plot has its own range of values.

Solution $\mathbf{r}_4^o(\omega_4^o, \mathbf{e}_4^o)$ is independent from fundamental solutions $\mathbf{r}_i(\omega_i, \mathbf{e}_i)$ with $i=1,2,3$ in Eqs.(14), and has the following properties. (1) The vortical component of velocity $\mathbf{u}_l$ is shifted relative to potential components of velocities $\mathbf{u}_h$, $\mathbf{u}_l$ and the interface perturbation $z^*$, whereas potential components of $\mathbf{u}_h$ and $\mathbf{u}_l$ are in anti-phase with one another. (2) The vortical field length-scale is infinitely large, with $\tilde{k}/k = 0$ and $\tilde{\lambda}/\lambda = \infty$, due to the null eigenvalue $\omega_4^o = 0$. (3) The pressure fields of $p_h$ and $p_l$ are in phase and decay away from the interface. The pressure fields are not symmetric and $|p_l|_{max(min)} \ll |p_h|_{max(min)}$. (4) The vortical component of the velocity is $\nabla \times \boldsymbol{\Psi}_l \neq 0$, and the vorticity is $\nabla \times \mathbf{u}_l \neq 0$. They change periodically in the $x$ direction and are uniform in the $z$ direction. (5) For $R \to \infty$, the values are $|\Psi/\Phi| \sim R$ and $|\Psi/\tilde{\Phi}| \sim 2$. For $R \to 1^+$, the values are $|\Psi/\Phi| \sim (R-1)$ and $|\Psi/\tilde{\Phi}| \sim (R-1)$, and $|\nabla \times \mathbf{u}_l|/|\Psi|k^2 = 1$ for any $R$.

### 3.3.4 Interface velocity for fundamental solutions

For fundamental solutions $\mathbf{r}_i$ with $i=1,2,3$ in Eqs.(16) the interface velocity in laboratory frame of reference is constant, $\tilde{\mathbf{V}} = \tilde{\mathbf{V}}_0$, as discussed in the foregoing.

For fundamental solutions $\mathbf{r}_4^o$ in Eqs.(16) in the laboratory frame of reference the interface velocity is $\tilde{\mathbf{V}} = \tilde{\mathbf{V}}_0 + \tilde{\mathbf{v}}$ with $\tilde{\mathbf{v}} \mathbf{n}_0 = -(\mathbf{u}_h \mathbf{n}_0 + \dot{\theta})\big|_{\theta=0^+}$. Due to the zero eigenvalue, $\omega_4^o = 0$, the value is $\dot{\theta} = 0$, whereas the value is $\mathbf{u}_h \mathbf{n}_0\big|_{\theta=0^+} = const$. Thus, for solution $\mathbf{r}_4^o$ with constant perturbed mass flux across the interface, $[\partial(\mathbf{u} \cdot \mathbf{n}_0)/\partial t] = 0$, the interface velocity $\tilde{\mathbf{V}} = \tilde{\mathbf{V}}_0 + \tilde{\mathbf{v}}$ is slightly shifted from its steady value $\tilde{\mathbf{V}}_0$. The shift is small and stationary, $|\tilde{\mathbf{v}}| \ll |\tilde{\mathbf{V}}_0|$ and $\dot{\tilde{\mathbf{v}}} = 0$.

### 3.3.5 Formal properties of the dynamic Landau's system
#### 3.3.5.1 Non-degeneracy of the dynamic Landau's system

To study formal properties of the dynamic Landau's system, we find for matrix $\mathbf{M} = \tilde{L}$ the associated matrices $\mathbf{S} = S_{\tilde{L}}$ and $\mathbf{P} = P_{\tilde{L}}^o$:

$$S_{\tilde{L}} = \begin{pmatrix} -R & -1 & 0 & i \\ 1 & -1 & 1-R & 0 \\ R & R & 0 & -2iR \\ 0 & 0 & 0 & 0 \end{pmatrix} \quad P_{\tilde{L}} = \begin{pmatrix} 0 & 0 & 1-R & 0 \\ 0 & 0 & 0 & -i/R \\ R & -1 & 0 & 0 \\ 1 & 1 & 0 & -i \end{pmatrix} \tag{16.3}$$

The equations $det(P_{\tilde{L}}^{-1} S_{\tilde{L}} - \omega I) = 0$ and $det \tilde{L} = 0$ yield the same eigenvalues. This implies that the dynamic Landau's system is non-degenerate. It has four fundamental solutions for four degrees of freedom, thus resolving the long-standing formal paradox.

### 3.3.5.2 Singular character of the classic Landau's system

An important consequence of the zero value $\omega_4^o = 0$ is the singular nature of the LDI and the classic Landau solution [Kadanoff et al. 1967, Barenblatt et al. 1962]. Indeed, differential equations Eqs.(13-16) govern the dynamics of slight perturbations $\mathbf{r}$ near the equilibrium state of the uniform flow fields in Eqs.(6,8-10), for which $\mathbf{r} = 0$. It is presumed that these departures are exponential in time, with $\mathbf{r}$ varying as $\mathbf{r} \sim e^{\omega t}$, and with characteristic time-scale being $\sim |\omega^{-1}|$. For $\omega = 0$ a formal substitution leads to $\mathbf{r} \sim e^{0 \cdot t} \sim const$. This implies that slight departures from the equilibrium dynamics are actually power-law functions of time, $\mathbf{r} \sim t^a$, rather than exponential functions, $\mathbf{r} \sim e^{\omega t}$ [Kadanoff et al. 1967, Barenblatt et al. 1962]. The requirement of asymptotic stability of fundamental solution $\mathbf{r}_4^o(\omega_4^o, \mathbf{e}_4^o)$ leads to condition $Re[a] < 0$ [Landau & Lifshitz 1987]. For $Re[a] < 0$, slight initial departures from the equilibrium, $\mathbf{r} \sim t^a$, vanish asymptotically with time, $\mathbf{r} \to 0$ for $t \to \infty$. To derive the exponent $a$, one has to account for nonlinear terms. We address this problem to future research.

Note that power-law functions are scale-invariant and describe the processes with no characteristic scales. Power-law dependence $\mathbf{r} \sim t^a$ also implies that the system may have a singularity and experience a phase transition [Kadanoff 2000, Kadanoff et al. 1967]. Note also that with given $\omega_4^o = 0$ solution $\mathbf{r}_4^o(\omega_4^o, \mathbf{e}_4^o)$ in Eqs.(16) can be viewed as a steady vortical field that is imposed at some instance of time and is neutrally stable, Figure 8,10. This vortical field can 'seed' the vortical field in Landau's solution $\mathbf{r}_1(\omega_1, \mathbf{e}_1)$ in Eqs.(14), and trigger the unstable dynamics.

### 3.3.6 Summary of properties of fundamental solutions for the dynamic Landau's system

For the dynamic Landau's system in Eqs.(15,16), there are four fundamental solutions $\mathbf{r}_i(\omega_i, \mathbf{e}_i)$ with $i = 1,...,4$, Eqs.(14,16), Table 2, Figure 5,7-10. Solution $\mathbf{r}_1(\omega_1, \mathbf{e}_1)$ with $C_1 \neq 0$ describes the Landau-Darrieus instability. For solution $\mathbf{r}_2(\omega_2, \mathbf{e}_2)$, the integration constant is $C_2 = 0$. Solution $\mathbf{r}_3(\omega_3, \mathbf{e}_3)$ corresponds to the unperturbed fields of the velocity, the pressure and the interface at any $C_3$. Solution $\mathbf{r}_4^o(\omega_4^o, \mathbf{e}_4^o)$ with $C_4 \neq 0$ is neutrally stable. It may lead to a power-law dynamics near the equilibrium and may seed the LDI.

For solutions $\mathbf{r}_i$ with $i = 1,2,3$ the interface velocity in laboratory frame of reference is constant, $\tilde{\mathbf{V}} = \tilde{\mathbf{V}}_0$, as discussed in the foregoing For solution $\mathbf{r}_4^o$, the interface velocity $\tilde{\mathbf{V}} = \tilde{\mathbf{V}}_0 + \tilde{\mathbf{v}}$ is slightly shifted from its steady value $\tilde{\mathbf{V}}_0$, and the shift is small, $|\tilde{\mathbf{v}}| \ll |\tilde{\mathbf{V}}_0|$, and stationary, $\dot{\tilde{\mathbf{v}}} = 0$.

## 4. Properties of interfacial dynamics in the conservative system and in the Landau's systems

In this section we focus on physical properties of the interfacial dynamics that are prescribed by fundamental solutions for the conservative system, and the classic and dynamic Landau's systems. We further find the new mechanisms of the interface stabilization, due to the inertial effects, and we model the influence of energy fluctuations on the interface dynamics.

### 4.1. Physical properties of fundamental solutions

#### 4.1.1 Stable, neutrally stable and unstable standing waves

We compare properties of fundamental solution $\mathbf{r}_{CD}$ for the conservative system, fundamental solution $\mathbf{r}_{LD}$ for the classic Landau's system, and fundamental solution with $\mathbf{r}_{DLD}$ for the dynamic Landau's system, where $\mathbf{r}_{CD} = (\mathbf{r}_1 + \mathbf{r}_2)/2$ in Eqs.(12), $\mathbf{r}_{LD} = \mathbf{r}_1$ in Eqs.(14), and $\mathbf{r}_{DLD} = \mathbf{r}_4^o$ in Eqs.(16). Solutions $\mathbf{r}_{CD}$, $\mathbf{r}_{LD}$, and $\mathbf{r}_{DLD}$ are standing waves; they have distinct properties, Tables 3-5.

The dynamics $\mathbf{r}_{CD}$ is stable, the dynamics $\mathbf{r}_{LD}$ is unstable, and the dynamics $\mathbf{r}_{DLD}$ is neutrally stable. For solution $\mathbf{r}_{CD}$ the velocity fields are potential; for solutions $\mathbf{r}_{LD}$ and $\mathbf{r}_{DLD}$ the potential and vortical components of the velocities and the interface perturbation are strongly coupled. There is zero shear at the interface for each of the dynamics $\mathbf{r}_{CD}$, $\mathbf{r}_{LD}$, $\mathbf{r}_{DLD}$. Solution $\mathbf{r}_{CD}$ conserves mass, momentum and energy at the interface. Solution $\mathbf{r}_{LD}$ conserves mass and momentum, and has zero perturbed mass flux at the interface. Solution $\mathbf{r}_{DLD}$ conserves mass and momentum, and has constant perturbed mass flux at the interface. The dynamics $\mathbf{r}_{CD}$ is non-degenerate (4 solutions, 4 degrees of freedom). The dynamics $\mathbf{r}_{LD}$ is degenerate (3 solutions, 4 degrees of freedom). The dynamics $\mathbf{r}_{DLD}$ is non-degenerate (4 solutions, 4 degrees of freedom), Tables 3-5.

For solutions $\mathbf{r}_{DLD}$ and $\mathbf{r}_{LD}$, potential and vortical components of the velocities are strong coupled with the interface perturbations. Yet, for solution $\mathbf{r}_{LD}$ the length-scale of the vortical field $\widetilde{\lambda} = 2\pi/\widetilde{k}$ depends on the density ratio, whereas for solution $\mathbf{r}_{DLD}$ the length-scale of the vortical field is independent of the density ratio and is infinitely large, $\widetilde{\lambda} = \infty$, Tables 3-5.

### 4.1.2 Other unstable solutions

The conservative system, the classic Landau system, the dynamic Landau system have a fundamental solution in common, $\mathbf{r}_3(\omega_3, \mathbf{e}_3)$ in Eqs.(12,14,16), which can be implemented for any integration constant $C_3$. This solution is the fastest unstable solution, as suggested by its eigenvalue $\omega_3$. This solution has the null perturbation fields of the velocity, pressure and interface, as suggested by its eigenvector $\mathbf{e}_3$, Figure 5.

Note that this solution is fully consistent with the classic works [Landau & Lifshitz 1987] and can be derived from the governing equations in the framework [Landau 1944]. The existence of this solution may imply that the velocity of the inertial frame of reference $\mathbf{V}_0$ is a free quantity, and it may differ from the velocity of the planar steady interface $\widetilde{\mathbf{V}}_0$. It may also imply that the interfacial dynamics may have flow fields that are uniform and are indistinguishable from those of the equilibrium state. While the solution is formally unstable, its dynamics may appear as being 'fully stabilized'! This apparent stabilization is achieved in the case of ideal incompressible fluids and at any density ratio. In realistic fluids various physical effects may influence the null perturbation fields of the velocity, pressure and interface associated with this solution, to be studied in the future research.

### 4.1.3 Other stable solutions

Our analysis identifies some new properties of stable fundamental solutions, which are the stable solutions that decay exponentially with time due to their real negative eigenvalues. These include solution $\mathbf{r}_4$ in Eqs.(12) for the conservative system, and solution $\mathbf{r}_2$ in Eqs.(14,16) for the classic and dynamic Landau's systems. As discussed in the foregoing, these stable solutions require a null integration constant in order to obey the boundary conditions at the outside boundaries of the domain at any times.

A somewhat similar requirement has been discussed in the study of the LDI by means of the Laplace transform [Zeldovich 1944]. This study interprets a non-zero vortical component of the velocity field away from the interface as vortical structures that are produced at the interface and are moved to the bulk. The study requires a null integration constant for the corresponding solution because these vortical structures have no influence on the development of the LDI [Zeldovich 1944]. In our study, we require the null integration constant in order for the exponentially decaying stable solutions to obey the boundary conditions at the outside boundaries of the domain at any time for any density ratio. We further note that for stable solutions decaying exponentially with time the requirement of zero integration constant directly follows from the solution structure. It thus should hold in realistic fluids, especially in case of fluids with contrasting densities, when the contribution of the vortical field to the dynamics is large.

### 4.2 Mechanisms of the interface stabilization and destabilization
#### 4.2.1 Interface stabilization for the conservative dynamics

To better understand the mechanism of the interface stabilization of the conservative dynamics, we consider the interface velocity. In the laboratory reference frame the interface velocity is $\widetilde{\mathbf{V}} = \widetilde{\mathbf{V}}_0 + \widetilde{\mathbf{v}}$, with $\widetilde{\mathbf{v}} \cdot \mathbf{n}_0 = -(\mathbf{u} \cdot \mathbf{n}_0 + \dot{\theta})\big|_{\theta=0^+}$.

For the conservative dynamics $\mathbf{r}_{CD}$, the values are $\mathbf{u} \cdot \mathbf{n}_0\big|_{\theta=0^+} \sim e^{\pm i\sqrt{R}t}$, $\dot{\theta}\big|_{\theta=0^+} \sim e^{\pm i\sqrt{R}t}$ leading to $(\mathbf{u} \cdot \mathbf{n}_0 + \dot{\theta})\big|_{\theta=0^+} \sim e^{\pm i\sqrt{R}t}$ and $\widetilde{\mathbf{v}} \cdot \mathbf{n}_0 \sim e^{\pm i\sqrt{R}t}$. Thus, the interface velocity experiences small stable oscillations near the steady value $\widetilde{\mathbf{V}}_0$, $(\widetilde{\mathbf{V}} - \widetilde{\mathbf{V}}_0) \cdot \mathbf{n}_0 \sim e^{\pm i\sqrt{R}t}$, as discussed in the foregoing

This suggests the inertial effect as the stabilization mechanism of the conservative dynamics $\mathbf{r}_{CD}$ Indeed, when the interface is slightly perturbed, the parcels of the heavy fluid and the light fluid follow the interface perturbation causing the change of momentum and energy of the fluid system. Yet, the dynamics is inertial. To conserve the momentum and energy, the interface should slightly change its velocity. This causes the reactive force occurs and stabilizes the dynamics. The resulting flow in the conservative dynamics is a superposition of two motions: the background motion of the fluids following the interface, whose velocity slightly oscillates near the steady value, and the stable oscillations of the interface perturbations, Figure 11.

#### 4.2.2 Interface destabilization for the classic Landau's dynamics

For the classic Landau-Darrieus dynamics $\mathbf{r}_{LD}$, the interface velocity is constant, $\widetilde{\mathbf{V}} = \widetilde{\mathbf{V}}_0$, as it is postulated by the boundary condition $[\mathbf{u} \cdot \mathbf{n}_0] = 0$, leading to $(\mathbf{u} \cdot \mathbf{n}_0 + \dot{\theta})\big|_{\theta=0^+} = 0$, $\widetilde{\mathbf{v}} = 0$. The constancy of the interface velocity prevents the reactive force to occur. The dynamics is destabilized. The LD unstable flow is a superposition of two motions – the background motion of the fluids following the interface with the constant velocity, and the growth of the interface perturbations, Figure 11.

Since in the classic LD dynamics, mass and momentum are conserved, for capturing the destabilization mechanism, we consider how the condition $[\mathbf{u} \cdot \mathbf{n}_0] = 0$ may influence the energy transport. Remarkably, for ideal incompressible fluids the solution $\mathbf{r}_{LD}$ is incompatible with the condition for energy balance at the perturbed interface [Abarzhi et al 2018]. Indeed, by substituting $[\mathbf{u} \cdot \mathbf{n}] = 0$ ($j_n = 0$) in the condition for the perturbed energy balance $[J_n(w + (\mathbf{J} \cdot \mathbf{j})/\rho^2)] = 0$, one obtains $[J_n(w+(\mathbf{J}\cdot\mathbf{j})/\rho^2)]=[J_n w]=0$ and, with $[J_n]=0$, reduces it further to $[w]=0$ [Abarzhi et al. 2015].

The enthalpy perturbations are $w = p/\rho + \delta e$, where $\delta e$ are the fluctuations of internal energy (in physics sense). In ideal incompressible fluids, free of energy sources, these fluctuations are zero, $\delta e = 0$, because $\dot{e}=0$, $\nabla e = 0$. Thus, with $w=p/\rho$ and with $\rho_h \neq \rho_l$, the condition for energy balance, $[w]=0$, contradicts the condition for momentum balance, $[p]=0$. We see that for ideal incompressible fluids, the classic Landau's dynamics, while conserving the energy (and the physics enthalpy) to the zeros order, $[W_0]=0$, requires energy imbalance at the interface to the first order. This imbalance is induced by the first order enthalpy perturbations, $[w]=[p/\rho]$. This is the work done by the fluid when a parcel of fluid of a unit mass expands its volume from $\rho_h^{-1}$ to $\rho_l^{-1}$ under pressure $p$.

In realistic fluids, this imbalance of the perturbed energy can be induced by energy fluctuations. The effect can be self-consistently derived from entropy conditions with account for chemical reactions. In ideal fluids, to evaluate the effect of the imbalance of perturbed energy on the interface stability, we may introduce an artificial energy flux and study a transition from stable to unstable dynamics with increase of its strength [Abarzhi et al. 2015].

### 4.3 Conservative system with effect of energy fluctuations

The theoretical framework of Landau 1944 describes the evolution of a phase boundary: On each side of the interface, the fluids are ideal and incompressible. They are transformed into one another at the interface. The process is accompanied by a mass flow across the interface. At the planar interface, the fluxes of mass, momentum and energy are fully balanced. For the perturbed interface, in some circumstances, the interface becomes unstable. The instability is accompanied by the appearance of a vortical field, and requires the constancy of the interface velocity, the complete balance of the perturbed momentum, and, for ideal incompressible fluids, an imbalance of perturbed energy, or energy fluctuations by the perturbed interface. Here we model the effect of energy fluctuations on the interface stability and the flow fields' structure [Abarzhi et al. 2015].

Mathematically, it is straightforward to verify that adding stationary energy source at the moving interface in the governing equations, such as $[(\tilde{\mathbf{j}}\cdot\mathbf{n})(W+\tilde{\mathbf{j}}^2/2\rho^2-\tilde{Q})]=0$, keeps the boundary conditions and the interfacial dynamics the same, except for the modification of the zero-order enthalpy as $W_0 \to \tilde{W}_0$, assuming that the fluids are ideal and incompressible. To influence the interface stability, the energy imbalance, or energy fluctuations, should be produced by the perturbed interface.

For realistic fluids, the flux of energy fluctuations occurs due to the difference in thermodynamic properties of the fluids [Kadanoff 2000, Landau & Lifshitz 1987]. It should be derived self-consistently from the condition for entropy conservation [Landau & Lifshitz 1987]. For ideal incompressible fluids, we model the effect of energy fluctuation by introducing an artificial energy flux produced by the perturbed interface. We modify the boundary conditions as [Abarzhi et al. 2015]:

$$[j_n]=0, \left[(p+2J_n j_n/\rho)\mathbf{n}_0\right]=0, \left[J_n(\mathbf{J}\cdot\boldsymbol{\tau}_1+\mathbf{j}\cdot\boldsymbol{\tau}_0)/\rho\right]=0, \left[J_n(w+(\mathbf{J}\cdot\mathbf{j})/2\rho^2)-j_n Q\right]=0 \quad (17)$$

This model Eqs.(17) can be used only for estimates of the effect of the perturbed energy imbalance, or the energy fluctuations, on the interface stability. A self-consistent study is addressed to the future.

### 4.3.1 Fundamental solutions

For the conservative system with energy fluctuations, with $Q_{h(l)} = q_{h(l)} V_h^2$ in Eqs.(17), the matrix M is $M = \widetilde{M}$:

$$\widetilde{M} = \begin{pmatrix} -R & -1 & -\omega+R\omega & i \\ 1 & -1 & 1-R & i\omega/R \\ R-R\omega & R+\omega & 0 & -2iR \\ q_h R+R\omega & q_l-R\omega & q_l\omega+R\omega-R^2\omega-q_h R\omega & i(-q_l+R^2) \end{pmatrix} \quad (18.1)$$

Its rank is 4. It has 4 eigenvalues $\omega_i$ and 4 corresponding eigenvectors $\mathbf{e}_i$. With $q_h - q_l = q$, its determinant is $\det\widetilde{M} = i(R-1)^2(\omega-R)(\omega+R)(\omega^2+R)+iq(\omega-R)((R+1)\omega^2+2R\omega-R(R-1))$. The parameter $q$, $q>0$, describes the strength of energy fluctuations (the perturbed energy imbalance).

With $\det\widetilde{M} = R \det M - q(R/(R-1))\det L$ and $\det\widetilde{M} = R\det M - (q/\omega)(R/(R-1))\det L$, the condition $\det\widetilde{M} = 0$ leads to $(R-1)\det M = q \det L$ and $(R-1)\det M = (q/\omega)\det\widetilde{L}$. Similarly to [Abarzhi et al. 2015], we scale the fluctuations strength as $q = f(R-1)^2 R/(R+1)$, where $f \geq 0$ is a constant, and reduce equation $\det\widetilde{M} = 0$ to

$$(\omega-R)\left[(\omega+R)(\omega^2+R)+f\left(\omega^2+\omega\frac{2R}{R+1}-\frac{R(R-1)}{R+1}\right)R\right]=0 \quad (18.2)$$

The dependence of fundamental solutions $\mathbf{r}_i = \mathbf{r}_i(\omega_i, \mathbf{e}_i)$ with $i=1,2,4$ on the density ratio $R$ and the fluctuation strength $f$ is cumbersome and not presented here. Solution $\mathbf{r}_3 = \mathbf{r}_3(\omega_3, \mathbf{e}_3)$ with $\omega_3 = R$ and $\mathbf{e}_3 = (0, i, 0, 1)$ is the same as for the conservative system, the classic Landau's system, and the dynamic Landau's system. Figure 12 illustrates the dependence of the eigenvalues on the fluctuation strength, and the dependence of the eigenvalues on the density ratio for weak and strong fluctuations.

### 4.3.2 Stability of fundamental solutions

In the limiting cases of weak ( $f \to 0$ ) and strong ( $f \to \infty$ ) fluctuations for fluids with very different ( $R \to \infty$ ) and with very similar ( $R \to 1^+$ ) densities in Eqs.(18), the properties of the dynamics are the following, Table 6, Figure 12.

#### 4.3.2.1 Weak fluctuations

For weak fluctuations and very different densities,

$$f \to 0 \ , \ R \to \infty : \omega_1 \sim i\sqrt{R} - 2f, \ \omega_2 \sim -i\sqrt{R} - 2f, \ \omega_3 \sim R, \ \omega_4 \sim -R . \qquad (18.3)$$

In Eqs.(18.3), solutions $\mathbf{r}_i$ with $i=1,2$ are stable and describe decaying oscillations of the flow fields. Solution $\mathbf{r}_3$ is the same as in the foregoing cases. Solution $\mathbf{r}_4$ is stable, Figure 12.

For weak fluctuations and similar densities,

$$f \to 0 \ , \ R \to 1^+ : \omega_1 \sim i + i(R-1)/2 - f/2, \ \omega_2 \sim -i - i(R-1)/2 - f/2,$$
$$\omega_3 \sim 1 + (R-1), \ \omega_4 \sim -1 - (R-1). \qquad (18.4)$$

In Eqs.(18.4), the stability of fundamental solutions is similar to the corresponding solutions in Eqs.(18.3), except for the difference in the density ratio, Figure 12.

#### 4.3.2.2 Strong fluctuations

For strong fluctuations and very different densities

$$f \to \infty \ , \ R \to \infty : \omega_1 \sim \sqrt{R}, \ \omega_2 \sim -\sqrt{R}, \ \omega_3 \sim R, \ \omega_4 \sim -fR . \qquad (18.5)$$

In Eqs.(18.5), fundamental solutions are similar to the corresponding solutions in the classic Landau system. Solution $\mathbf{r}_1$ is unstable and corresponds to the Landau-Darrieus instability. Solution $\mathbf{r}_2$ is formally stable. Solution $\mathbf{r}_3$ is the same as before. Solution $\mathbf{r}_4$, is stable,, Figure 12.

For strong fluctuations and similar densities

$$f \to \infty \ , \ R \to 1^+ : \omega_1 \sim (R-1)/2, \ \omega_2 \sim -1 - (R-1), \ \omega_3 \sim 1 + (R-1), \ \omega_4 \sim -f \qquad (18.6)$$

In Eqs.(18.6), the stability of fundamental solutions is similar to the corresponding solutions in Eqs.(18.5), expect for the difference in the density ratio, Figure 12.

### 4.3.3 Flow fields of fundamental solutions

#### 4.3.3.1 Weak fluctuations

In Eqs.(18), in case of weak fluctuations and very different densities, solutions $\mathbf{r}_i(\omega_i, \mathbf{e}_i)$ with $i=1,2$ describe slightly decaying oscillations of the flow fields, Figure 12. At the same time, the condition $\text{Re}[\omega_{1(2)}] < 0$ leads to an increase in the vortical velocity component far from the interface; thus the values of $C_{1(2)}, C_1 = C_2^*$ should be zero $C_{1(2)} = 0$ in order to satisfy the boundary conditions at the outside boundaries at all time. Solution $\mathbf{r}_3(\omega_3, \mathbf{e}_3)$ corresponds to unperturbed flow fields in the entire domain at any time for any $C_3$ as before. For solution $\mathbf{r}_4(\omega_4, \mathbf{e}_4)$, due to $\text{Re}[\omega_4] < 0$, the integration constant must be zero, $C_4 = 0$, in order to satisfy the boundary conditions at the outside boundaries. In Eqs.(18), in case of weak fluctuations and very similar densities, for solutions $\mathbf{r}_i(\omega_i, \mathbf{e}_i)$ with $i = 1, 2, 4$, the corresponding constants $C_i$ should be zero, $C_i = 0$ in order to satisfy all the boundary conditions. As before, solution $\mathbf{r}_3(\omega_3, \mathbf{e}_3)$ corresponds to unperturbed flow fields at any time for any $C_3$.

### 4.3.3.2 Strong fluctuations

In Eqs.(18), in case of strong fluctuations and very different densities, fundamental solutions $\mathbf{r}_i(\omega_i, \mathbf{e}_i)$ with $i = 1, 2, 3$ are similar to the corresponding solutions in the classic Landau system. Specifically: (1) For solution $\mathbf{r}_1(\omega_1, \mathbf{e}_1)$ the interface perturbations are strongly coupled with vortical and potential components of the velocity fields. This solution is unstable. (2) For solution $\mathbf{r}_2(\omega_2, \mathbf{e}_2)$, the interface perturbation and vortical and potential velocity components are also strongly coupled. While this solution is formally stable, its velocity field has a vortical structure that decays exponentially in time yet grows away from the interface because of $\text{Re}[\omega_2] < 0$. In order for solution $\mathbf{r}_2(\omega_2, \mathbf{e}_2)$ to satisfy the boundary condition $\mathbf{u}_l|_{z \to +\infty} = 0$ at all times, the integration constant must be $C_2 = 0$. (3) Solution $\mathbf{r}_3(\omega_3, \mathbf{e}_3)$ corresponds to unperturbed flow fields in the entire domain at any time and for any $C_3$, as before. (4) For the formally stable solution $\mathbf{r}_4(\omega_4, \mathbf{e}_4)$, due to $\text{Re}[\omega_4] < 0$, the integration constant must be $C_4 = 0$ in order to satisfy the boundary condition at the outside boundaries of the domain at any time. In Eqs.(18), in case of strong fluctuations and similar densities, the solutions are similar to those with very different densities. These similar features include the development of LD-type of instability for solution $\mathbf{r}_1(\omega_1, \mathbf{e}_1)$, the choice of $C_2 = 0$ for solution $\mathbf{r}_2(\omega_2, \mathbf{e}_2)$ with $\text{Re}[\omega_2] < 0$, the unperturbed flow fields for solution $\mathbf{r}_3(\omega_3, \mathbf{e}_3)$, and the choice of $C_4 = 0$ for solution $\mathbf{r}_4(\omega_4, \mathbf{e}_4)$ with $\text{Re}[\omega_4] < 0$.

### 4.3.4 Interface velocity for fundamental solutions

Properties of fundamental solutions for the conservative dynamics with effect of energy fluctuations indicate that for $f > 0$ the interface velocity in laboratory frame of reference is constant, $\widetilde{\mathbf{V}} = \widetilde{\mathbf{V}}_0$ for these solutions in Eqs.(18), either in case of strong $f \to \infty$ or weak $f \to 0^+$ fluctuations. Indeed, solution $\mathbf{r}_3$ has the null perturbed fields of the velocity and the interface at any integration constant $C_3$. For stable solutions $\mathbf{r}_{2(4)}$ the integration constants should be zero to satisfy the boundary conditions at the outside boundaries of the domain, $C_{2(4)} = 0$. For solution $\mathbf{r}_1$ the integration constant should be zero in case of weak fluctuations, $f \to 0^+$, $f \neq 0$, to satisfy the boundary conditions at the outside boundaries of the domain; whereas in case of strong fluctuations $f \to \infty$ the solution approaches the classic LD solution $\mathbf{r}_1 \to \mathbf{r}_{LD}$ and has constant interface velocity $\tilde{\mathbf{V}} = \tilde{\mathbf{V}}_0$, as discussed in the foregoing.

In the limit of zero fluctuations, $f \equiv 0$, the interface velocity experiences stable oscillations near the steady value $\tilde{\mathbf{V}}_0$. These oscillations enable the inertial stabilization mechanism of the conservative dynamics to occur, as found in the foregoing.

### 4.3.5 Formal properties of the conservative system with effect of energy fluctuations

To study formal properties of the dynamic Landau's system, we find for matrix $\mathbf{M} = \tilde{M}$ the associated matrices $\mathbf{S} = S_{\tilde{M}}$ and $\mathbf{P} = P_{\tilde{M}}$

$$S_{\tilde{M}} = \begin{pmatrix} -R & -1 & 0 & i \\ 1 & -1 & 1-R & 0 \\ R & R & 0 & -2iR \\ q_h R & q_l & 0 & i(-q_l + R^2) \end{pmatrix}, \quad P_{\tilde{M}} = \begin{pmatrix} 0 & 0 & 1-R & i \\ 0 & 0 & 0 & -i/R \\ R & -1 & 0 & 0 \\ -R & R & -q_l - R + q_h R + R^2 & 0 \end{pmatrix} \quad (18.7)$$

Equations $det(P_{\tilde{M}}^{-1} S_{\tilde{M}} - \omega I) = 0$ and $det \tilde{M} = 0$ yield the same eigenvalues $\omega = \{\omega_i\}, i = 1.,..4$. The conservative system with effect of energy fluctuations is non-degenerate, and has four fundamental solutions for 4 degrees of freedom.

### 4.3.6 Summary of properties of fundamental solutions for the conservative system with effect of energy fluctuations

#### 4.3.6.1 Outline of properties of fundamental solutions

For strong and weak fluctuations and for fluids with similar and contrasting densities, there are four fundamental solutions $\mathbf{r}_i(\omega_i, \mathbf{e}_i)$ with $i = 1,2,3,4$, Eqs.(18), Table 6, Figure 12. Solution, $\mathbf{r}_3(\omega_3, \mathbf{e}_3)$, corresponds to unperturbed flow fields for any $C_3$ and any $f$. In the case of strong fluctuations, $f \to \infty$, solution $\mathbf{r}_1(\omega_1, \mathbf{e}_1)$ describes (for $C_1 \neq 0$) unstable dynamics of LD-type, whereas for solutions $\mathbf{r}_2(\omega_2, \mathbf{e}_2)$ and $\mathbf{r}_4(\omega_4, \mathbf{e}_4)$ the constants are $C_{2(4)} = 0$. In case of weak fluctuations, $f \to 0$, for solutions $\mathbf{r}_1(\omega_1, \mathbf{e}_1)$, $\mathbf{r}_2(\omega_2, \mathbf{e}_2)$, and $\mathbf{r}_4(\omega_4, \mathbf{e}_4)$ the constants are $C_1 = C_2 = C_4 = 0$. For $f > 0$ the interface velocity in laboratory frame of reference is constant, $\tilde{\mathbf{V}} = \tilde{\mathbf{V}}_0$ for these solutions.

At $f \equiv 0$, solutions $\mathbf{r}_1(\omega_1, \mathbf{e}_1)$ and $\mathbf{r}_2(\omega_2, \mathbf{e}_2)$ with non-zero $C_1, C_2$ describe stable oscillations of the flow fields for the conservative dynamics, accompanied by stable oscillations of the interface velocity near the steady value $\tilde{\mathbf{V}}_0$.

#### 4.3.6.2 Transition from the stable to the unstable dynamics

Transition from the stable to unstable dynamics is illustrated in Figure 12 showing eigenvalues $\omega_i(R, f)$ with $i = 1,2,3,4$ in Eqs.(17,18) for a fixed $R$ as a function of $f$, as well as the in cases of small and large $f$. For very small $f$, the values $\mathrm{Im}[\omega_{1(2)}]$ and $\omega_{3(4)}$ are indistinguishable from the corresponding values in the conservative dynamics. For very large $f$, eigenvalues $\omega_i$ with $i = 1,2,3$ are indistinguishable from the corresponding values in the classic Landau system.

The properties of the flow fields of fundamental solutions for the conservative system with energy fluctuations indicate that when the value $f$ changes from $0$ to $\infty$ in system Eqs.(17,18), an abrupt transition may occur to the unstable dynamics. Critical value $f = f_{cr}$, at which the transition is expected, can be estimated from condition $\omega = 0$ in Eqs.(17,18) leading to $f_{cr} = (R+1)/(R-1)$, $q = q_{cr} = R(R-1)$, and $(Q_h - Q_l)_{cr} = R(R-1)V_h^2$.

The flow dynamics may behave as follows: At $f \equiv 0$ the flow fields experience stable oscillations near the uniform. For $0 < f < f_{cr}$, the flow fields are uniform flow fields. For $f > f_{cr}$ the dynamics becomes unstable. For large fluctuations, $f \to \infty$, the unstable dynamics fully resembles the properties of the Landau-Darrieus instability.

In realistic fluids, other mechanisms are possible that account for the influences of entropy flux and of thermodynamic properties of the fluids on energy fluctuations produced by the perturbed interface [Kadanoff 2000, Landau & Lifshitz 1987], as well as for nonlinear effects [Sivashinsky 1983].

## 5  Comparison with other studies

In this section we discuss connections of our work to the seminal studies in the theory and the experiment of combustion. On the basis of our analysis, we propose new experiments and diagnostics, and discuss chemically reactive systems.

### 5.1  Landau theoretical framework and theory of combustion

While the theoretical framework [Landau 1994] describes the dynamics of a phase boundary broadly defined, the LDI is traditionally associated with the theory of combustion. Landau 1944 considers the evolution of a flame in premixed combustion by treating the unburned gas as a heavy fluid, the burned gas - as a light fluid, both fluids - as ideal and incompressible, and the interface – as a discontinuity.

A premixed flame is a self-sustained wave of an exothermic reaction in a gaseous mixture [Sivashinsky 1983, Williams 1965, Zeldovich 1944]. The wave is initiated by a spark, propagates through the mixture, and produces combustion products. A freely propagating flame is a multi-scale, multi-parameter, multi-physics, and multi-chemistry system. Its evolution is the classical long-standing problem [Peters 2000]. For planar and steady flames, the interface dynamics is well captured by the rigorous theories [Mayo et al 1990, Frank-Kamenetsky 1969, Zeldovich & Frank-Kamenetsky 1938, Mallard & Chatelier 1883]. For unsteady and curved flames, the interaction of the flame with the hydrodynamic flow that it produces is still a challenge [Peters 2000, Liberman 1994, Sivashinsky 1983, Williams 1965].

Landau's result on the unconditional instability of the flame front contradicts some experiments, because stable flames exist and are observed in a laboratory [Clanet & Searby 1998, Buckmaster 1979, Markstein 1951]. Several approaches have been developed to explain these observations. Theoretical research has focused on bridging the gap between the idealistic framework and the realistic conditions, and on augmenting the theory [Landau 1944] with physical effects typical for flames in premixed combustion [Clavin & Searby 2016, Peters 2000, Liberman et al. 1994, Clavin 1985, Searby et al. 1983, Williams 1971, Sivashinsky 1983, Istratov & Librovich 1966, Zeldovich 1944]. These effects have included the finite thickness of the flame and the finite curvature of the interface [Markstein 1951], the molecular diffusion and the thermal expansion [Clavin & Williams 1982], the dynamics of a reaction zone inside a thin flame [Matalon & Matkowsky 1982] and the kinematic viscosity [Frankel & Sivashinsky 1982], as well as assumptions of a gradual change of the flow quantities across the interface [Williams 1971]. These fundamental works have found that various physical effects can stabilize the development LDI at small scales [Sivashinsky 1983]. The seminal work [Clavin 1985] has summarized these studies, reconciled the theory and the experiment, and found that in the premixed combustion the flame stability depend on the interplay of compressibility, mass diffusion and heat diffusion.

Our results agree with the Landau's results [Landau 1944] and thus substantiate that the seminal theories of premixed combustion are accurate and complete, within the range of their applicability [Peters 2000, Clavin 1985, Sivashinsky 1983, Matalon & Matkowsky 1982, Williams 1971, Markstein 1951].

Particularly, by applying proper transformations of variables, one can directly match our results and the classical results [Peters 2000, Landau & Lifshitz 1987]. This includes the governing equations, the boundary conditions, the structure of the solution, the characteristic length scale of the vortical field, and the instability growth-rate. Furthermore, the flow fields in Figures 8-9 can be obtained with some effort from the classical works [Peters 2000, Landau & Lifshitz 1987, Landau 1944, Zeldovich 1944].

On the side of experiments, the implementation and diagnostics of the Landau-Darrieus instability is an extreme challenge. It is remarkable that the first experimental study of the Landau-Darrieus instability [Clanet & Searby 1998] has been reported nearly 60 years after the theoretical work has been published [Landau 1944]. These experiments have measured the LDI growth-rate in the compressible reactive fluids (gases) for a relative thick interface (Markstein parameter 4.0) in the presence of the acceleration (gravity) under the influence of the acoustic parametric forcing. The direct quantitative comparison of these experimental results and our theory is challenge because the experimental conditions strongly depart from our theoretical approximation [Clanet & Searby 1998].

### 5.2    New experiments and diagnostics

The existing experiments and simulations of the interfacial dynamics are focused on the evolution of the perturbation amplitude [Peters 2000, Clanet & Searby 1998, Clavin 1985, Sivashinsky 1983]. Our theoretical analysis, in addition to identifying the evolution of the perturbation amplitude, finds that the flow is highly sensitive to the interfacial boundary conditions and that the properties of interfacial transports at microscopic scales can be linked to the flow fields at macroscopic scales. Note that Figures 3-6,8-10 present the perturbations of the flow fields. In order to obtain the flow fields, the corresponding zero-order component should be added to the perturbations.

Figure 13 illustrates the velocity streamlines $\mathbf{S}_{h(l)}$ of the heavy (light) fluids for the conservative dynamics and the classic Landau's dynamics at some density ratio and some instance of time. Each plot has its own range of values. The streamlines are defined as $(d\mathbf{S}_{h(l)}/dt) \times \mathbf{v}_{h(l)}$, where the heavy (light) fluid velocity is $\mathbf{v}_{h(l)} = \mathbf{V}_{h(l)} + \mathbf{u}_{h(l)}$. The integration constants for solutions $\mathbf{r}_{CD}$ and $\mathbf{r}_{LD}$ are chosen such that the amplitude of the interface perturbation is same for the conservative dynamics and the classic Landau's dynamics and equals to $|\bar{z}| = |kZ| = 1/3$. Figure 13 illustrates slight deviations of the velocity streamlines from straight lines for both conservative dynamics and the classic Landau's dynamics. For the conservative dynamics, the deviations of the velocity streamlines are clearly seen in the heavy fluid; they are also present in the light fluid though they are somewhat less pronounced due to the large value of the uniform velocity of the light fluid. For the classic Landau's dynamics, the deviations of the velocity streamlines are clearly seen in both fluids. For a given value of the perturbation amplitude, the fluid motion is more intense in the classic Landau's dynamics (due to the presence of the vortical components of the velocity of the light fluid) when compared to that in the conservative dynamics. The results in Figure 4, 8 and 13 are fully consistent with one another.

Note that due to the small values of the perturbed velocities, $|\mathbf{u}_{h(l)}| \ll |\mathbf{V}_{h(l)}|$, the properties of the flow fields in the fluid bulk may be a challenge to precisely diagnose. For addressing these challenges, advanced experimental technologies may be applied [Orlov et al. 2010]. New experiments with new implemented diagnostics can be conducted on the basis of our results to better understand the interfacial dynamics. Here we outline some parameters of a new experiment.

According to our results, in the experiment, the interface velocity is sub-sonic, the dynamics is observed from a fat field, and its characteristic length scale $1/k = \lambda/2\pi$ and time scale $\tau = 1/kV_h = \lambda/2\pi V_h$ are greater than those that are induced by the processes of dissipation and diffusion and by a finite thickness of the interface. These scales are set by the initial conditions. To select the density ratio for the experimental fluid system, we notice that in the classical Landau's solution for a given $\lambda$ the length-scale $\tilde{\lambda}$ of the vortical field is a non-monotone function of the density ratio $R$ with $\tilde{\lambda}/\lambda = k/\tilde{k} = R(1+R)/(-R + \sqrt{-R + R^2 + R^3})$. The minimum value of $\tilde{\lambda}/\lambda$ is achieved at $R = 2 + \sqrt{5} \approx 4.24$. Thus, the vortical field can be easier to diagnose when the fluids density ratio is $R = 2 + \sqrt{5} \approx 4.24$ and the ratio of uniform velocities is $V_l/V_h = 2 + \sqrt{5} \approx 4.24$.

In this case, the Landau solution has the following characteristics:

$$R = 2 + \sqrt{5}, \quad k/\tilde{k} = 2 + \sqrt{5}, \quad \omega = 1, \quad \mathbf{e}_1 = \left( \frac{-i(-1+\sqrt{5})}{2}, \frac{i(1+\sqrt{5})}{2}, \frac{-2i}{1+\sqrt{5}}, 1 \right)^T \quad (19)$$

The numerical values are $R = 4.24$, $\tilde{\lambda}/\lambda = 4.24$, $\omega = 1$ and $\mathbf{e}_1 = (-0.62i, 1.6i, 0.62i, 1)$. The choice of the integration constant for the solution is suggested by the requirement that the perturbations of the flow fields are small. By evaluating the amplitude of the interface perturbation for which linear theory is valid as $|\bar{z}| = 1/3$, we estimate the integration constant as $|C| = 5.4 \times 10^{-1}$. By implementing in experiments the fluid density $R = 4.24$ and the corresponding initial conditions, one can further diagnose the flow fields in a vicinity of the interface and in the bulk, and compare with the theoretical solutions.

The experimental outcome may include the following scenarios. (1) The interface is stable; the perturbed flow fields are the same as those of the conservative solution; they slightly oscillate near the equilibrium. This can be interpreted as the absence of energy fluctuations and/or as their weak effect. (2) The interface is unstable; the flow fields are the same as in the classical Landau solution. This can be interpreted as the existence of the LDI and the applicability of the Landau theoretical framework for the experimental system. (3) The interface is unstable; the flow fields depart qualitatively from those of the classical LDI and from those of the conservative dynamics with strong energy fluctuations. This can be interpreted as that the approximation of ideal incompressible fluids is not a valid approximation for the fluid system. It may also suggest that the fluid system experiences a hydrodynamic instability that is different from the LDI. If the instability growth-rate of this instability agrees with the results of the seminal theories of combustion, it may give some grounds to name those instabilities in honor of those, who have studied its properties [Peters 2000, Clanet & Searby 1998, Liberman et al. 1994, Clavin 1985, Searby et al. 1983, Sivashinsky 1983, Matalon & Matkowsky 1982, Williams 1971, Markstein 1951].

### 5.3 Chemically reactive systems

In reactive fluids it is generally well understood that chemically reactive systems can be hydrodynamically unstable [Ilyin et al. 2018a, Mayo et al. 1990]. Yet, it is a challenge to construct a model experiment or a simulation that cleanly displays significant chemical reaction instability at a simple interface. The atomistic simulations can study the energy transport in a reactive system and the effect of chemical reaction on the interface stability. Atomistic simulations of combustible systems find that even for simplest chemical reactions realistic combustion is very complex and is accompanied by a chain of processes strongly influencing reaction constants and energy transports. Further investigations are required to fully understand the properties of chemistry-induced instabilities at atomistic and continuous scales [Ilyin et al. 2018]. Further advancements of traditional combustion models are required to better understand and reliably model realistic flames and their stability. Our general method can serve to expand the theory of combustion beyond diffusive approximations.

## 6 Discussion

We have studied in a far field approximation the dynamics of the interface separating incompressible ideal fluids with different densities and having the interfacial mass flux, Eqs.(1-3). We have developed and applied the general matrix method to solve the boundary value problem for the system of governing equations linearized near the equilibrium state of the uniform flow fields, Eqs.(1-11). We have formally derived the fundamental solutions, have found their eigenvalues and eigenvectors, and have directly linked the interface stability to the structure of the flow fields, Figure 1-13, Table 1-6. Our theoretical framework is consistent with and generalizes the classic approach [Landau & Lifshitz 1987] and assumes sharp changes of the flow quantities at the interface and the negligible effects of dissipation, diffusion, compressibility, surface tension, and the interface thickness. We have identified extreme sensitivity of the interfacial dynamics to the interfacial boundary conditions, Eqs.(12-18). We have found the new properties of the interface dynamics for the conservative and the Landau's systems, and the new mechanisms of the interface stabilization and destabilization, Eqs.(12-18), Figure 1-13.

Particularly, the conservative dynamics is stable, and is stabilized by the reactive force and the inertial effects. The stable solutions are traveling waves; their interference results in the stably oscillating standing waves. The stable flow is a superposition of two motions – the background motion of the fluids following the interface, whose velocity slightly oscillates near the steady value, and the stable oscillations of the interface perturbations, Eqs.(12), Figure 1,2,4,11. For the classic Landau's dynamics, the interface velocity is constant, the reactive force is absent, and the dynamics is unstable. The unstable solution is the standing wave with the increasing amplitude. The LD unstable flow is a superposition of two motions – the background motion of the fluids following the interface with the constant velocity, and the growth of the interface perturbations, Eqs.(14), Figure 1,7,8,11. For the stable conservative dynamics, the velocity fields are potential in the fluids' bulk, Figure 4. For the classic unstable LD dynamics, the potential and vortical components of the velocity fields are strongly coupled with the interface perturbation, Figure 8. In either case, there is the null velocity shear at the interface. In either case, the velocity streamlines slightly deviate from straight lines, though for the unstable Landau's dynamics the deviations are stronger when compared to those in the conservative dynamics, Figure 13.

Our results clearly indicate that the problem of the Landau-Darrieus instability remains a source of inspiration for theory research. Landau 1944 analyzed the interface stability by applying compatibility conditions and identifying 2 eigenvalues of the dynamics [Landau 1944]. Over the past 70 years, the theoretical framework of Landau 1944 has been extended to incorporate the properties of realistic environments. This has been done, for instance, in the seminal papers [Peters 2000, Clavin 1985, Sivashinsky 1983, Matalon & Matkowsky 1982, Williams 1971, Markstein 1951]. Our work focuses on general conditions of stability of the interface with the interfacial mass flux, considers incompressible ideal fluids from a far field, and treats the interface as a discontinuity, Eqs.(1-16). We have developed and employed the general matrix method and have found the formal fundamental solutions for the linearized dynamics directly linking the interface stability to the structure of the flow fields', Figure 1 – 13. To our knowledge, this detailed and rigorous consideration has not been performed before.

By applying our general matrix method, we have found that the conservative system is non-degenerate, and has 4 fundamental solutions associated with 4 independent degrees of freedom, Eqs.(12), Table 1,3, Figure 2-6. These include the two stable traveling waves, whose interference results in two stably oscillating standing waves, and the two other solutions – the unstable solution with the null perturbation fields of the velocity and pressure, and the stable solution with the null integration constant. We have further found that the classic Landau system is degenerate and has only 3 fundamental solutions, Eqs.(14), Table 2-4, Figure 5,7-9. These include the unstable standing wave corresponding to the LDI, the stable solution with the null integration constant, and the unstable solution with the null perturbation fields of the velocity and pressure, which is the same as for the conservative dynamics. The degeneracy is associated with the static nature of the special postulated condition at the interface [Landau 1944], and can be eliminated with the use of a more general condition.

When the degeneracy is eliminated by modifying the postulated condition of zero perturbed mass flux by a constant mass flux across the interface, Eqs.(15), the 4th solution appears, Eqs.(16), Table 2,5, Figure 10. This solution is neutrally stable. It has a null eigenvalue, and its vortical field can seed the LDI. The null eigenvalue implies that the perturbations of the flow fields can be scale-invariant power-law functions of time rather than scale-dependent exponential functions. Furthermore, by noting that the interface is a phase boundary at which one fluid is transformed into another, this result is consistent with the linear response theory of phase transitions, where thermodynamic fluctuations are known to be power-laws rather than exponentials [Landau & Lifshitz 1987, Kadanoff et al. 1967, Kadanoff 2000].

We have identified the new stabilization mechanism of the conservative dynamics. It is due to the inertial effects that cause the reactive force to occur and, by influencing the interface velocity, to balance the change of momentum produced by the interface perturbation. Note that this inertial stabilization mechanism leads to the new hydrodynamic instability of the accelerated interface that develops when the acceleration is directed from the heavy fluid to the light fluid, and when the acceleration magnitude exceeds a threshold [Abarzhi et al. 2018, Ilyin et al. 2018b,c]. Both in the inertial and accelerated conservative dynamics, the flow has potential velocity fields in the fluid bulk and is shear free at the interface [Abarzhi et al. 2018, Ilyin et al. 2018b,c]. In the LDI this inertial stabilization mechanism is absent, due to the perfect constancy of the interface velocity. Furthermore, in ideal incompressible fluids, the development of the LDI is associated with an imbalance of perturbed energy, due to the energy fluctuations by the perturbed front. To model the effect of energy of fluctuations (i.e., the imbalance of the perturbed energy), we have introduced an artificial energy flux, and have found that the LDI may develop when the fluctuations are strong. The structure of the flow fields of the solutions for the conservative dynamics with effect of energy fluctuations suggests a possibility of an abrupt transition to unstable dynamics with the increase of fluctuations strength over a critical value, Figure 12. In realistic systems, the energy imbalance can be induced by chemical reactions and by difference in thermodynamic properties of the fluids [Abarzhi et al. 2018, Ilyin et al. 2018a-c].

To our knowledge, these properties of the conservative dynamics and the Landau's dynamics have not been discussed in other studies [Abarzhi et al. 2018]. Traditionally, it is believed that an interface separating nearly ideal incompressible fluids and having an interfacial mass flux is a subject to the LDI at large scales, and that in realistic environment the LDI is a challenge to implement because the effects of dissipation, diffusion, and finite interface thickness stabilize the small scales [Peters 2000]. Our analysis is fully consistent with these results: For fluids with similar densities, the contribution of vortical fields to the dynamics is small, and the fluxes induced in realistic fluids by the stabilizing effects may dominate the dynamics and define the interface stability [Abarzhi et al. 2015]. For fluids with very different densities, a more careful consideration is required. Our analysis yields qualitative and quantitative characteristics of the dynamics that have not been measured before, and that require new diagnostics, Figure 1-13 [Abarzhi et al. 2018, Abarzhi et al. 2015].

One such experiment can be a study of the dynamics of fluids with very different densities, with diagnostics of the flow fields near the interface and in the fluids' bulk, and with the measurements of the interface evolution, including the interface velocity as a whole and the interface perturbation growth-rate. By comparing the observations with our benchmarks, one can further identify the fundamentals of the interfacial dynamics in realistic environments, and elaborate new approaches for the flow control, Figure 1-13. Several questions may frame these studies: Can the LDI unconditionally develop at the large scales? How can the dynamics be stabilized - by inertial effect, by dissipation and diffusion, or by their combination? How strong should energy fluctuations be to destabilize the flow? Can these fluctuations be induced by chemical reactions? How are the properties of chemistry-induced instabilities compared to those of the LDI?

Existing experimental and numerical studies of the interface stability are focused on the growth of the perturbation amplitude [Abarzhi et al. 2018]. Our analysis derives the amplitude growth-rate, and finds that the dynamics is highly sensitive to interfacial boundary conditions. According to our theory, by measuring at macroscopic scales the flow fields in the bulk and at the interface, one can capture the properties of the interfacial transport at microscopic scales, Figure 1-13. This information is particularly important in systems where experimental parameters can be difficult to control and where data are a challenge to obtain. They include, for instance, supernovae in astrophysics, laser ablation in plasma physics, plumes in the solar atmosphere, as well as convection in planetary interiors, and industrial applications [Abarzhi et al. 2013]. Furthermore, by directly connecting the macroscopic observations far from the interface to the microscopic properties at the interface, one may get a better understanding of connection between the Lagrangian and Eulerian dynamics away from equilibrium and potentially achieve a better control of complex transport processes involving phase transitions [Abarzhi 2010].

Our approach can be employed in a broad range of problem with interfacial dynamics. Particularly, our theoretical framework can be expanded to other multiphase flows, such as ablative Rayleigh-Taylor instabilities, detonation to deflagration transition, convection in stellar and planetary interiors, and stability of liquid-liquid and liquid-vapor interface, [Abarzhi et al. 2018, Ilyin et al. 2018a-c, Hurricane et al. 2014, Anisimov et al. 2013, Abarzhi 2010, Meshkov 2006, Bell et al. 2004, Piriz & Portugues 2003, Remington et al. 2000, Stein & Norlund 2000, Bodner et al. 1998, Sanz 1994, Prosperetti & Plesset 1984, Williams 1961]. When augmented with linear response theory, it can also serve to analyze dynamic processes in fluids experiencing phase transitions and super-critical fluids [Kadanoff 2000, Landau & Lifshitz 1987, Kadanoff et al. 1967]. Our approach can self-consistently account for the effects of compressibility, viscosity, heat conduction, mass ablation and mass diffusion, and thus enable a unified treatment of the interfacial dynamics in realistic environments [Landau & Lifshitz 1987]. We address these studies to the future.

## 7. Conclusion

To conclude, we have analyzed from a far field the dynamics of the interface separating incompressible ideal fluids with different densities and with mass flux across it. We have found fundamental solutions of the dynamics and directly linked the interface stability to the flow fields' structure, associating rigorous mathematical attributes to physical observables. Our approach is consistent with the spirit of Landau's approach and identifies the new properties of the interface evolution and the new mechanisms of the interface stabilization and destabilization.

## 8. Acknowledgments

The authors thank for support the University of Western Australia in the AUS, the National Science Foundation in the USA, the Summer Undergraduate Research Fellowship Program at California Institute of Technology in the USA, and the Japan Society for the Promotion of Science in Japan. SIA expresses her deep gratitude to Prof Leo P Kadanoff for comments and remarks and for discussions on singularities and similarities.

## 10. Tables

Table 1: Attributes of fundamental solutions for the conservative system (1,2,3,4) with asterisk marking a quantity

|  | $\mathbf{r}_1$ | $\mathbf{r}_2$ | $\mathbf{r}_3$ | $\mathbf{r}_4$ |
|---|---|---|---|---|
| $\text{Re}[\omega]$ | 0 | 0 | >0 | <0 |
| $\text{Im}[\omega]$ | >0 | <0 | 0 | 0 |
| $C$ | * | * | * | 0 |
| $\mathbf{u}_h$ | * | * | 0 | 0 |
| $\mathbf{u}_l$ | * | * | 0 | 0 |
| $\nabla \times \mathbf{u}_l$ | 0 | 0 | 0 | 0 |
| $z^*$ | * | * | 0 | 0 |

Table 2: Attributes of fundamental solutions for the classic Landau's system (1,2,3) and the dynamic Landau's system (1,2,3,4) with asterisk marking a quantity.

|  | $\mathbf{r}_1$ | $\mathbf{r}_2$ | $\mathbf{r}_3$ | $\mathbf{r}_4^o$ |
|---|---|---|---|---|
| $\text{Re}[\omega]$ | >0 | <0 | >0 | 0 |
| $\text{Im}[\omega]$ | 0 | 0 | 0 | 0 |
| $C$ | * | 0 | * | * |
| $\mathbf{u}_h$ | * | * | 0 | * |
| $\mathbf{u}_l$ | * | * | 0 | * |
| $\nabla \times \mathbf{u}_l$ | * | * | 0 | * |
| $z^*$ | * | * | 0 | * |

Table 3: Eigenvalues of standing waves

| $\mathbf{r}_{CD}$ | $\omega_{CD} = \pm i\sqrt{R}$ |
|---|---|
| $\mathbf{r}_{LD}$ | $\omega_{LD} = \left(-R + \sqrt{(R^3 + R^2 - R)}\right)/(1+R)$ |
| $\mathbf{r}_{DLD}$ | $\omega_{DLD} = 0$ |

Table 4: Properties of the conservative dynamics (CD) and the classic Landau-Darrieus (LD) dynamics

|  | CD | LD |
|---|---|---|
| Conservation properties | Conserves mass, momentum and energy at the interface | Conserves mass and momentum, has zero perturbed mass flux at the interface |
| Interface velocity | Slight stable oscillations near the constant value | Constant value (postulate) |
| Flow field | Potential velocity fields | Vortical structures in the light fluid |
| Inerfacial shear | Shear-free | Shear-free |
| Formal properties | Non-degenerate; 4 fundamental solutions and 4 degrees of freedom | Degenerate; 3 fundamental solutions and 4 degrees of freedom |
| Stability | Stable; stabilized by inertial effects | Unstable |

Table 5: Properties of the classic Landau-Darrieus (LD) and dynamic Landau-Darrieus (DLD) systems

|  | LD | DLD |
|---|---|---|
| Conservation properties | Conserves mass and momentum, has the null perturbed interfacial mass flux | Conserves mass and momentum, has constant perturbed mass flux at the interface |
| Interface velocity | Constant value (postulate) | Slight shift near the constant value |
| Flow field | The vortical field length-scale depends on the density ratio and is large and finite when compared to the perturbation wavelength | The vortical field length-scale is independent of the density ratio and is infinite when compared to the perturbation wavelength |
| Inerfacial shear | Shear-free | Shear-free |
| Formal properties | Degenerate; 3 fundamental solutions and 4 degrees of freedom | Non-degenerate; 4 fundamental solutions and 4 degrees of freedom |
| Stability | Unstable | Neutrally stable |

Table 6: Attributes of fundamental solution for the conservative system with energy fluctuations with asterisk marking a quantity.

| | $f \equiv 0$ | $f > 0$ | $0 < f < f_{cr}$ | $f_{cr} < f$ |
|---|---|---|---|---|
| **$\mathbf{r}_1$** | | | | |
| $\text{Re}[\omega]$ | 0 | <0 | <0 | >0 |
| $C$ | * | 0 | 0 | * |
| **$\mathbf{r}_2$** | | | | |
| $\text{Re}[\omega]$ | 0 | <0 | <0 | <0 |
| $C$ | * | 0 | 0 | 0 |
| **$\mathbf{r}_3$** | | | | |
| $\text{Re}[\omega]$ | >0 | >0 | >0 | >0 |
| $C$ | * | * | * | * |
| **$\mathbf{r}_4$** | | | | |
| $\text{Re}[\omega]$ | <0 | <0 | <0 | <0 |
| $C$ | 0 | 0 | 0 | 0 |

## 11. Figure captions

Figure 1: Schematics of the flow dynamics in a far field approximation (not to scale). Blue color marks the planar (dashed line) interface and the perturbed (solid line) interface.

Figure 2: Dependence of the eigenvalues on the density ratio for the conservative system.

Figure 3: Fundamental solution $\mathbf{r}_1$ for the conservative system at some density ratio and some instance of time in the motion plane. The solution is the stable traveling wave. Plots of the interface perturbation, the perturbed velocity vector fields, the perturbed velocity streamlines, and the perturbed pressure.

Figure 4: Fundamental solution $\mathbf{r}_{CD} = (\mathbf{r}_1 + \mathbf{r}_2)/2$ for the conservative system at some density ratio and some instance of time in the motion plane. The solution is the stable standing wave and is resulted from the interference of the stable traveling waves. Plots of the interface perturbation, the perturbed velocity vector fields, the perturbed velocity streamlines, and the perturbed pressure.

Figure 5: Fundamental solution $\mathbf{r}_3$ for the conservative system as well as for the classic Landau's system, dynamic Landau's system and the conservative system with energy fluctuations at some density ratio and some instance of time in the motion plane. (a) Plots of the interface perturbation, the perturbed velocity vector fields, the perturbed velocity streamlines, and the perturbed pressure. (b) Plots of the perturbed velocity vortical component, the vorticity and the interface perturbation.

Figure 6: Fundamental solution $\mathbf{r}_4$ for the conservative system at some density ratio and some instance of time in the motion plane. (a) Plots of the interface perturbation, the perturbed velocity vector fields, the perturbed velocity streamlines, and the perturbed pressure. (b) Plots of the perturbed velocity vortical component, the vorticity and the interface perturbation.

Figure 7: Dependence of the eigenvalues on the density ratio for the classic and the dynamic Landau systems. The classic Landau system has three eigenvalues (1,2,3). The dynamic Landau system has four eigenvalues: the first three are identical to the classic system eigenvalues; the fourth behaves as shown.

Figure 8: Fundamental solution $\mathbf{r}_1$, $\mathbf{r}_{LD} = \mathbf{r}_1$ for the classic Landau's system at some density ratio and some instance of time in the motion plane. (a) Plots of the interface perturbation, the perturbed velocity vector fields, the perturbed velocity streamlines, and the perturbed pressure. (b) Plots of the perturbed velocity vortical component, the vorticity and the interface perturbation.

Figure 9: Fundamental solution $\mathbf{r}_2$ for the classic Landau's system at some density ratio and some instance of time in the motion plane. (a) Plots of the interface perturbation, the perturbed velocity vector fields, the perturbed velocity streamlines, and the perturbed pressure. (b) Plots of the perturbed velocity vortical component, the vorticity and the interface perturbation.

Figure 10: Fundamental solution $\mathbf{r}_4^o$, $\mathbf{r}_{DLD} = \mathbf{r}_4^o$, for the dynamic Landau's system at some density ratio and some instance of time in the motion plane. (a) Plots of the interface perturbation, the perturbed velocity vector fields, the perturbed velocity streamlines, and the perturbed pressure. (b) Plots of the perturbed velocity vortical component, the vorticity and the interface perturbation.

Figure 11: Schematics of the flow dynamics in a far field approximation (not to scale) for the conservative dynamics (left) and the classic Landau's dynamics (right) in the inertial reference frame. Blue color marks the planar interface (dashed) and the perturbed interface (solid). For the conservative dynamics the blue double arrows mark the oscillations of the interface perturbations (solid) and the interface velocity as a whole (dashed) with the latter occurring due to inertial effects and causing the reactive force to occur. For the classic Landau's dynamics the single blue arrows mark the growth of the interface perturbations; the velocity of the interface is postulated constant.

Figure 12: Dependence of the eigenvalues on the fluctuations strength for the conservative system with energy fluctuations (a) at some density ratio; (b) top - for a broad range of fluctuations' strength at some density ratio and bottom – for (left) weak fluctuations and (right) strong fluctuations as a function on density ratio.

Figure 13: Velocity streamlines for the conservative (left) and the classic Landau's (right) dynamics.

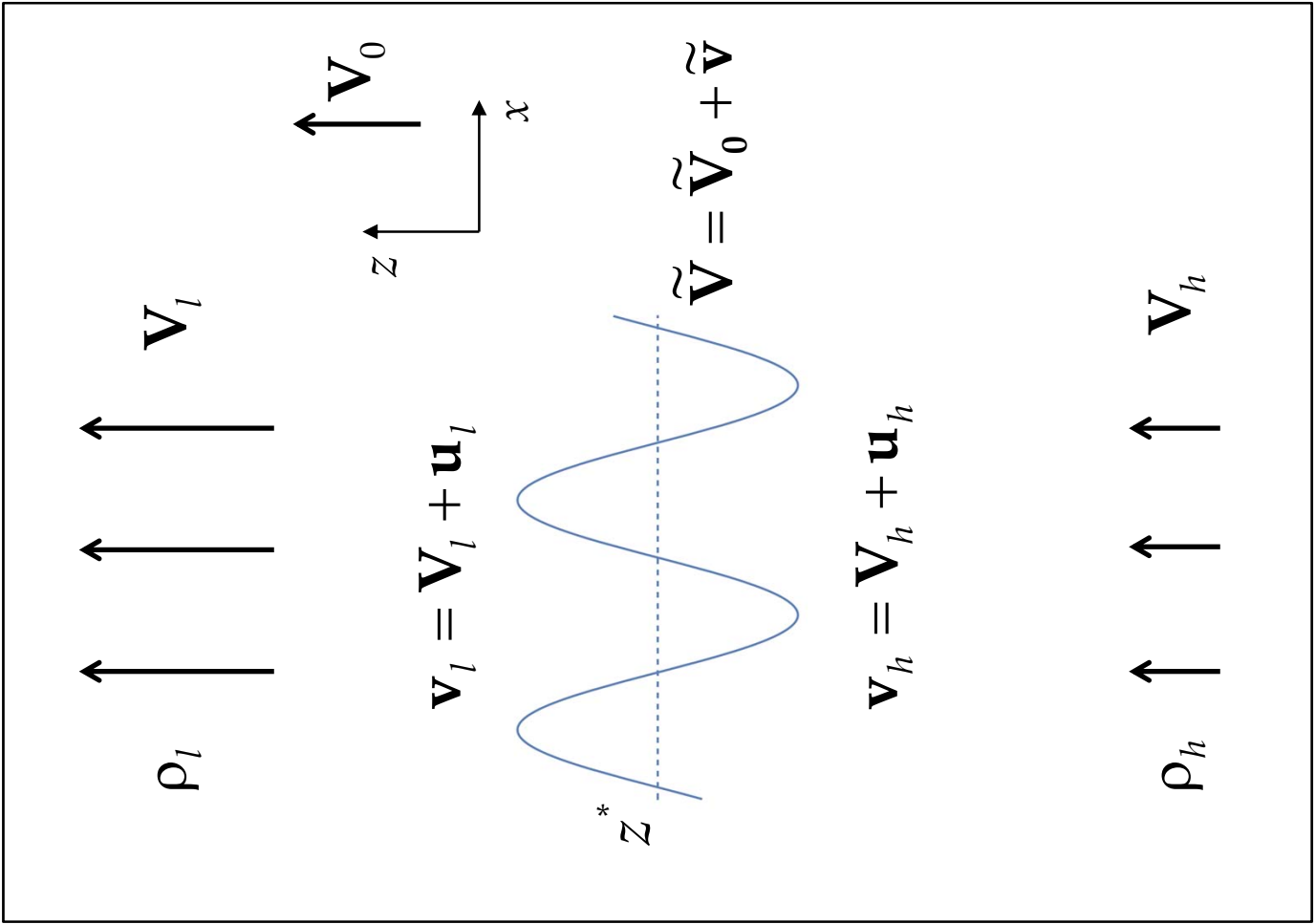

Fig. 1

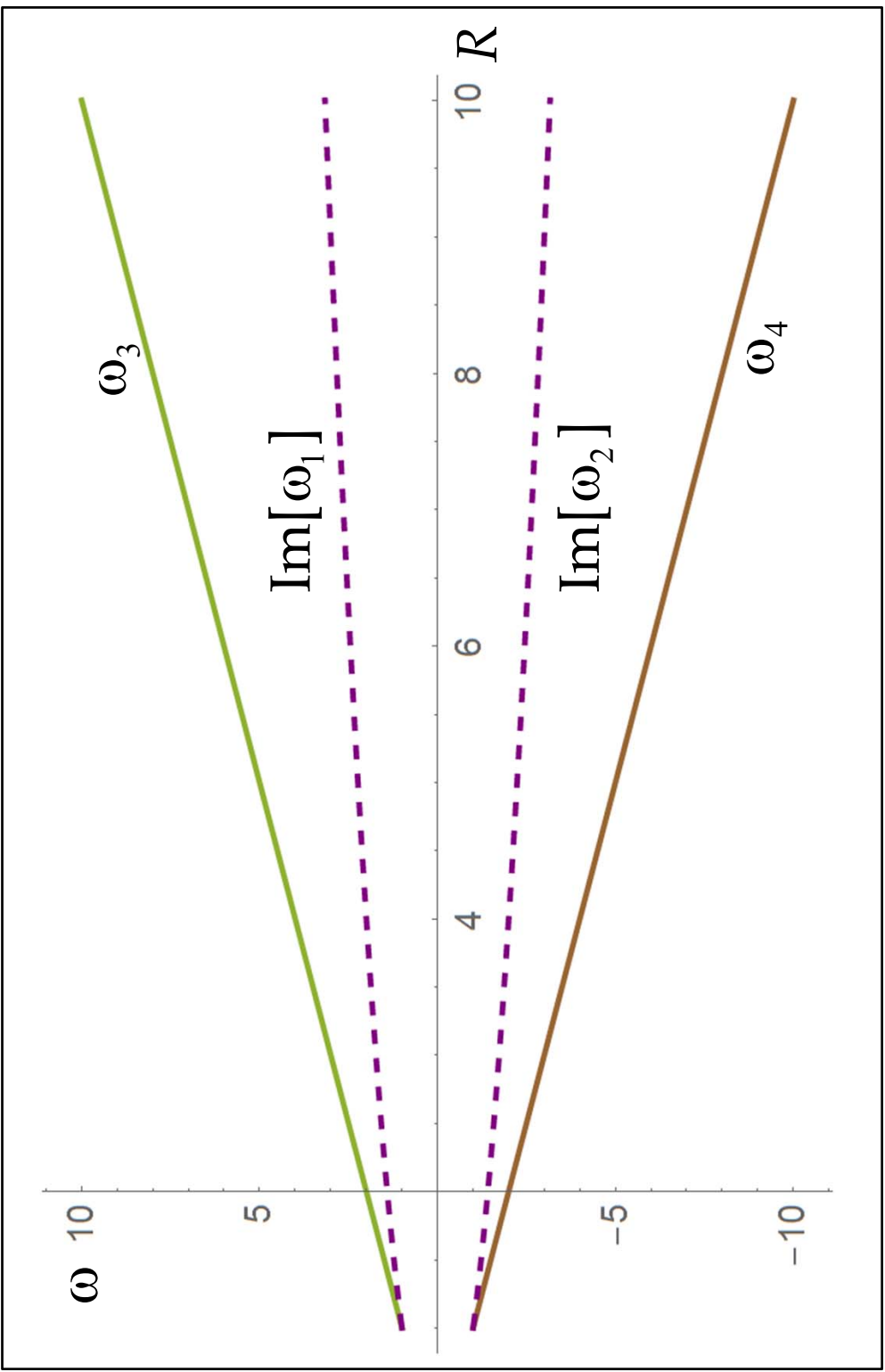

Fig. 2

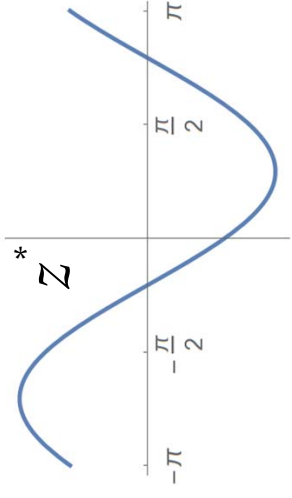
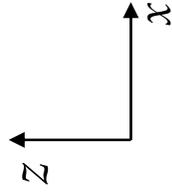
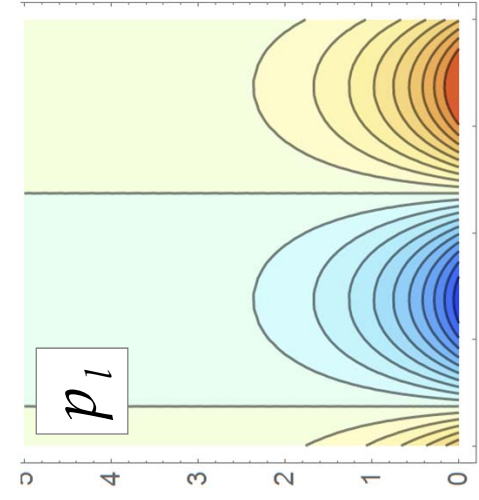
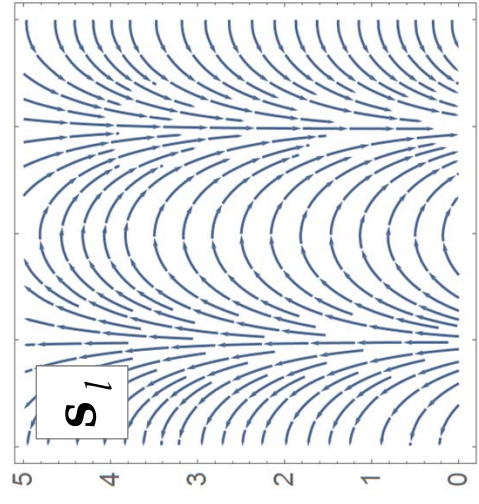
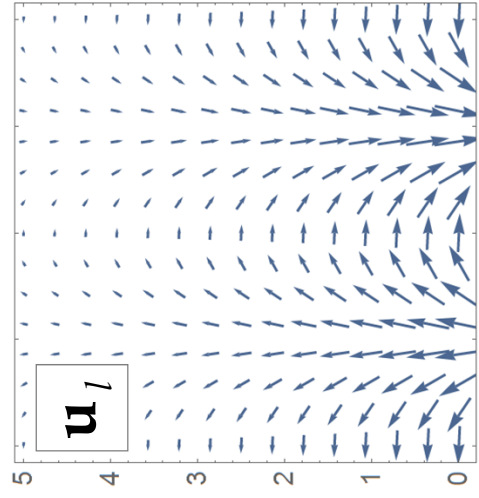
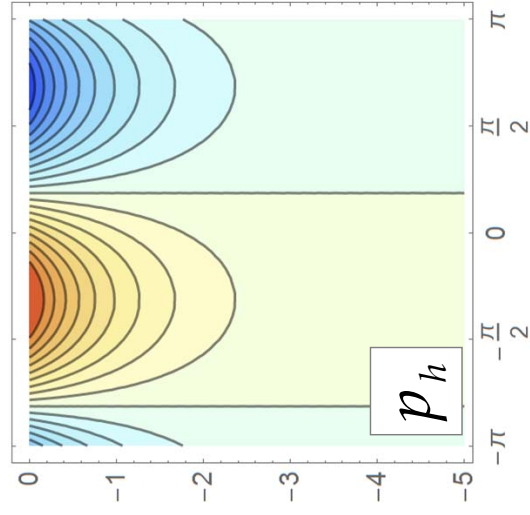
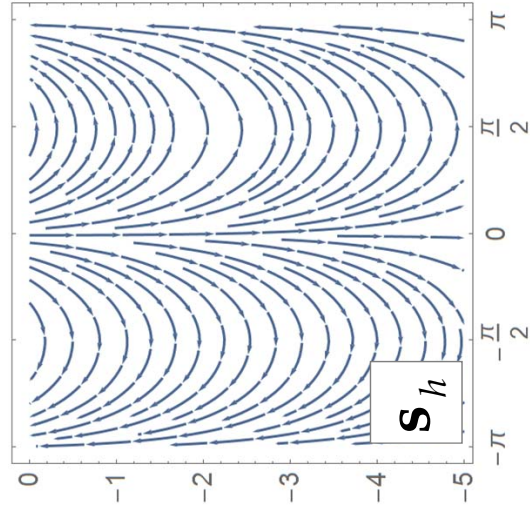
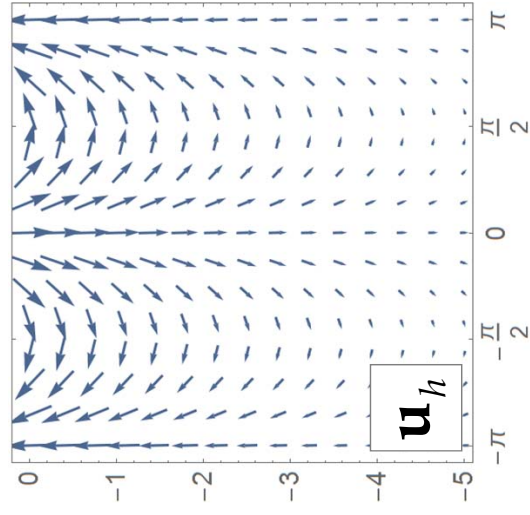

Fig. 3

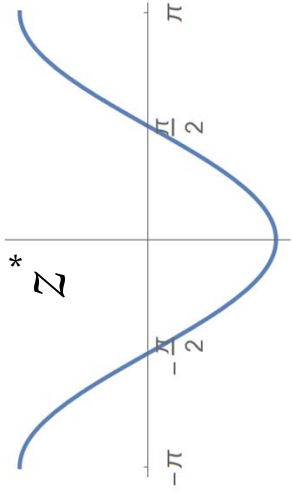

*CD*

$\mathbf{r}_{CD} = (\mathbf{r}_1 + \mathbf{r}_2)/2$

$R = 2$
$t = \pi/2$

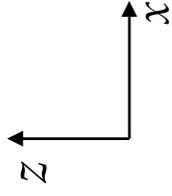

$z^*$

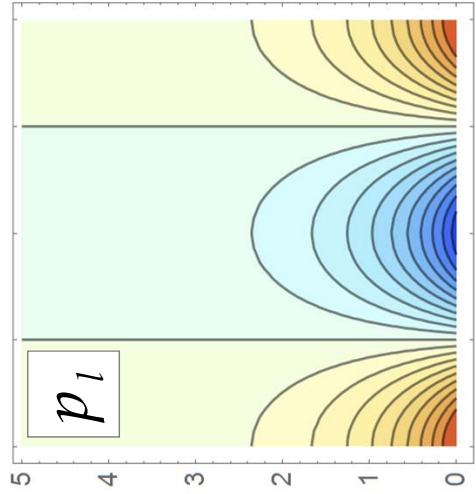

$\mathbf{u}_l$

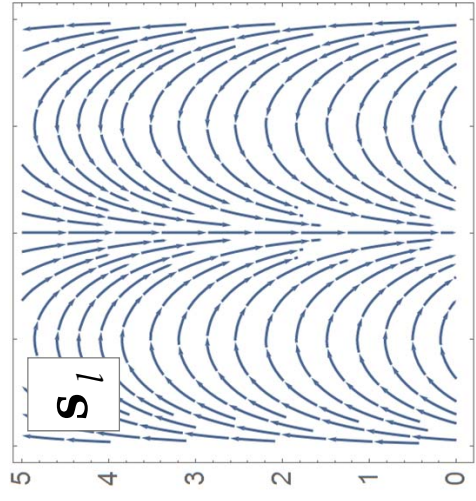

$\mathbf{s}_l$

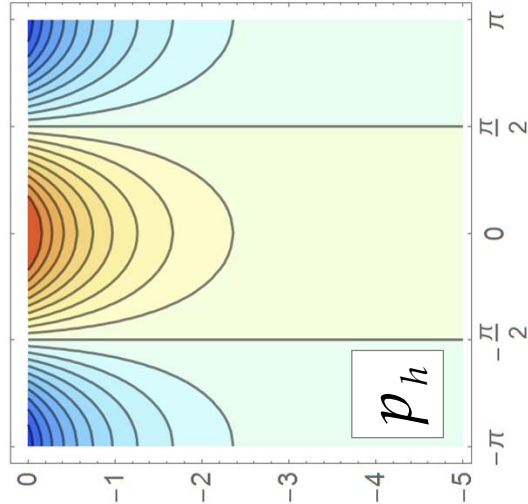

$p_l$

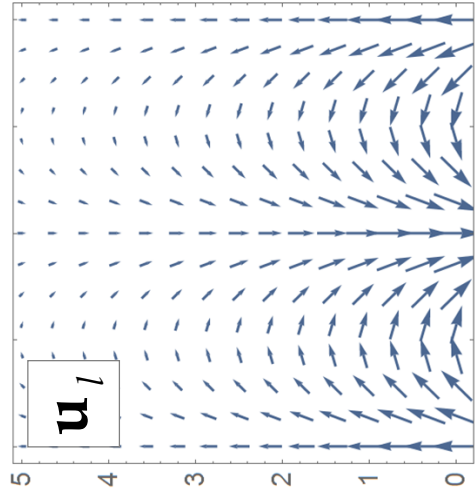

$\mathbf{u}_h$

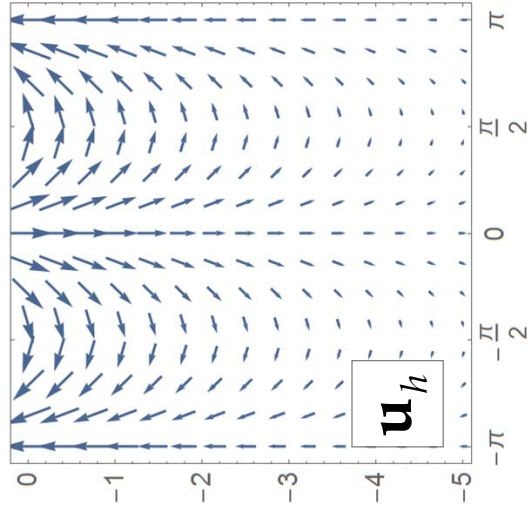

$\mathbf{s}_h$

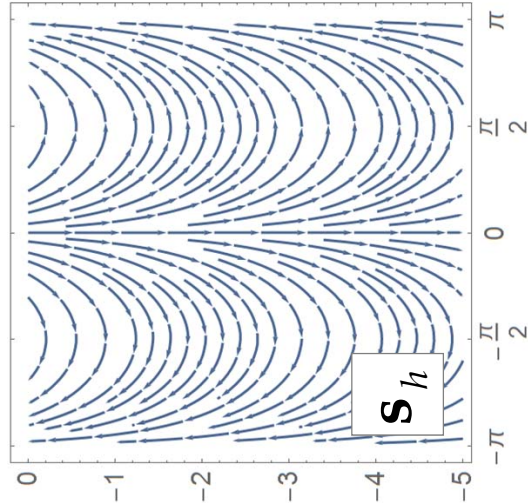

$p_h$

Fig. 4

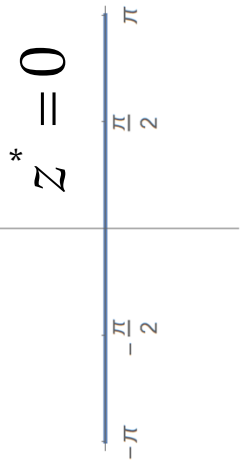

$\mathbf{r}_3(\omega_3, \mathbf{e}_3)$

$R = 2$
$t = \pi/2$

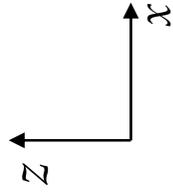

$z^* = 0$

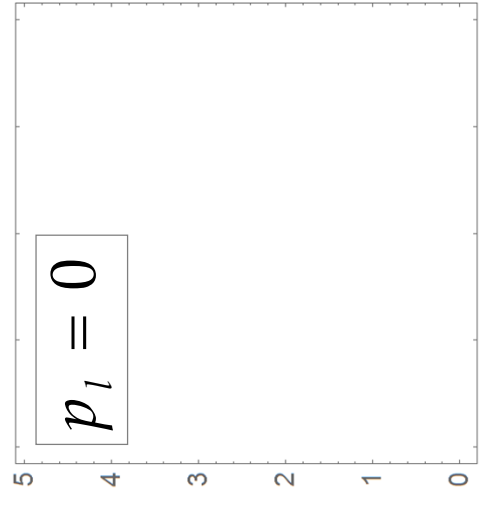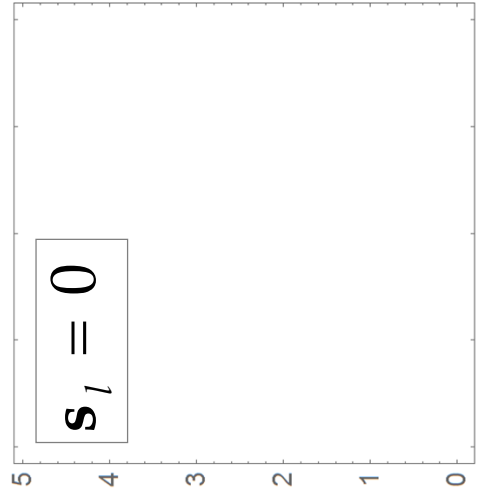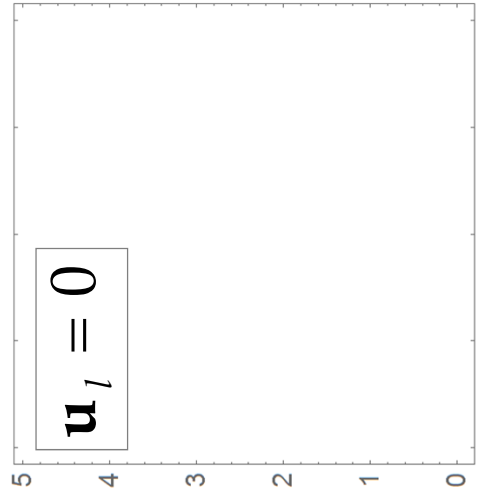

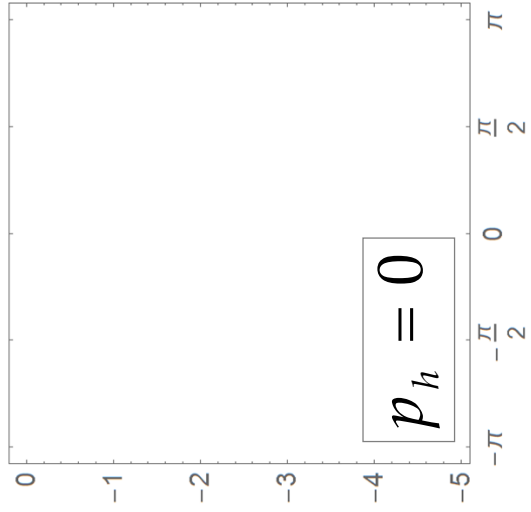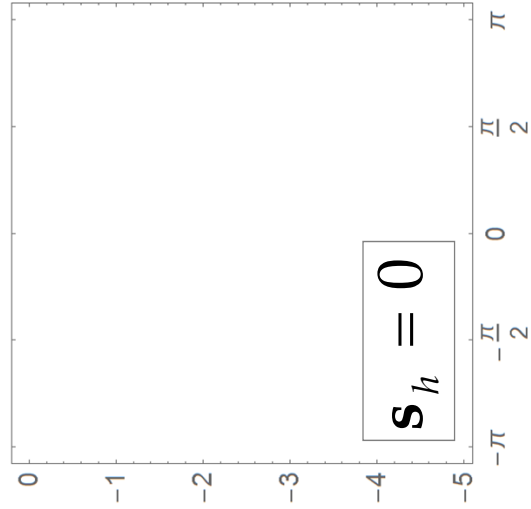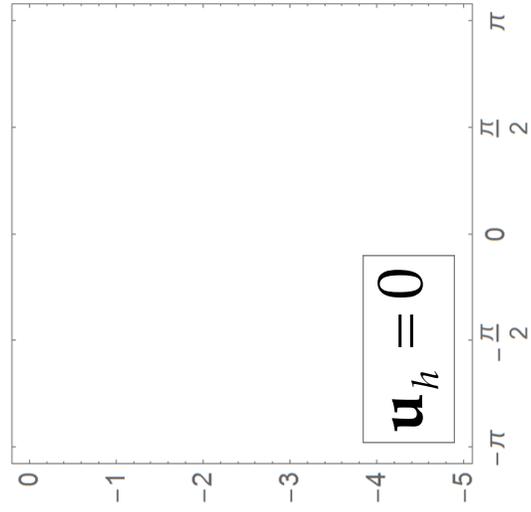

Fig. 5a

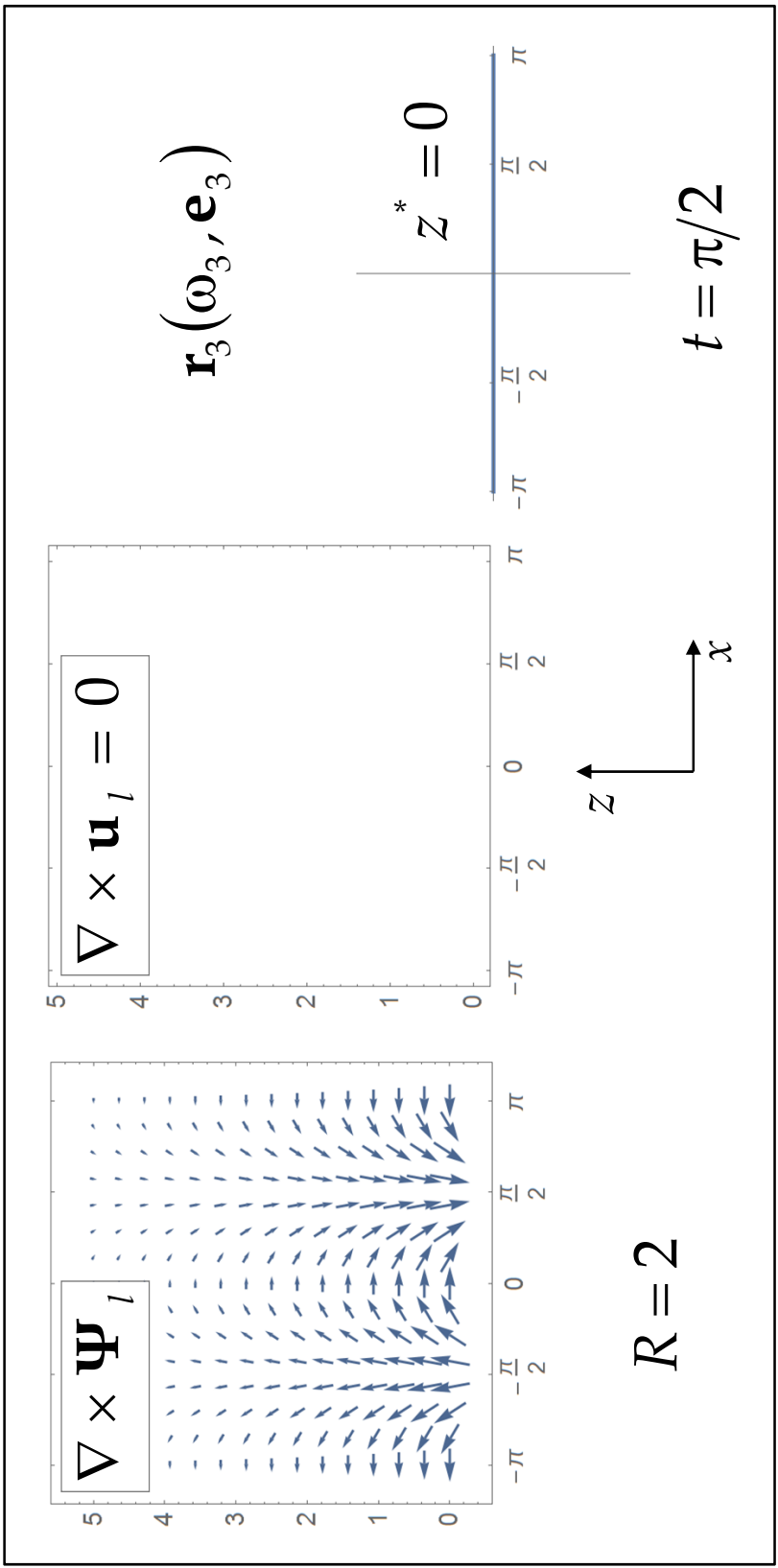

Fig. 5b

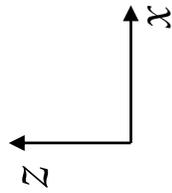
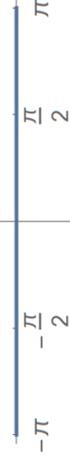
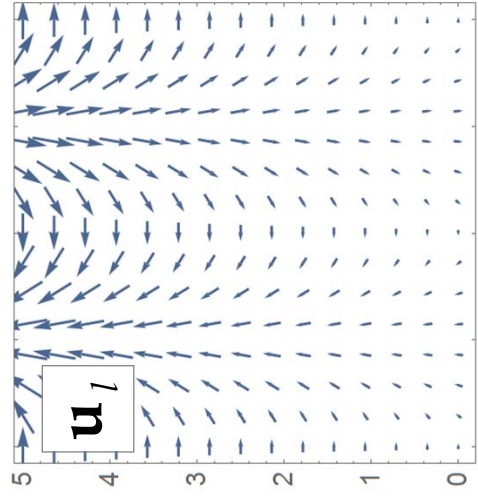
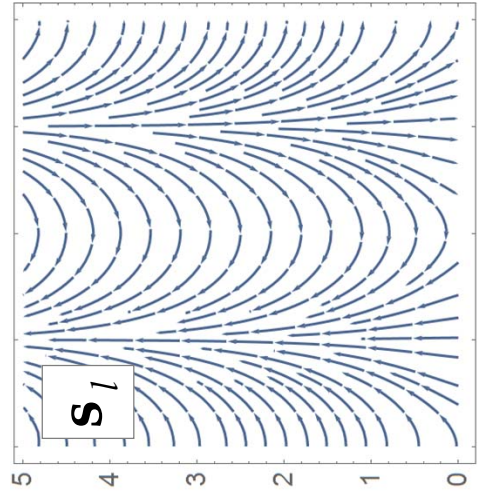
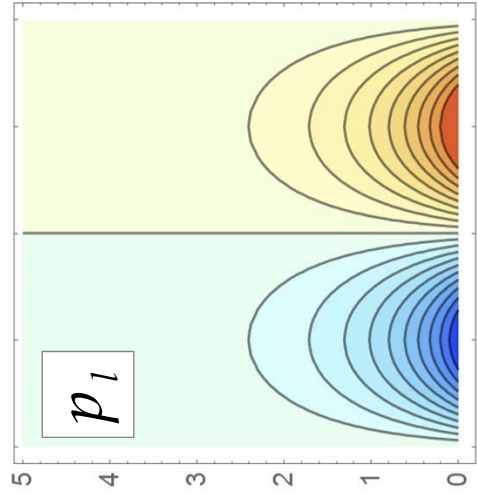
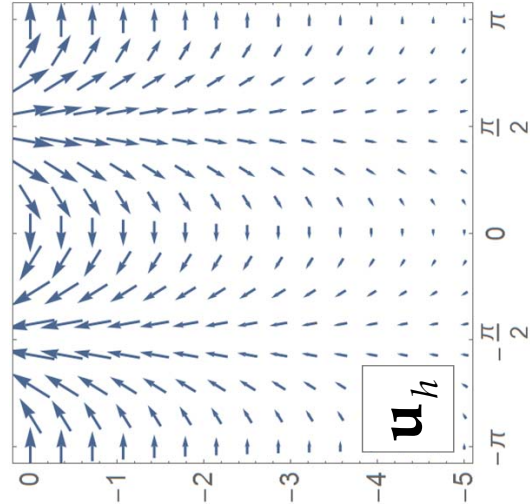
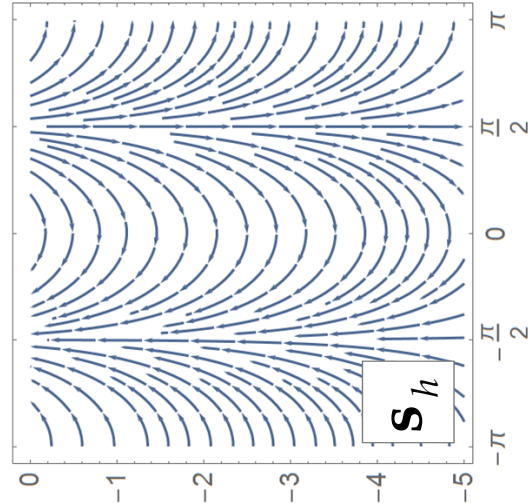
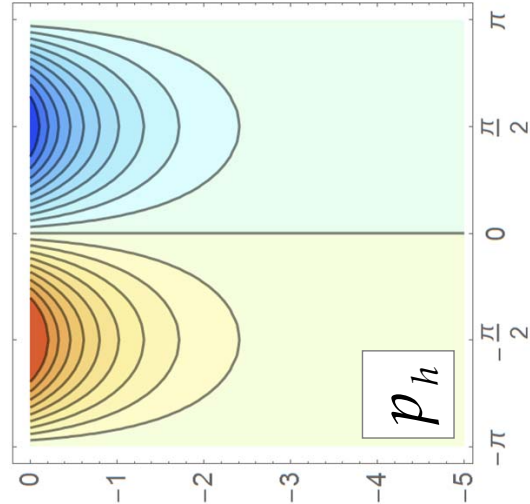

Fig. 6a

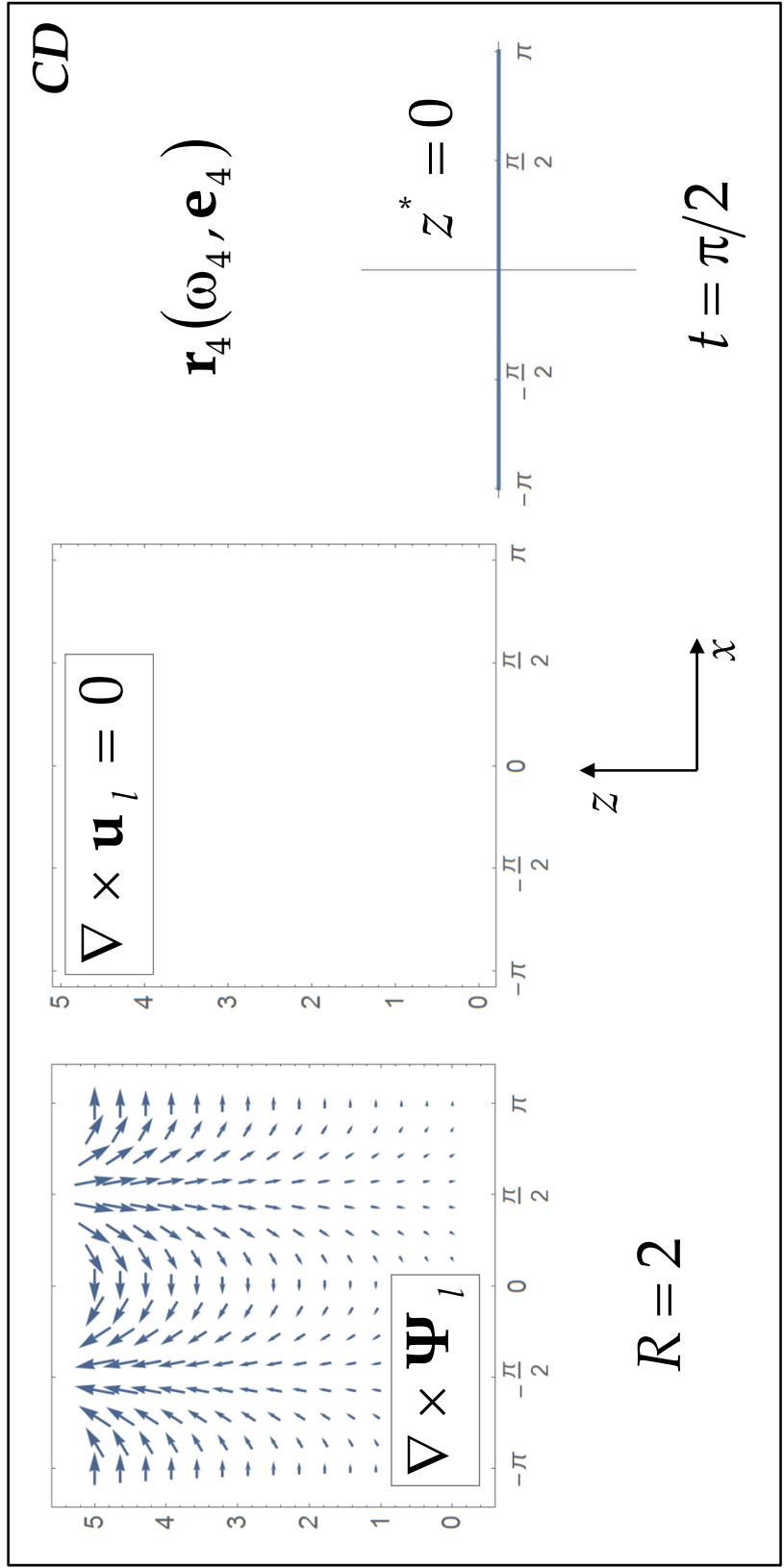

Fig. 6b

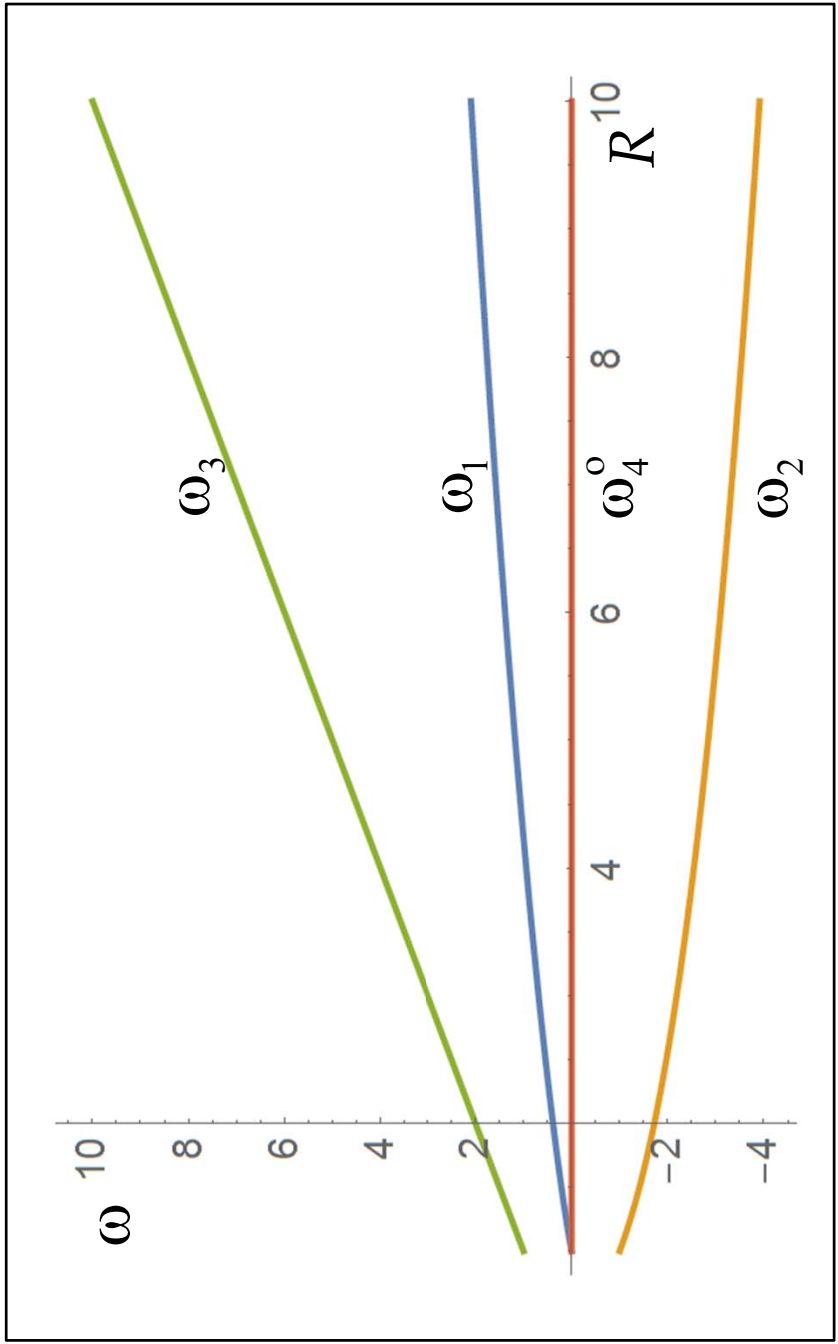

Fig. 7

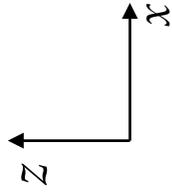
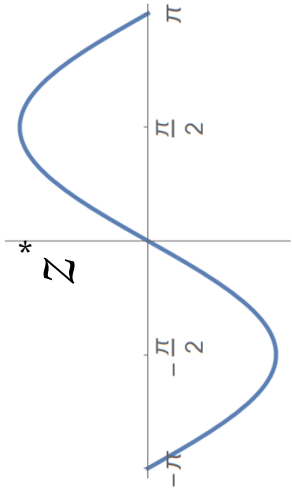
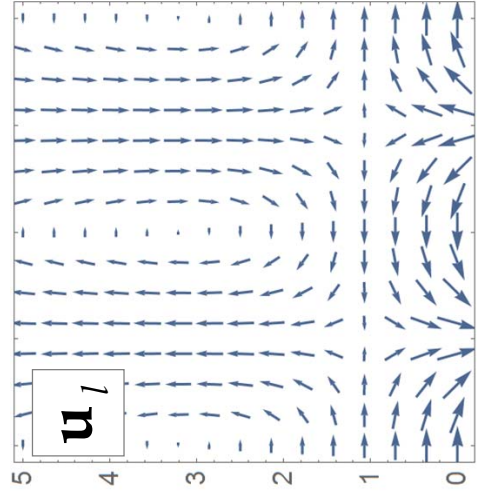
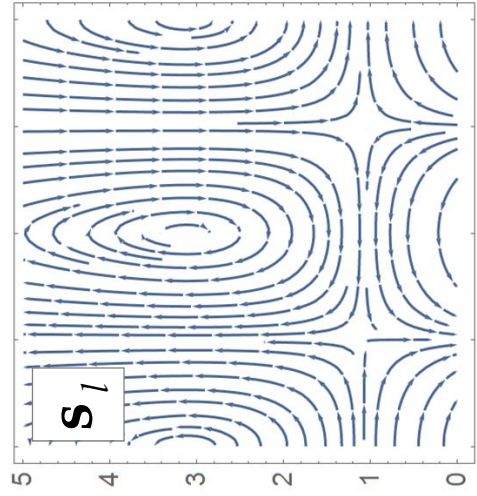
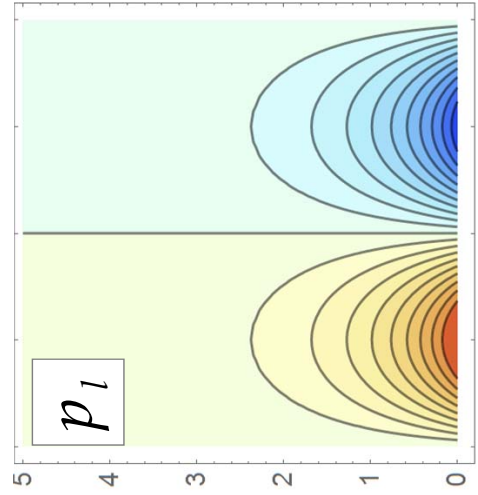
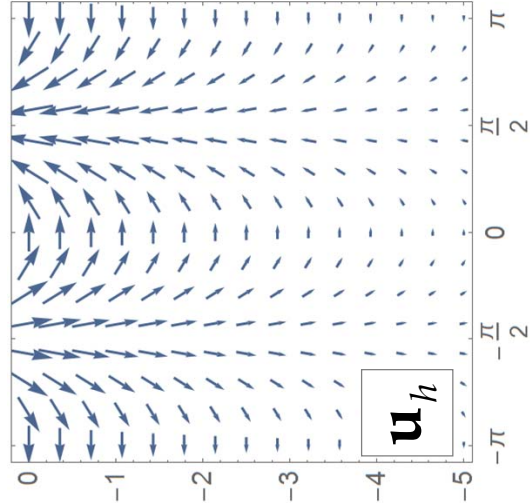
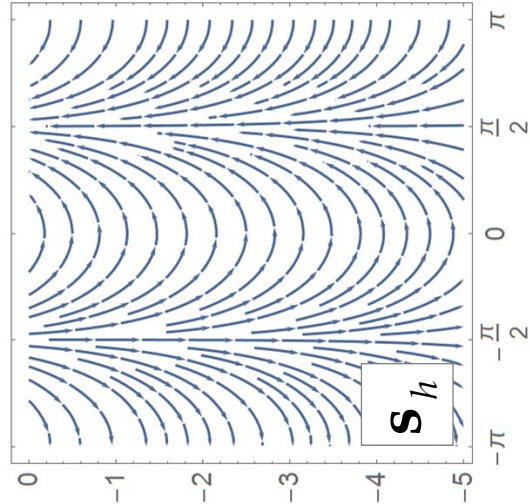
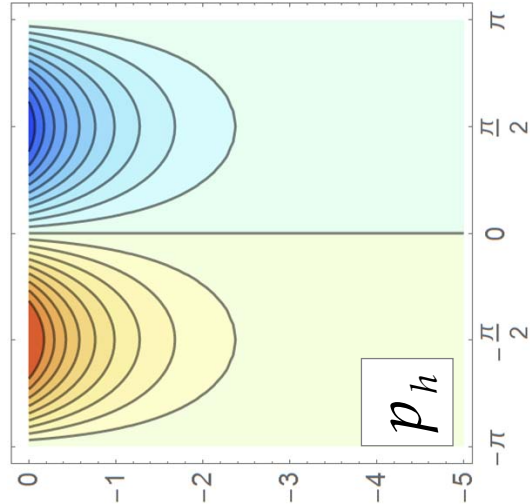

Fig. 8a

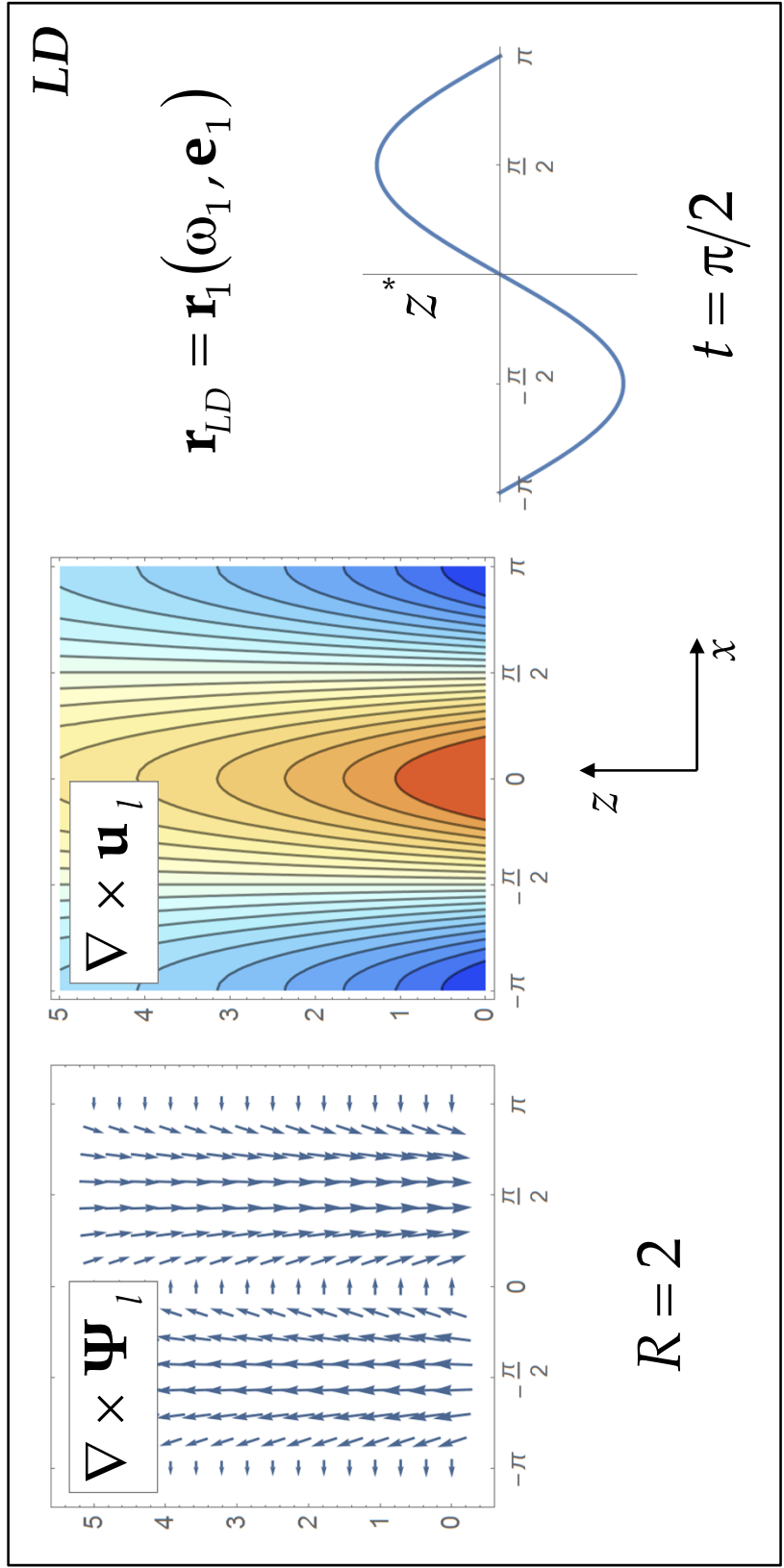

Fig. 8b

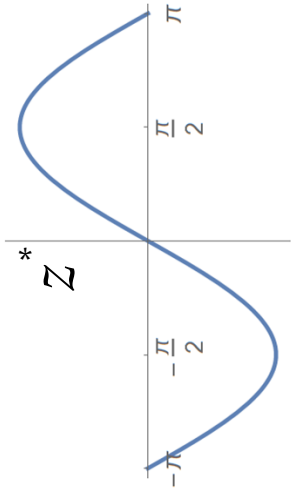
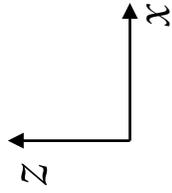

*LD*     $\mathbf{r}_2(\omega_2, \mathbf{e}_2)$     $R = 2$     $t = \pi/2$

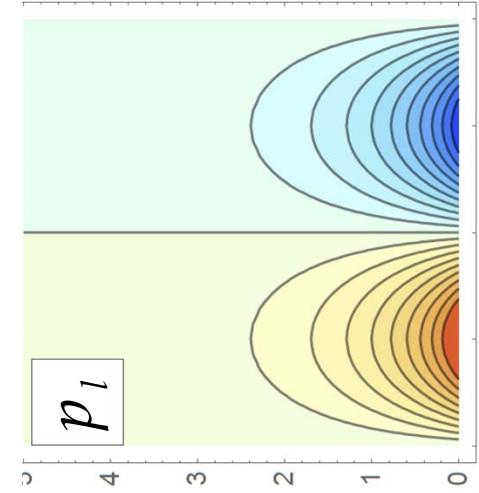
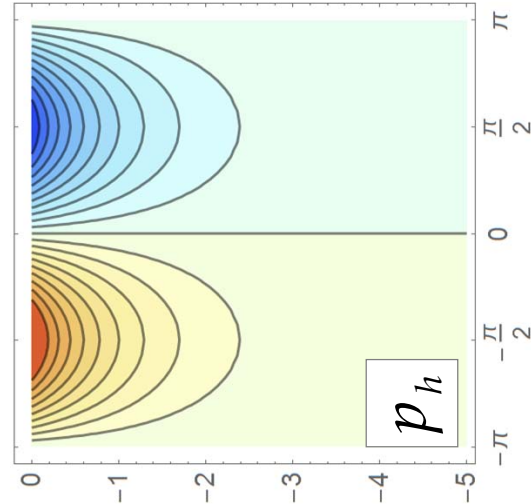
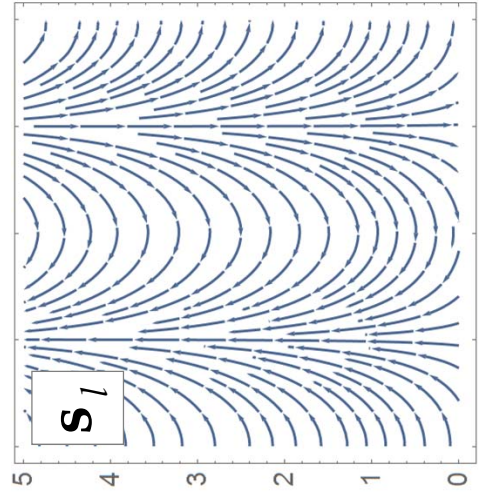
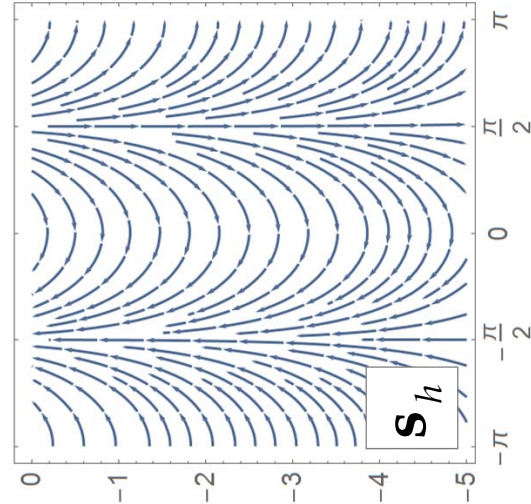
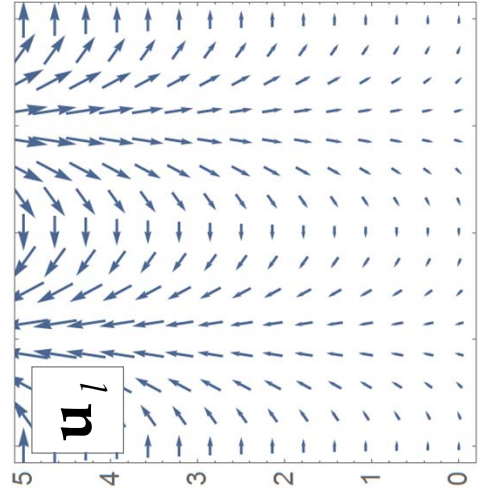
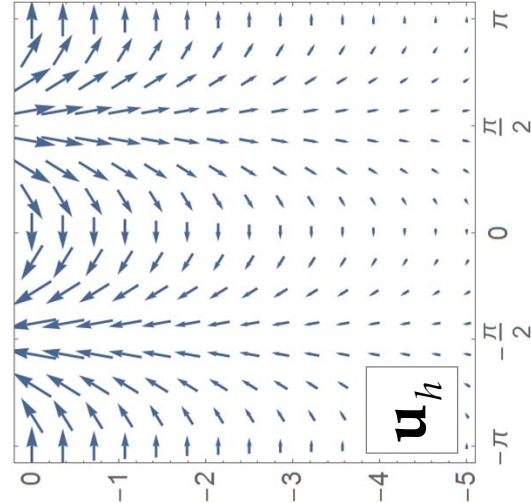

Fig. 9a

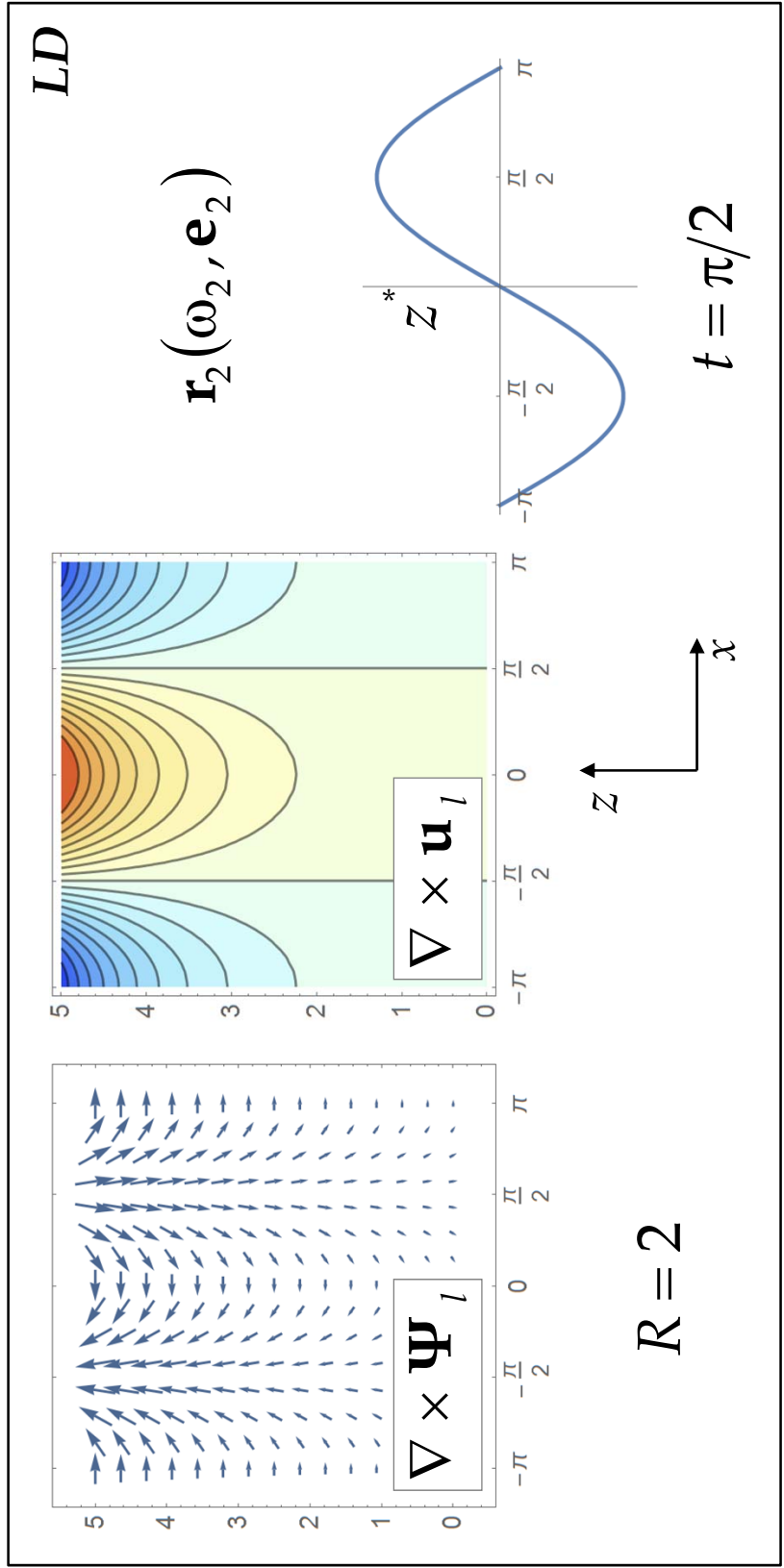

Fig. 9b

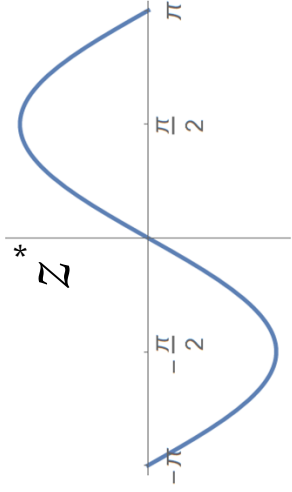
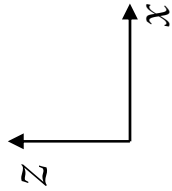
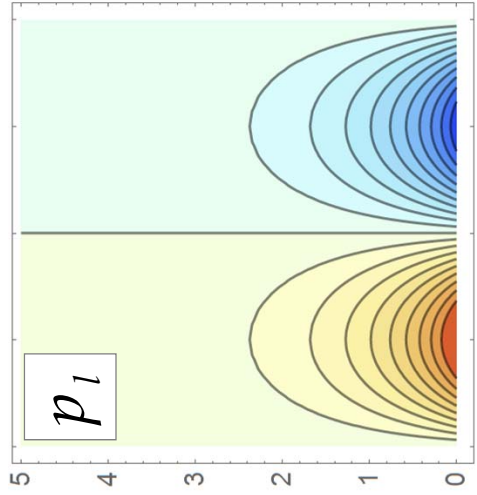
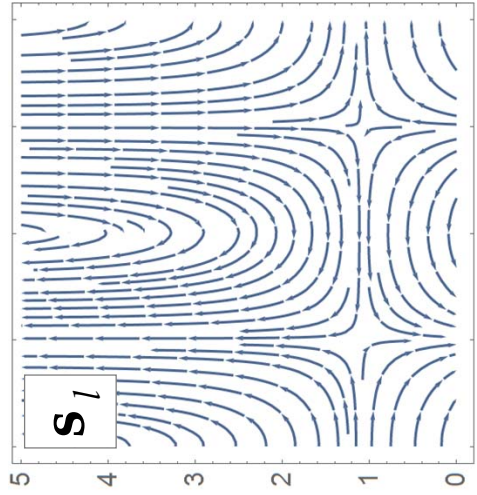
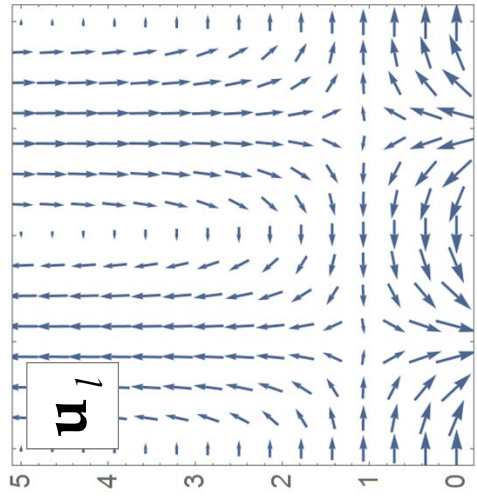
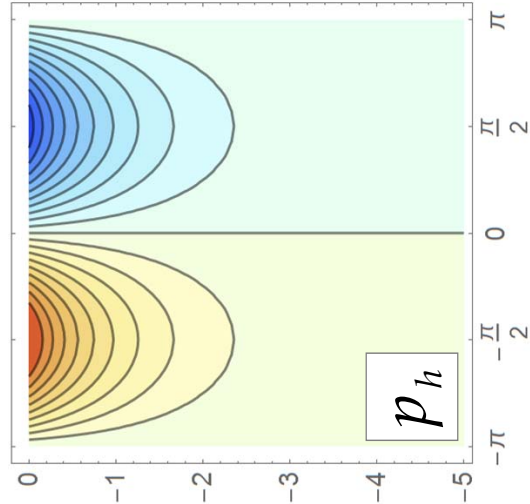
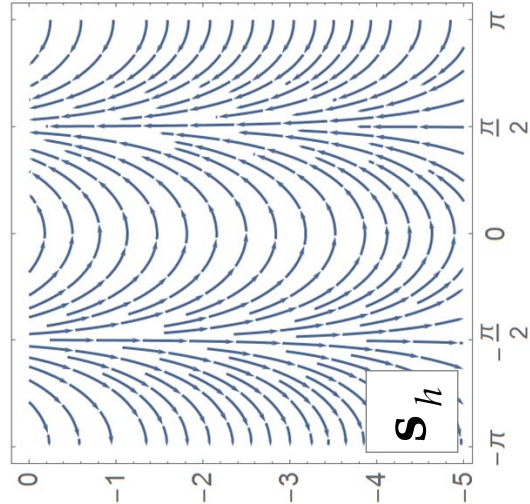
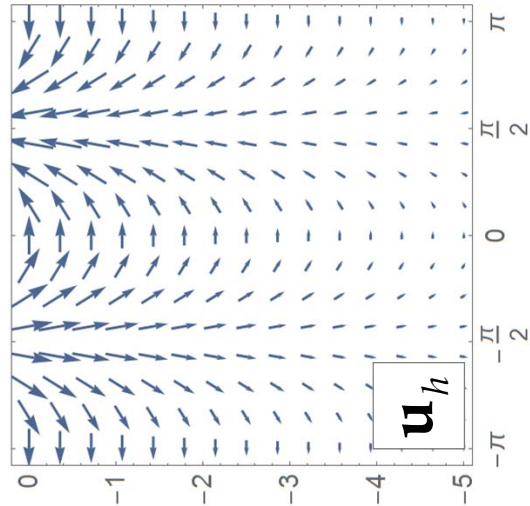

Fig. 10a

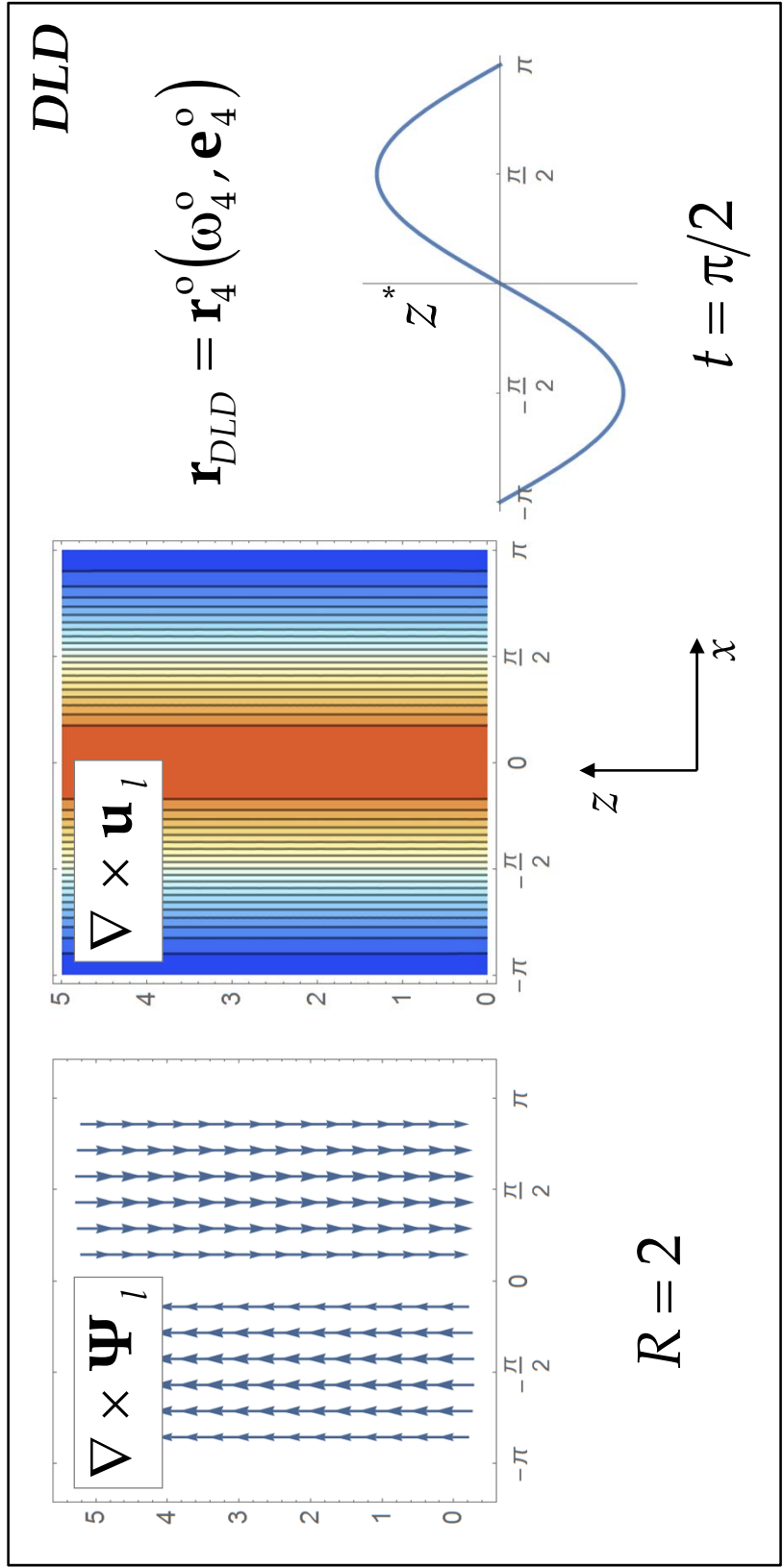
Fig. 10b

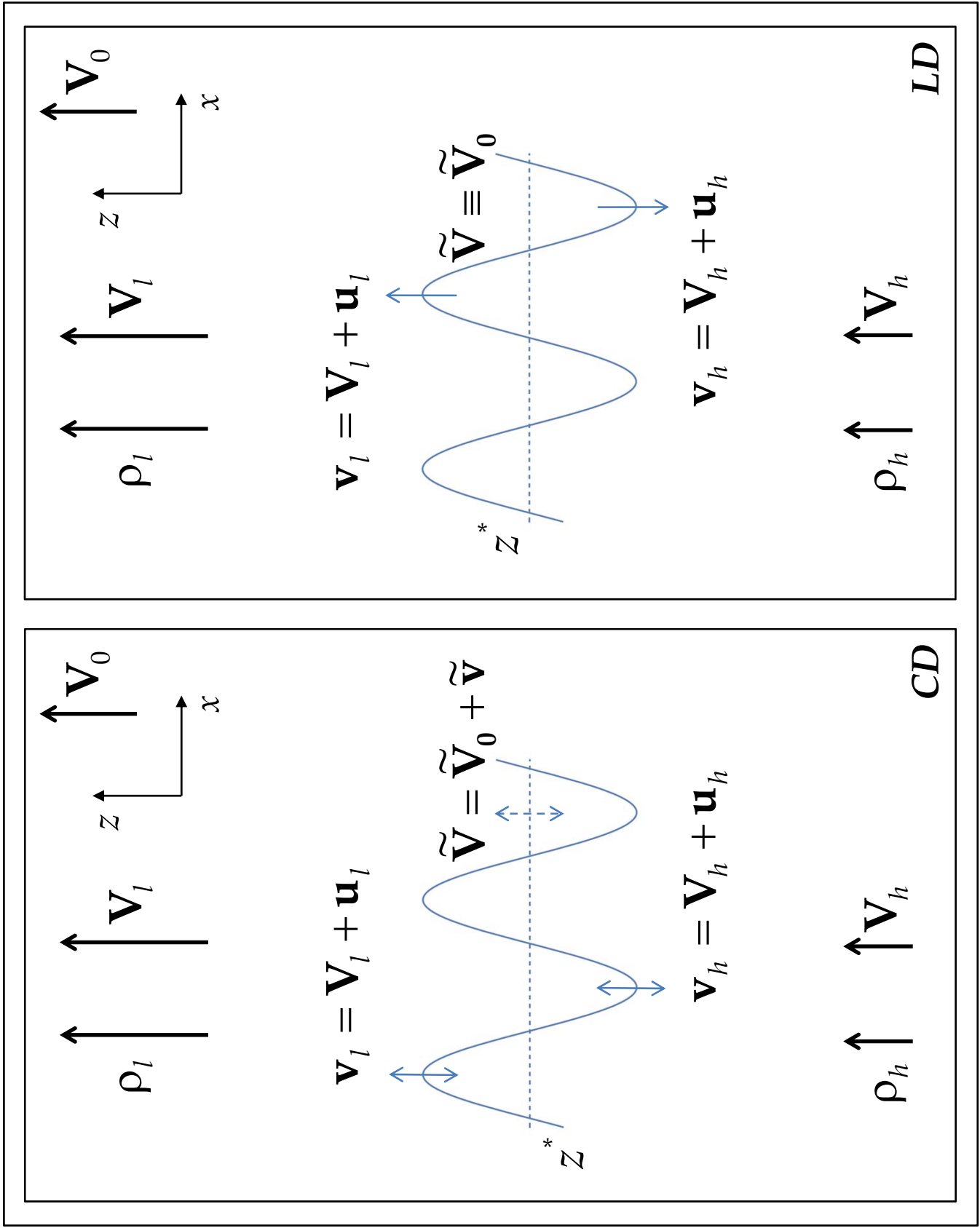

Fig. 11

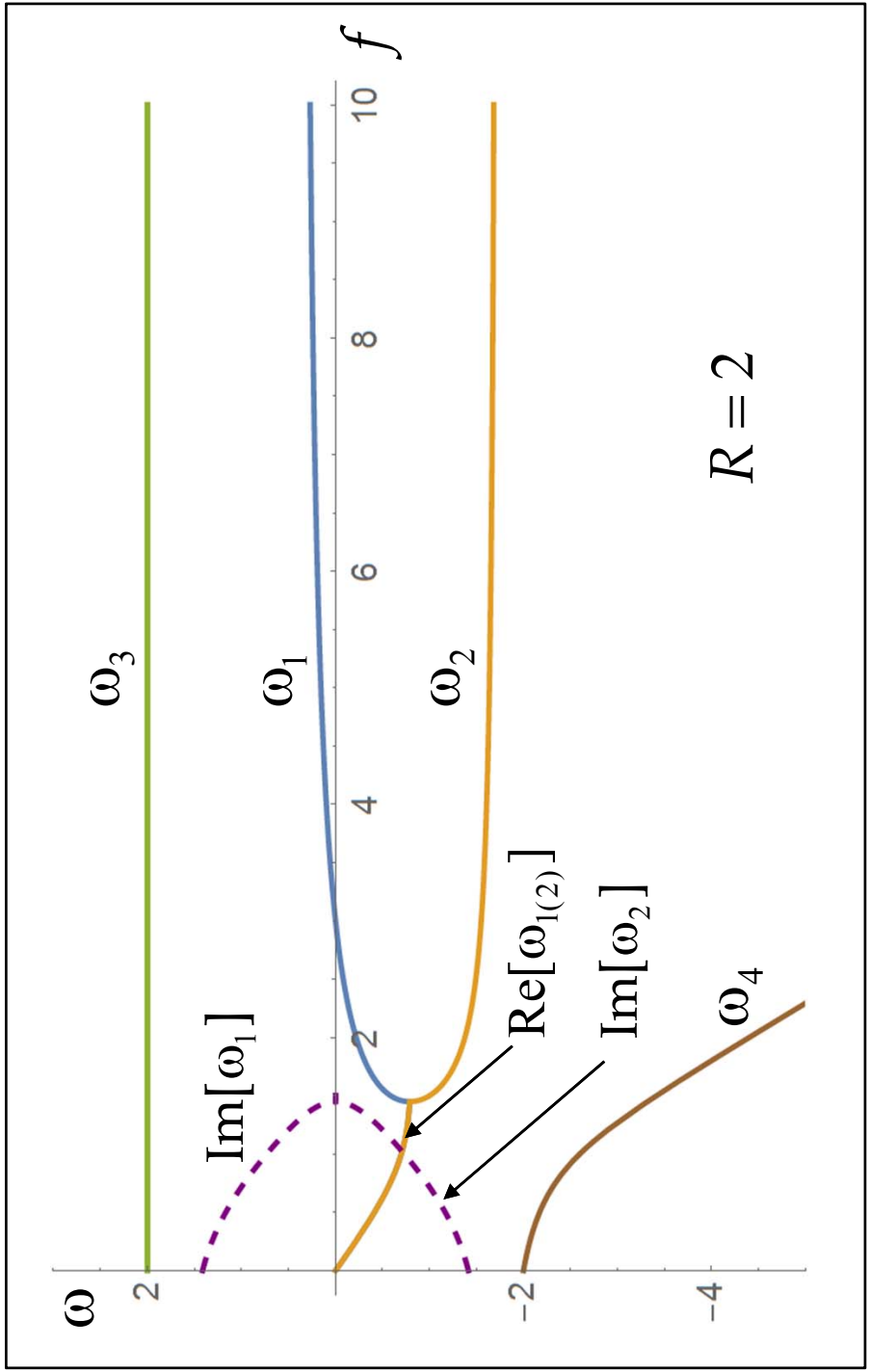

Fig. 12a

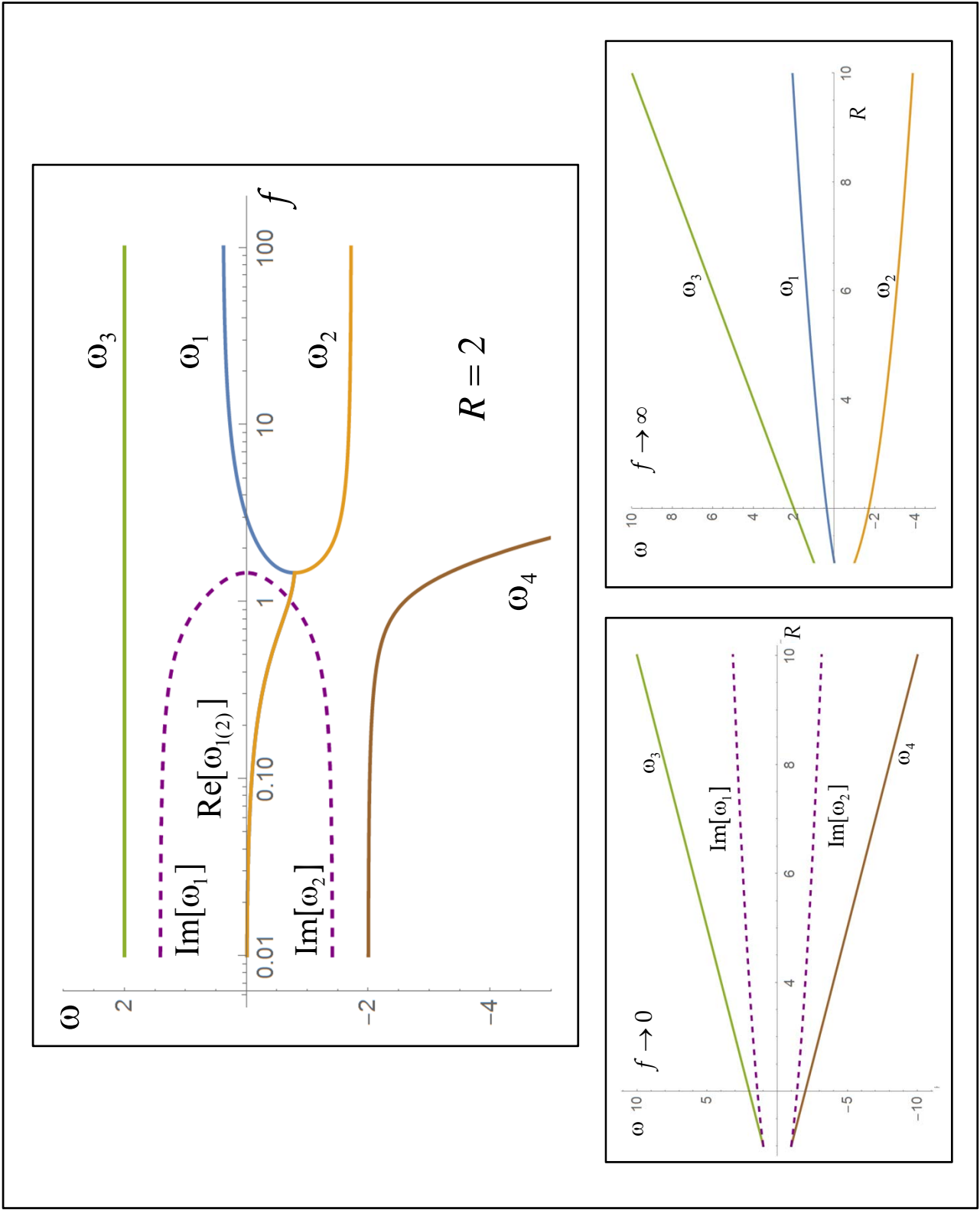

Fig. 12b

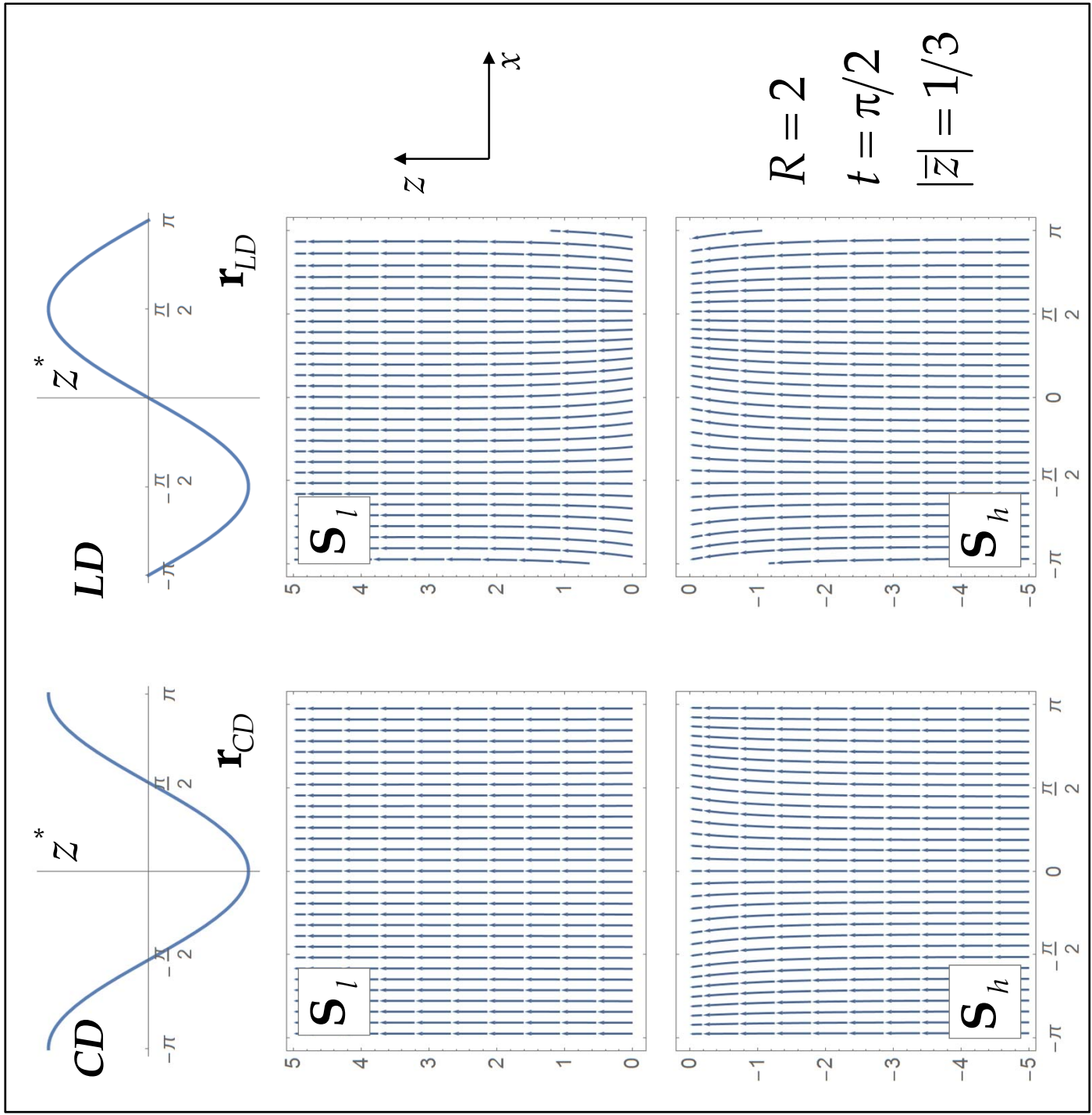

Fig. 13